\documentclass[aps,prd,showpacs,preprintnumbers,nofootinbib,superscriptaddress,groupedaddress,10pt]{revtex4}
\usepackage{graphics}
\usepackage{epsfig}
\usepackage{bm}
\usepackage{appendix}
\providecommand{\abs}[1]{\lvert#1\rvert}

\usepackage{amssymb}
\usepackage{color}
\usepackage{float}
\usepackage{amsmath}
\usepackage[utf8]{inputenc}
\usepackage{dcolumn}

\def\doi{http://doi.org}

\def\be{\begin{eqnarray}}
\def\ee{\end{eqnarray}}

\begin{document}

\title{Quasinormal modes for massive charged scalar fields in Reissner-Nordstr\"om dS black holes: anomalous decay rate}

\author{P. A. Gonz\'{a}lez}
\email{pablo.gonzalez@udp.cl} \affiliation{Facultad de
Ingenier\'{i}a y Ciencias, Universidad Diego Portales, Avenida Ej\'{e}rcito
Libertador 441, Casilla 298-V, Santiago, Chile.}

\author{Eleftherios Papantonopoulos}
\email{lpapa@central.ntua.gr} \affiliation{Physics Division,
National Technical University of Athens, 15780 Zografou Campus,
Athens, Greece.}

\author{Joel Saavedra}
\email{joel.saavedra@ucv.cl} \affiliation{Instituto de
F\'{i}sica, Pontificia Universidad Cat\'olica de Valpara\'{i}so,
Casilla 4950, Valpara\'{i}so, Chile.}

\author{Yerko V\'asquez}
\email{yvasquez@userena.cl}
\affiliation{Departamento de F\'isica, Facultad de Ciencias, Universidad de La Serena,\\
Avenida Cisternas 1200, La Serena, Chile.}

\date{\today }

\begin{abstract}

The anomalous decay rate of the quasinormal modes occurs when the longest-lived modes are the ones with higher angular number. Such behaviour has been recently studied in different static spacetimes, for uncharged scalar and fermionic perturbations, being observed in both cases. In this work we consider the propagation of  charged  massive scalar fields in the background of Reissner-Nordstr\"om-de Sitter black holes and we mainly study the effect of the scalar field charge in the spectrum of quasinormal frequencies, as well as, its effect on the anomalous decay rate. 
Mainly, we show that the anomalous behaviour is present for massive charged scalar fields as well, and a critical value of scalar field mass exists, beyond which the behaviour is inverted. However, there is also a critical value of the parameter $qQ$ of the charge of the scalar field and of the charge of the black hole, which increases when the cosmological constant increases, and beyond the critical value the anomalous behaviour of the decay rate could be avoided for the fundamental mode.

\end{abstract}

%\pacs{04.40.-b, 95.30.Sf, 98.62.Sb}

\maketitle

\flushbottom

\clearpage

\tableofcontents
\newpage

\section{Introduction}

The recent observation of gravitational waves (GWs)~\cite{Abbott:2016blz}-\cite{TheLIGOScientific:2017qsa} produced during the  collision of two compact objects, provided a new understanding of the gravitational interaction and astrophysics in extreme-gravity conditions. It is expected that the detection of the ringdown phase, which is governed by a series of damped oscillatory modes at early times, named quasinormal modes (QNMs) \cite{Regge:1957td, Zerilli:1970wzz, Kokkotas:1999bd, Nollert:1999ji, Berti:2009kk, Konoplya:2011qq},  may potentially contain unexpected  behaviour due to new physics at late times. Thus, the detection of overtones from the ringdown signal allows for precision measurements of the characteristic parameters of compact objects like the mass, charge and angular momentum and it will also  provide vital information on the nature and physics of the near-horizon region of compact objects, like black holes (BHs) and also it will give  information on their stability. Additionally, the spectrum of QNMs is independent of the initial conditions of the perturbation and depends only on the fundamental constants of the system.

Different studies of the quasinormal frequencies (QNFs) found that when a Schwarzschild or Kerr black hole is perturbed by a probe massless scalar field then the longest-lived modes are those with lower angular number $\ell$. This can be understood from the fact that in a physical system  the more energetic modes with higher angular number $\ell$ would have faster decaying rates. However it was found that, if the probe scalar field is massive, then in the case of a light scalar field  the longest-lived QNMs are those with a high angular number $\ell$, whereas for a heavy scalar field the longest-lived modes are those with a low angular number $\ell$ \cite{Konoplya:2004wg, Konoplya:2006br, Dolan:2007mj, Tattersall:2018nve}. Such different behaviour occurs because if the probe scalar field has  small  mass, the fluctuations of the field can maintain the QNMs to live longer even if the angular number $\ell$ is large. This anomalous behaviour is characterized by a critical value of the mass of the scalar field, and this anomalous decay rate for small scalar mass was recently discussed in \cite{Lagos:2020oek}.

The propagation of scalar fields  in the Schwarzschild black hole background in the presence of a positive cosmological constant was studied  in \cite{Aragon:2020tvq} with the aim to investigate  the effect of the  cosmological constant on the anomalous decay of QNMs.  The QNMs in the background of a Schwarzschild de Sitter (dS) black hole are characterized by the photon sphere family of modes (oscillatory modes) and by the dS  family modes consisting of purely imaginary QNFs, for small scalar masses. For the photon sphere modes it was shown that the presence of the cosmological constant leads to the decrease of the real oscillation frequency and to a slower decay when the mass of the scalar field increases for a fixed angular harmonic number. Furthermore, it was shown  the existence of an anomalous decay rate of QNMs, i.e, the absolute values of the imaginary part of the QNFs diminishes when the angular harmonic numbers increases if the mass of the scalar field is smaller than a critical value. The presence of the  cosmological constant introduces a new scale and its effect is to shift the values of the critical masses i.e. when the cosmological constant increases the value of the critical mass also increases.

For the dS  family modes it was shown that for a fixed value of the black hole mass, the purely imaginary QNFs can be dominant depending on the scalar field mass and the angular harmonic number. Also, a faster decay is observed when the $\ell$ parameter increases, as well as, when the scalar field mass increases up to the point where the QNFs acquire a real part and  the decay is stabilized with the frequency of the oscillations to increase as  the scalar field mass increases.  This analysis showed  that the dS family does not present an anomalous behaviour of the QNFs, for a wide range of scalar field masses. In the case that the cosmological constant is negative,  it was found that in Schwarzschild-AdS black holes backgrounds  the decay rate of the QNMs always exhibits an anomalous behaviour independent of the scalar mass. The introduction of  fermionic fields in the background of Schwarzschild-dS black hole shown that there is an anomalous decay rate behaviour  which depends on the chirality \cite{Aragon:2020teq} and also  there is an anomalous decay rate for accelerating black holes \cite{Destounis:2020pjk}.

The study of the anomalous decay of QNMs was extended to $f(R)$ modified gravity in \cite{Aragon:2020xtm} where black hole solutions in  asymptotically de Sitter $f(R)$ modified gravity theories were considered as the background. The motivation of \cite{Aragon:2020xtm} was to investigate the competing effects of the cosmological term and the non-linear curvature terms which appear in $f(R)$ function, on the behaviour of QNMs and QNFs generated by massless or massive scalar field perturbations. These non-linear curvature terms were characterized by a parameter $\beta$ which  appears in the metric function of the resulted black hole  solution. For small $\beta$, i.e small deviations from the Schwarzschild-dS black hole, the anomalous
behaviour in the QNMs is present and the critical value of the mass
of the scalar field depends on the parameter $\beta$  while for large $\beta$, i.e large deviations the anomalous
behaviour and the critical mass does not appear.

The anomalous decay rate of the QNMs was also studied in  charged spacetimes, in the presence of Reissner-Nordstr\"om-dS (RNdS) and Reissner-Nordstr\"om-AdS (RNAdS) black holes. It was shown in \cite{Fontana:2020syy} that the anomalous decay rate behaviour of the scalar field perturbations is present for every charged
geometry in the photon sphere modes, with the existence of a critical scalar field mass whenever
$\Lambda  \geq 0$. In general, this critical value of mass increases with the raise of the black hole charge. Also the dominant mode/family for the
massless and massive scalar field in these geometries were studied in \cite{Fontana:2020syy}  showing a non-trivial dominance of the spectra
that depends on the black hole mass and charge.

 From the above discussion we can point out that in the Anti-de Sitter (AdS) spacetime the neutral or charged black holes  when they are perturbed by massive scalar fields, the  decay rate of the resulting QNMs have an anomalous behaviour. However, in this spacetime there is no a critical scalar field mass which signals the anomalous behaviour of the QNMs decay rate. In a model in (2+1)-dimensional Coulomb like AdS black holes from non-lineal electrodynamics \cite{Aragon:2021ogo} it was shown that there is no any anomalous decay rate, i.e the longest-lived modes are the ones with smallest angular number independent of the value of the scalar field mass. Then one may think that in an AdS spacetime background, to have an anomalous behaviour of the decay rate of QNMs we have to consider a spacetime with dimensions greater than three.

Recently there were many studies of the decay rate of QNMs of the perturbations of massless scalar fields in the exterior of RNdS black holes. These studies were connected with the validity of the strong cosmic censorship (SCC) conjecture \cite{Penrose:1964wq,Penrose:1969pc}. In \cite{Cardoso:2017soq} three families of modes were identified, the photon sphere family, the dS family whose existence and time scale is closely related to the dS horizon, and a third family which dominates for near-extremally-charged black holes and which is also present in asymptotically flat spacetimes. Also, it was established the dominant mode/family.  Then, massive scalar fields were considered and the dominant mode/family was established in \cite{Fontana:2020syy}.

In this work we consider the propagation of charged massive scalar fields in the background of RNdS black holes. Our main motivation is to study the influence of the scalar field charge on each family of modes, discuss  the dominant mode/family and see the effects of the anomalous decay rate of QNMs on each family of modes. Perturbations of charged massive scalar fields in the background of RNdS black holes were also studied in \cite{Cardoso:2018nvb} in the context of SCC. We will also study the superradiance effect, which is known to generate an instability of RNdS black holes under massive charged scalar perturbations with vanishing angular momentum, $\ell= 0$ \cite{Zhu:2014sya} and recently it was also studied in \cite{Destounis:2019hca}.

 We will show that the introduction of a scalar field charge modifies the spectrum of the QNFs, while  the purely imaginary modes (dS and near extremal (NE) modes) acquire a real part, while for the complex ones  two branches  for the modes appear, with a frequency of oscillation negative (negative branch) and positive (positive branch). We will also show that the anomalous decay rate behaviour is possible for massive charged scalar field, and there exists a critical value of the scalar field mass,  beyond which the behaviour of the decay rate is inverted. However, there is a critical value of the parameter $qQ$ which is increasing  when the cosmological constant is increasing. Beyond this critical value the anomalous behaviour of the decay rate could be avoided for the fundamental mode (negative branch).

The work is organized as follows. In Section \ref{QNM} we derive the effective potential of the charged scalar perturbations of the RNdS black hole. In
in Section \ref{QM} we compute the QNMs for charged scalar fields in a RNdS black hole and we study their spectrum, the superradiant modes, and the anomalous behaviour of the decay rate. Our concluding remarks are in  Section \ref{conclusion}.

\section{Charged scalar perturbations}
\label{QNM}
%%%%%%%%%%%%%%%%%%%%%%%%%%%%%%%%%%%%%%%%%%%%%%%%%%%%%%%%5

The RNdS black hole
 is the
 solution of the equations of motion that arise from the Einstein-Hilbert action with a positive cosmological constant
\begin{equation}
    S=\frac{1}{16\pi G}\int d^4x\sqrt{-g}(R-2\Lambda+ F_{\mu\nu}F^{\mu\nu})\,,
\end{equation}
where $G$ is the Newton constant, $R$ is the Ricci scalar, $\Lambda$ is the cosmological constant, and $F_{\mu \nu} F^{\mu \nu}$ represents the electromagnetic Lagrangian. The RNdS black hole is described by the metric
\begin{equation}
    ds^2=-f(r)dt^2+\frac{dr^2}{f(r)}+r^2 d\Omega^2 \,,
    \label{metric}
\end{equation}
where $d \Omega^2= d\theta^2+sin^2\theta d\phi^2$ , $f(r)=1-\frac{2M}{r}+\frac{Q^{2}}{r^{2}}-\frac{\Lambda r^2}{3}$, $M$ is the black hole mass, $Q$ the electric charge and $\Lambda>0$.

In order to study charged scalar perturbations in the background of  the metric (\ref{metric})
we consider the Klein-Gordon equation for charged massive scalar fields
\begin{equation}
\label{KGE}
\frac{1}{\sqrt{-g}}(\partial_{\mu}-iqA_{\mu})(\sqrt{-g}g^{\mu\nu}(\partial_{\nu}-iqA_{\nu})\psi)=m^2\psi ,
\end{equation}
plus suitable boundary conditions for a black hole geometry. In the above expression $m$ represents the mass and $q$ the charge of the scalar field $\psi$. Due to the spherical symmetry, the Klein-Gordon equation can be written as
\begin{equation}
\frac{d}{dr}\left(r^2 f(r)\frac{dR}{dr}\right)+\left(\frac{r^2(\omega+qA_t(r))^2}{f(r)}-\kappa^2-m^{2}r^2 \right) R(r)=0\,, \label{radial}
\end{equation}%
by means of the ansatz $\psi =e^{-i\omega t} Y(\Omega) R(r)$, where $-\kappa^2=-\ell(\ell+1)$, with $\ell = 0,1,2,..$, represents the eigenvalue of the Laplacian on the two-sphere, and $A_t=-\frac{Q}{r}$. Then, redefining $R(r)$ as $R(r)=\frac{F(r)}{r}$, and by using the tortoise coordinate $r^*$ given by $dr^*=\frac{dr}{f(r)}$, the Klein-Gordon equation can be written as
\begin{equation}
 \label{ggg}
 \frac{d^{2}F(r^*)}{dr^{*2}}-V_{eff}(r)F(r^*)=-\omega^{2}F(r^*)\,,
 \end{equation}
that corresponds to a one-dimensional Schr\"{o}dinger-like equation
 with an effective potential $V_{eff}(r)$, which is parametrically described by an effective potential $V_{eff}(r^*)$, given by
 \begin{widetext}
  \begin{eqnarray}\label{pot}
 \nonumber V_{eff}(r)&=&\frac{f(r)}{r^2} \left(\kappa^2 +  m^2r^2+ f^\prime(r)r \right) -2\omega q A_t(r)-q^2A_t(r)^2~.\\
 \end{eqnarray}
 \end{widetext}

\section{Quasinormal modes}
\label{QM}
%%%%%%%%%%%%%%%%%%%%%%%%%%%%%%%%%%%%%%%%%%%%%%%%
There are several methods for calculating the QNFs, like the well known Mashhoon method, Chandrasekhar-Detweiler method, Wentzel-Kramers-Brillouin (WKB) method, Frobenius method, method of continued fractions, (with the Nollert improvements), asymptotic iteration method (AIM) and improved AIM. For an extensive review, see \cite{Konoplya:2011qq}. Here, in order to solve numerically the differential equation (\ref{radial}) we consider the pseudospectral Chebyshev method \cite{Boyd}. Firstly, it is convenient to perform the change of variable  $y=(r-r_H)/(r_{\Lambda}-r_H)$ in order to bound the value of the radial coordinate to the range $[0,1]$, and the radial equation (\ref{radial}) becomes
\begin{equation} \label{r}
f(y) R''(y) + \left( \frac{2 \left(    r_{\Lambda}- r_H  \right) f(y)}{r_H+\left( r_{\Lambda}-r_H \right) y } + f'(y) \right) R'(y)+ \left( r_{\Lambda}-r_H  \right)^2 \left( \frac{(\omega+qA_t(y))^2}{f(y)}- \frac{ \ell(\ell+1)}{\left( r_H + \left( r_{\Lambda}-r_H \right)y \right)^2} -m^2  \right) R(y)=0\,,
\end{equation}
where the prime means derivative with respect to $y$. Now, the event horizon is located at $y=0$ and the spatial infinity at $y=1$. So, in order to propose an ansatz for the field, we analyze the behaviour of the differential equation at the horizon and at infinity. In the neighborhood of the horizon the function $R(y)$ behaves as
\begin{equation}
\label{horizon}
R(y)=C_1 y^{-\frac{i (\omega+qA_t(0)) (r_{\Lambda}-r_H)}{f'(0)}}+C_2 y^{\frac{i (\omega+qA_t(0)) (r_{\Lambda}-r_H)}{f'(0)}} \,,
\end{equation}
where: i) the first term represents an ingoing wave, and ii) the second term represents an outgoing wave near the black hole horizon. So, imposing the requirement of only ingoing waves on the horizon, we fix $C_2=0$. On the other hand, at the cosmological horizon the function $R(y)$ behaves as
\begin{equation}
R(y)= D_1 (1-y)^{-\frac{i (\omega+qA_t(1)) (r_{\Lambda}-r_H)}{f'(1)} }+D_2 (1-y)^{\frac{i (\omega+qA_t(1)) (r_{\Lambda}-r_H)}{f'(1)} } \,.
\end{equation}
So, imposing that there is only ingoing waves at the cosmological horizon requires $D_1=0$. Therefore, an ansatz for $R(y)$ is $R(y) = y^{-\frac{i (\omega+qA_t(0)) (r_{\Lambda}-r_H)}{f'(0)}}(1-y)^{\frac{i (\omega+qA_t(1)) (r_{\Lambda}-r_H)}{f'(1)} } F(y)$, and by inserting this expression in Eq. (\ref{r}), we obtain the differential equation for the function $F(y)$. Now, to use the pseudospectral method, $F(y)$ is expanded in a complete basis of functions $\{\varphi_i(y)\} $: $F(y)=\sum_{i=0}^{\infty} c_i \varphi_i(y)$, where $c_i$ are the coefficients of the expansion, and we choose the Chebyshev polynomials as the complete basis, which are defined by $T_j(x)= \cos (j \cos^{-1}x)$, where $j$ corresponds to the grade of the polynomial. The sum must be truncated until some $N$ value, therefore the function $F(y)$ can be approximated by
\begin{equation}
F(y) \approx \sum_{i=0}^N c_i T_i (x)\,.
\end{equation}
Thus, the solution is assumed to be a finite linear combination of the Chebyshev polynomials,
that are well defined in the interval $x \in [-1,1]$. Due to $y \in [0,1]$, the coordinates $x$ and $y$ are related by $x=2y-1$.

Then, the interval $[0,1]$ is discretized at the Chebyshev collocation points $y_j$ by using the so-called Gauss-Lobatto grid, where
\begin{equation}
    y_j=\frac{1}{2}[1-\cos(\frac{j \pi}{N})]\,, \,\,\,\, j=0,1,...,N \,.
\end{equation}
The corresponding differential equation is then evaluated at each collocation point. So, a system of $N+1$ algebraic equations is obtained, which corresponds to a generalized eigenvalue problem and it can be solved numerically to obtain the QNMs spectrum, by employing the built-in Eigensystem [ ] procedure in Wolfram’s Mathematica \cite{WM}.

In this work, we use a value of $N$ in the interval [100 - 270] 
which depends on the convergence of $\omega$ to the desired accuracy.  We will use an accuracy of eight decimal places for the majority of the cases. In addition, to ensure the accuracy of the results, the code was executed for several increasing values of $N$ stopping when the value of the QNF was unaltered. Also, the complete parameter space associated to the models is $M\geq 0$, $\Lambda>0$, $Q\geq 0$, $\ell= 0, 1, 2, ...$, $m\geq 0$, and $q\geq 0$. Here, the regions of the parameter space explored is $\Lambda M^2=10^{-4}, 0.01, 0.04$, $Q/M=0.1, 0.5, 0.999, 1, 1.004$, a discrete set of values of $\ell$ in the interval [0, 100], for $qQ$ we consider a discrete set of values in the interval [0, 5], for $mM$ a discrete set of values in the interval [0, 1], and for $qM$ a discrete set of values in the interval [0, 10]. Also, we work in a space of parameters such that there are two positive real roots for the lapse function.

\subsection{Quasinormal Spectrum}

In this subsection we will discuss the uncharged and charged quasinormal spectrum.

\subsubsection{Uncharged scalar field}

\begin{figure}[H]
\begin{center}
\includegraphics[width=0.48\textwidth]{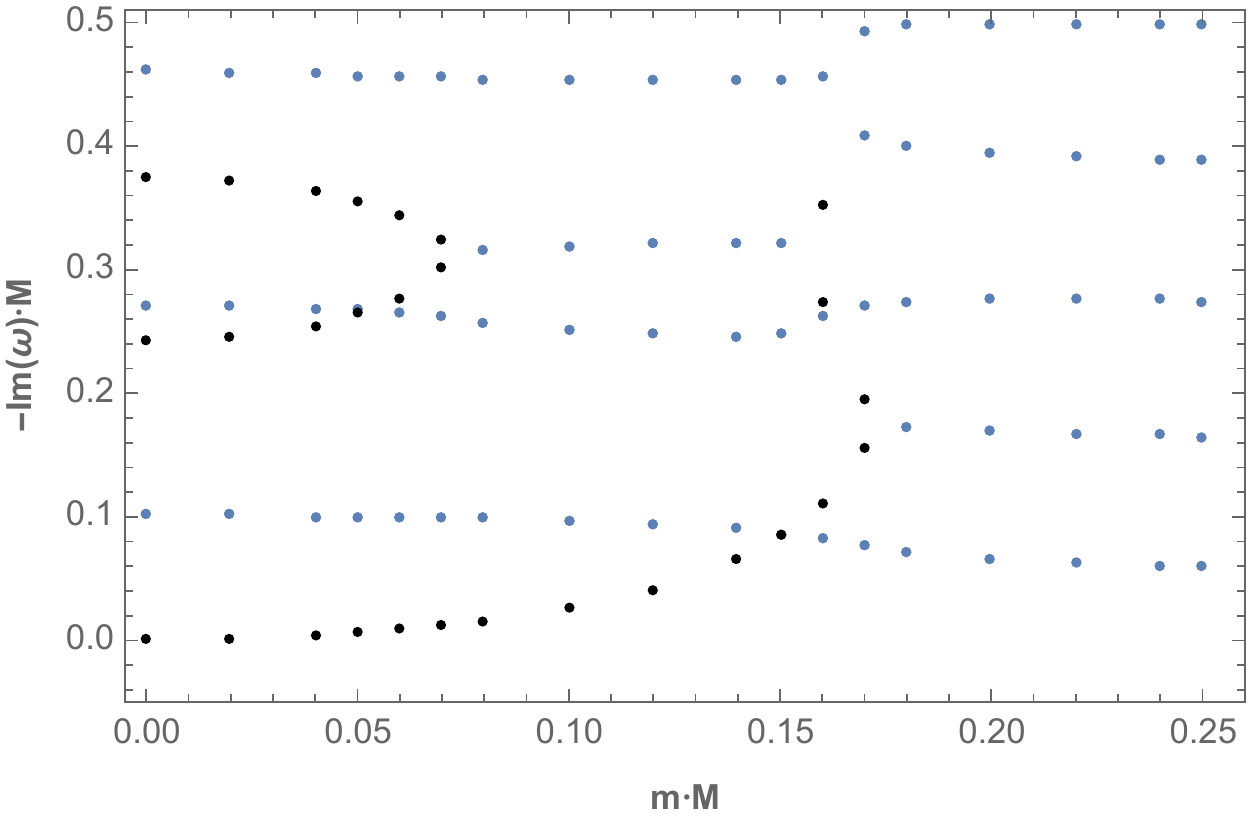}
\includegraphics[width=0.48\textwidth]{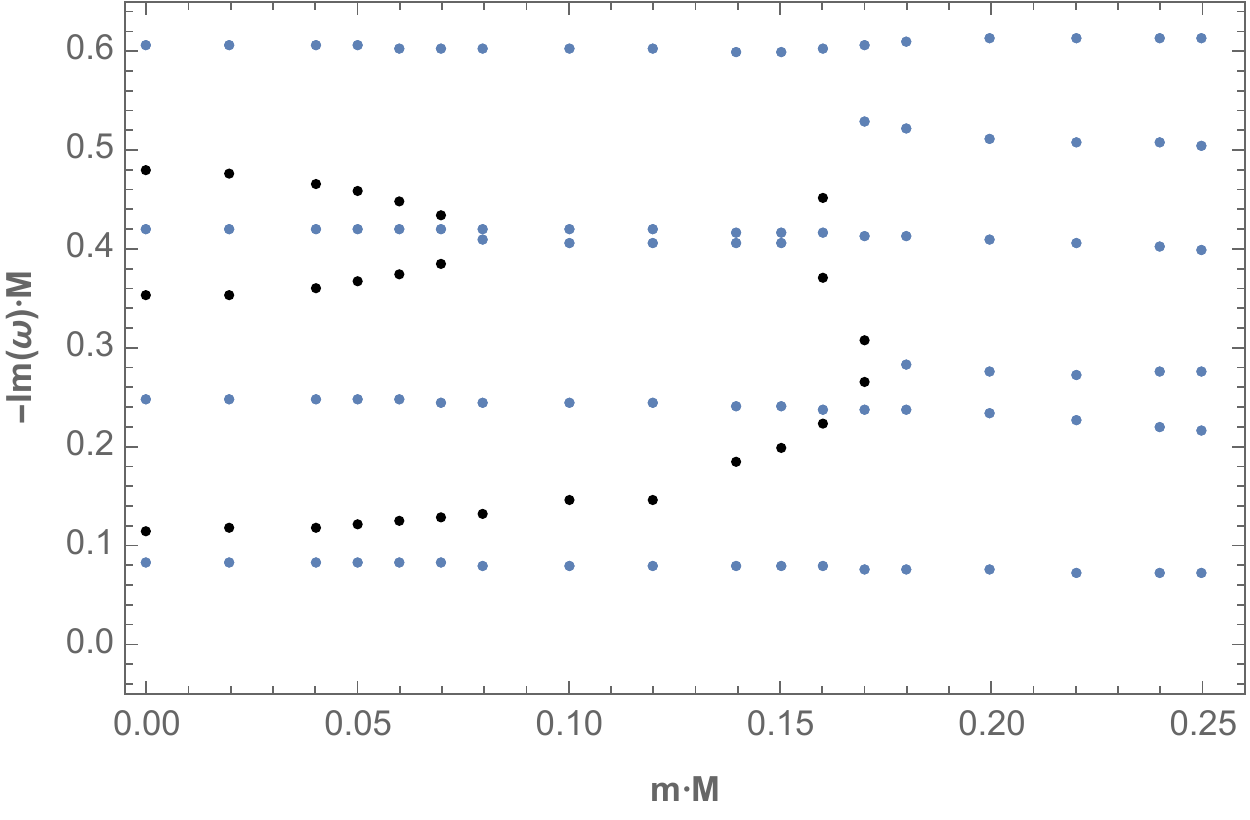}
\end{center}
\caption{The behaviour of the imaginary part of the quasinormal frequencies $Im(\omega) M$ 
as a function of the uncharged scalar field mass $m M$ for different overtone numbers with $Q/M=1/10$, $\Lambda M^2=0.04$, and $\ell= 0$ (left panel), $\ell=1$ (right panel).  Black points for purely imaginary QNFs and blue points for complex QNFs.}
\label{PotL004}
\end{figure}

In Fig. \ref{PotL004} we show the QNFs for a fixed value of $\Lambda M^2$ and $Q/M$, for uncharged scalar field. In this figure we can recognize two families, for zero mass, a family of complex QNFs given by the blue points, and a purely imaginary family given by the black points. The purely imaginary modes belong to the family of de Sitter modes, in that they continuously
approach those of empty de Sitter in the limit that the black hole vanishes,
while the complex ones are those of the RN black hole, in the sense
that they limit to the modes of the asymptotically
at RN black hole in that limit, and this family corresponds to the photon sphere modes.

As it was shown in  \cite{Cardoso:2017soq}, by excluding in the analysis the $\ell=0$ mode, the dS family is dominant  for small values of the cosmological constant (when $\Lambda M^2 < 0.02$), and  in the opposite regime the PS modes are dominant. In order to show this behaviour, we show in Table \ref{dF1} and \ref{dF2}, the dominant modes. We can observe that such behaviour in the dominant family modes is recovered.  Also, we include the scalar field mass in order to visualize if there is an influence of the scalar field mass in the dominant family modes. We can observe that for small values of the scalar field mass the behaviour in the dominant family modes is the same. However, for a bigger value of the scalar field mass the PS family dominates. Another effect that it is possible to observe in Tables \ref{dF1} and \ref{dF2} is the existence of an anomalous behaviour of the decay rate of the quasinormal modes, which occurs when the longest-lived modes are the ones with higher angular number for the PS family, and the existence of a critical mass where beyond this value the behaviour is inverted, that is, the longest-lived modes are the ones with small angular number, which was analyzed in detail in \cite{Fontana:2020syy}.

\begin{table}[H]
\caption {The QNFs $\omega M$ for massive  uncharged scalar fields in the background of RN dS black holes with $\Lambda M^2= 0.01 $, $Q/M=1/10$, and different values of $\ell$.}
\label {dF1}\centering
\scalebox{0.65} {
\begin{tabular} {  | c | c | c | c | c | c | c |  }
\hline
$\ell$ & $m M= 0$ & $m M =1/2500$ & $m M=1/1000$ & $m M=0.1$ &  $m M=0.5$ & $m M=1$   \\\hline
$0$ &
$0$ &
$-8.5005119*10^{-7} I$ &
$-5.3129970*10^{-6} I$ &
$0.0600620007 - 0.0816569100 i$ &
$0.372753995 - 0.035037656 i$ &
$0.742995963 - 0.036775235 i$ \\
${}$ &
$0.10447782 - 0.10454954 i$ &
$0.10447786 - 0.10454939 I$ &
$-0.10447809 - 0.10454860 I$ &
$-0.0600620007 - 0.0816569100 i$&
$-0.372753995 - 0.035037656 i$ &
$-0.742995963 - 0.036775235 i$ \\
${}$ &
$-0.10447782 - 0.10454954 i$ &
$-0.10447786 - 0.10454939 I$ &
$0.10447809 - 0.10454860 I$ &
$-0.105601028 - 0.098841884 i$ &
$0.378493098 - 0.107674493 i$ &
$0.744477192 - 0.110389865 i$ \\
${}$ &
$-0.119131058 I$ &
$-0.119132503 I$ &
$-0.119140089 I$ &
$0.105601028 - 0.098841884 i$ &
$-0.378493098 - 0.107674493 i$ &
$-0.744477192 - 0.110389865 i$ \\\hline
$1$ &
$-0.057704317 I$ &
$-0.057705205 I$ &
$-0.057709870 I$ &
$0.281931797 - 0.092160540 i$ &
$-0.421072795 - 0.040311060 i$ &
$-0.761198717 - 0.035483704 i$ \\
${}$ &
$0.27747429 - 0.09463684 I$ &
$0.27747436 - 0.09463680 I$ &
$0.27747473 - 0.09463659 I$ &
$-0.281931797 - 0.092160540 i$ &
$0.421072795 - 0.040311060 i$ &
$0.761198717 - 0.035483704 i$ \\
${}$ &
$-0.27747429 - 0.09463684 I$ &
$-0.27747436 - 0.09463680 I$ &
$-0.27747473 - 0.09463659 I$ &
$0.053140491 - 0.140839507 i$ &
$0.421690770 - 0.112918764 i$ &
$0.763931772 - 0.106773051 i$ \\
${}$ &
$-0.17411414 I$ &
$-0.17411536 I$ &
$-0.17412173 I$ &
$-0.053140491 - 0.140839507 i$ &
$-0.421690770 - 0.112918764 i$ &
$-0.763931772 - 0.106773051 i$ \\\hline
$2$ &
$-0.46049358 - 0.09291697 I$ &
$0.46049363 - 0.09291695 I$ &
$-0.46049389 - 0.09291687 I$ &
$0.463557970 - 0.091955911 i$ &
$0.539893260 - 0.068079368 i$ &
$0.805405488 - 0.035095589 i$ \\
${}$ &
$0.46049358 - 0.09291697 I$ &
$-0.46049363 - 0.09291695 I$ &
$0.46049389 - 0.09291687 I$ &
$-0.463557970 - 0.091955911 i$ &
$-0.539893260 - 0.068079368 i$ &
$-0.805405488 - 0.035095589 i$ \\
${}$ &
$-0.115457458 I$ &
$-0.115458353 I$ &
$-0.115463053 I$ &
$-0.052435386 - 0.199186353 i$ &
$0.466495684 - 0.148085701 i$ &
$-0.809449712 - 0.106660058 i$ \\
${}$ &
$-0.23149589 I$ &
$-0.23149703 I$ &
$-0.23150305 I$ &
$0.052435386 - 0.199186353 i$ &
$-0.466495684 - 0.148085701 i$ &
$0.809449712 - 0.106660058 i$
\\\hline
$10$ &
$1.9313201 - 0.0919205 I$ &
$1.9313201 - 0.0919205 I$ &
$1.9313202 - 0.0919205 I$ &
$-1.93210166 - 0.09186396 i$ &
$1.95088060 - 0.09050747 i$ &
$-2.00983345 - 0.08627089 i$ \\
${}$ &
$-1.9313201 - 0.0919205 I$ &
$-1.9313201 - 0.0919205 I$ &
$-1.9313202 - 0.0919205 I$ &
$1.93210166 - 0.09186396 i$ &
$-1.95088060 - 0.09050747 i$ &
$2.00983345 - 0.08627089 i$ \\
${}$ &
$1.9269343 - 0.2760096 I$ &
$1.9269343 - 0.2760096 I$ &
$1.9269344 - 0.2760096 I$ &
$1.92769479 - 0.27584227 i$ &
$1.94597276 - 0.27182329 i$ &
$-2.00341932 - 0.25924105 i$ \\
${}$ &
$-1.9269343 - 0.2760096 I$ &
$-1.9269343 - 0.2760096 I$ &
$-1.9269344 - 0.2760096 I$ &
$-1.92769479 - 0.27584227 i$ &
$-1.94597276 - 0.27182329 i$ &
$2.00341932 - 0.25924105 i$  \\\hline
$100$ &
$-18.484286 - 0.091860 i$ &
$-18.484286 - 0.091860 i$ &
$-18.484286 - 0.091860 i$ &
$-18.484368 - 0.091859 i$ &
$-18.486336 - 0.091844 i$ &
$18.492485 - 0.091798 i$ \\
${}$ &
$18.484286 - 0.091860 i$ &
$18.484286 - 0.091860 i$ &
$18.484286 - 0.091860 i$ &
$18.484368 - 0.091859 i$ &
$18.486336 - 0.091844 i$ &
$-18.492485 - 0.091798 i$ \\
${}$ &
$-18.483825 - 0.275581 i$ &
$-18.483825 - 0.275581 i$ &
$-18.483825 - 0.275581 i$ &
$-18.483907 - 0.275579 i$ &
$-18.485874 - 0.275535 i$ &
$-18.492021 - 0.275396 i$ \\
${}$ &
$18.483825 - 0.275581 i$ &
$18.483825 - 0.275581 i$ &
$18.483825 - 0.275581 i$ &
$18.483907 - 0.275579 i$ &
$18.485874 - 0.275535 i$ &
$18.492021 - 0.275396 i$\\\hline
\end{tabular}}
\end{table}

\begin{table}[H]
\caption {The QNFs $\omega M$ for massive  uncharged scalar fields in the background of RN dS black holes with $\Lambda M^2= 0.04 $, $Q/M=1/10$, and different values of $\ell$.}
\label {dF2}\centering
\scalebox{0.65} {
\begin {tabular} { | c | c | c | c | c | c |c |}
\hline
$\ell$ & $m M= 0$  & $m M=1/2500$ & $mM=1/1000$ & $m M=0.1$ &  $m M=0.5$ & $mM=1$   \\\hline
$0$ &
$0$ &
$-3.5596829*10^{-7} I$ &
$-2.22482731*10^{-6} i$ &
$-0.026278371 I$ &
$-0.267854426 - 0.052616360 i$ &
$0.536203691 - 0.053167845 i$ \\
${}$ &
$0.082248297 - 0.101449188 I$ &
$0.082248381 - 0.101449135 I$ &
$0.0822488200 - 0.1014488582 i$ &
$-0.086104082 - 0.097603948 I$ &
$-0.267854426 - 0.052616360 i$ &
$-0.536203691 - 0.053167845 i$ \\
${}$ &
$-0.082248297 - 0.101449188 I$ &
$-0.082248381 - 0.101449135 I$ &
$-0.0822488200 - 0.1014488582 i$ &
$0.086104082 - 0.097603948 I$ &
$0.272344951 - 0.159059042 i$ &
$0.538018507 - 0.159637531 i$ \\
${}$ &
$-0.24325499 I$ &
$-0.24325606 I$ &
$- 0.243261671 i$ &
$0.07023918 - 0.25138489 I$ &
$-0.272344951 - 0.159059042 i$ &
$-0.538018507 - 0.159637531 i$ \\\hline

$1$ &
$0.22532155 - 0.08215986 I$ &
$0.22532162 - 0.08215983 I$ &
$-0.225321971 - 0.082159682 i$ &
$-0.22951759 - 0.08040506 I$ &
$-0.334862242 - 0.057628643 i$ &
$0.568424089 - 0.052683138 i$ \\
${}$ &
$-0.22532155 - 0.08215986 I$ &
$-0.22532162 - 0.08215983 I$ &
$0.225321971 - 0.082159682 i$ &
$0.22951759 - 0.08040506 I$ &
$0.334862242 - 0.057628643 i$ &
$-0.568424089 - 0.052683138 i$ \\
${}$ &
$-0.115250221 I$ &
$-0.115250648 I$ &
$-0.1152528890 i$ &
$-0.14516899 I$ &
$-0.331303166 - 0.168488614 i$ &
$-0.570792599 - 0.158453452 i$ \\
${}$ &
$-0.21593099 - 0.24781626 I$ &
$0.21593101 - 0.24781621 I$ &
$0.215931144 - 0.247815912 i$ &
$0.21746445 - 0.24429791 I$ &
$0.331303166 - 0.168488614 i$ &
$0.570792599 - 0.158453452 i$
\\\hline
$2$ &
$0.38180841 - 0.07887236 I$ &
$-0.38180845 - 0.07887235 I$ &
$0.381808681 - 0.078872294 i$ &
$0.38450827 - 0.07826155 I$ &
$-0.449718749 - 0.066162216 i$ &
$0.633426884 - 0.054199945 i$ \\
${}$ &
$-0.38180841 - 0.07887236 I$ &
$0.38180845 - 0.07887235 I$ &
$-0.381808681 - 0.078872294 i$ &
$-0.38450827 - 0.07826155 I$ &
$0.449718749 - 0.066162216 i$ &
$-0.633426884 - 0.054199945 i$ \\
${}$ &
$-0.23084162 I$ &
$-0.23084205 I$ &
$-0.230844327 i$ &
$0.37473638 - 0.23657636 I$ &
$-0.435865076 - 0.195113817 i$ &
$-0.634204244 - 0.162335250 i$ \\
${}$ &
$0.37266985 - 0.23829238 I$ &
$-0.37266988 - 0.23829236 I$ &
$0.372670054 - 0.238292214 i$ &
$-0.37473638 - 0.23657636 I$ &
$0.435865076 - 0.195113817 i$ &
$0.634204244 - 0.162335250 i$
\\\hline

$10$ &
$-1.6197944 - 0.0771937 I$ &
$1.61979445 - 0.07719373 i$ &
$1.61979450 - 0.07719373 i$ &
$1.6204523 - 0.0771602 I$ &
$1.63623058 - 0.07636258 i$ &
$-1.68539834 - 0.07396874 i$ \\
${}$ &
$1.6197944 - 0.0771937 I$ &
$-1.61979445 - 0.07719373 i$ &
$-1.61979450 - 0.07719373 i$ &
$-1.6204523 - 0.0771602 I$ &
$-1.63623058 - 0.07636258 i$ &
$1.68539834 - 0.07396874 i$ \\
${}$ &
$1.6172051 - 0.2316855 I$ &
$-1.61720514 - 0.23168548 i$ &
$1.61720519 - 0.23168548 i$ &
$1.6178562 - 0.2315848 I$ &
$-1.63348008 - 0.22919027 i$ &
$1.68226900 - 0.22198395 i$ \\
${}$ &
$-1.6172051 - 0.2316855 I$ &
$1.61720514 - 0.23168548 i$ &
$-1.61720519 - 0.23168548 i$ &
$-1.6178562 - 0.2315848 I$ &
$1.63348008 - 0.22919027 i$ &
$-1.68226900 - 0.22198395 i$  \\\hline

$100$ &
$15.513303 - 0.077096 I$ &
$15.5133030 - 0.0770960 i$ &
$15.5133030 - 0.0770960 i$ &
$15.513372 - 0.077096 I$ &
$15.5150232 - 0.0770869 i$ &
$15.5201837 - 0.0770595 i$ \\
${}$ &
$-15.513303 - 0.077096 I$ &
$15.5133030 - 0.0770960 i$ &
$-15.5133030 - 0.0770960 i$ &
$-15.513372 - 0.077096 I$ &
$-15.5150232 - 0.0770869 i$ &
$-15.5201837 - 0.0770595 i$ \\
${}$ &
$15.513030 - 0.231289 I$ &
$15.5130305 - 0.2312892 i$ &
$15.5130305 - 0.2312892 i$ &
$15.513099 - 0.231288 I$ &
$-15.5147505 - 0.2312618 i$ &
$15.5199104 - 0.2311796 i$ \\
${}$ &
$-15.513030 - 0.231289 I$ &
$-15.5130305 - 0.2312892 i$ &
$-15.5130305 - 0.2312892 i$ &
$-15.513099 - 0.231288 I$ &
$15.5147505 - 0.2312618 i$ &
$-15.5199104 - 0.2311796 i$

\\\hline

\end {tabular}}
\end{table}

\begin{figure}[H]
\begin{center}
\includegraphics[width=0.32\textwidth]{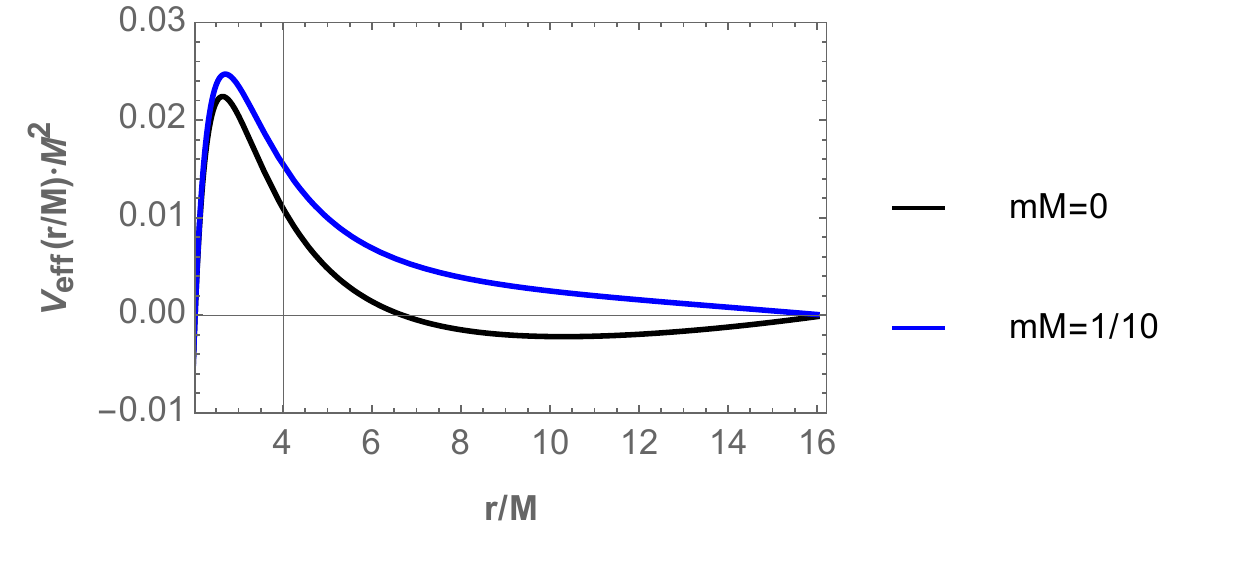}
\includegraphics[width=0.32\textwidth]{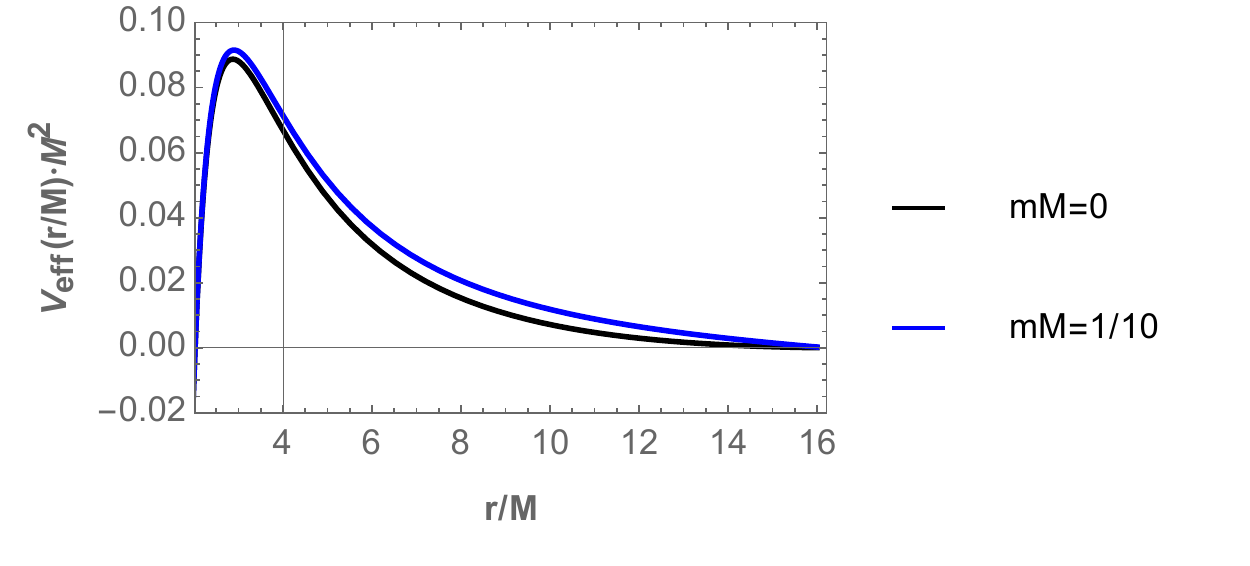}
\includegraphics[width=0.32\textwidth]{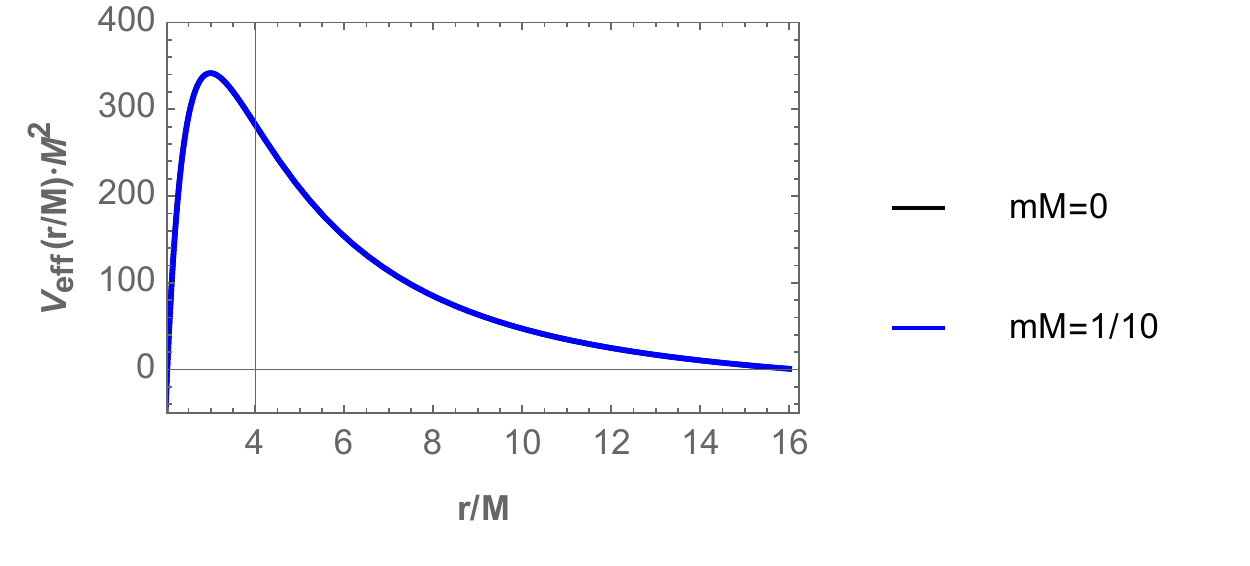}
\end{center}
\caption{The effective potential $V_{eff} M^2$ as a function of $r/M$, for uncharged scalar fields with $\ell=0$ (left panel), $\ell=1$ (center panel), and $\ell=100$ (right panel) with $\Lambda M^2=0.01$, and $Q/M=1/10$. Here $r_H/M\approx 2.023$, and $r_{\Lambda}/M\approx 16.218$.}
\label{Potell001}
\end{figure}

\begin{figure}[H]
\begin{center}
\includegraphics[width=0.32\textwidth]{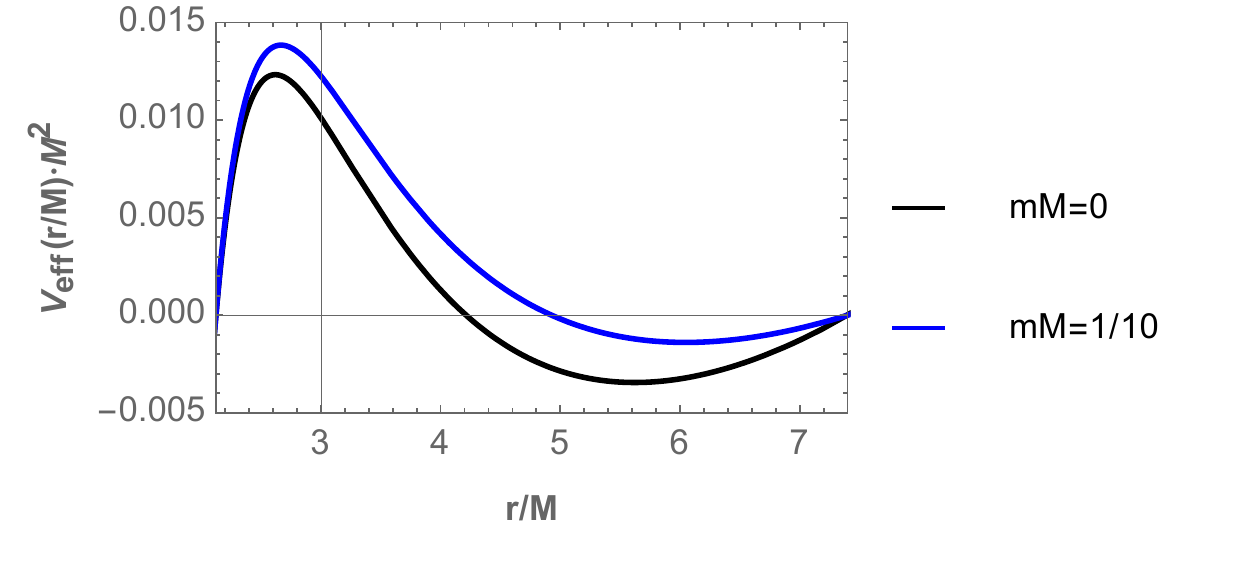}
\includegraphics[width=0.32\textwidth]{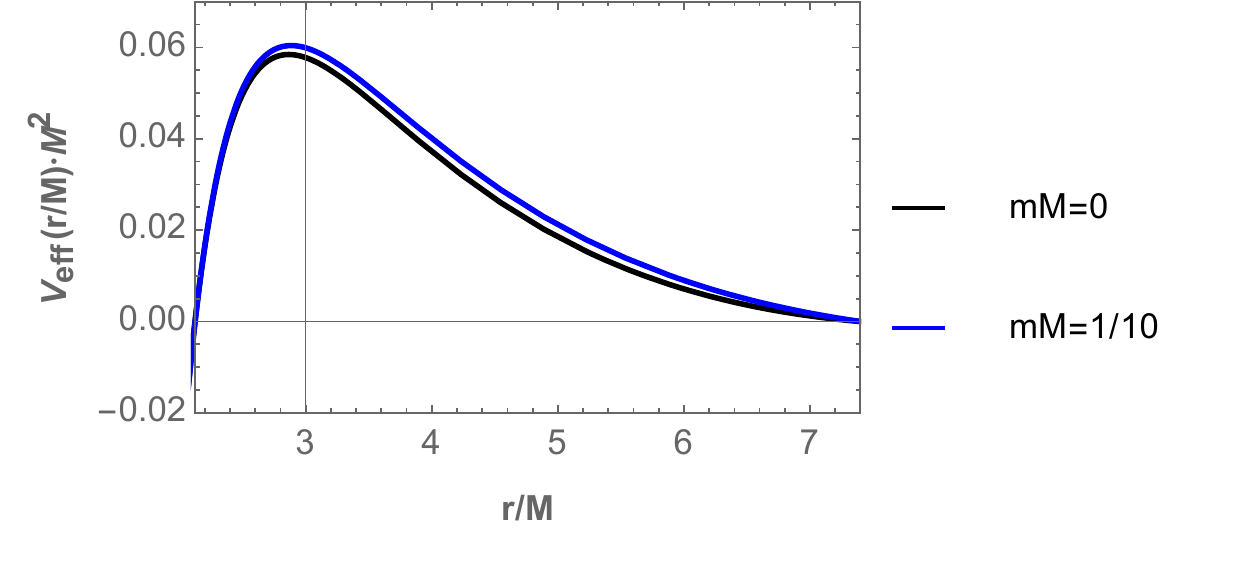}
\includegraphics[width=0.32\textwidth]{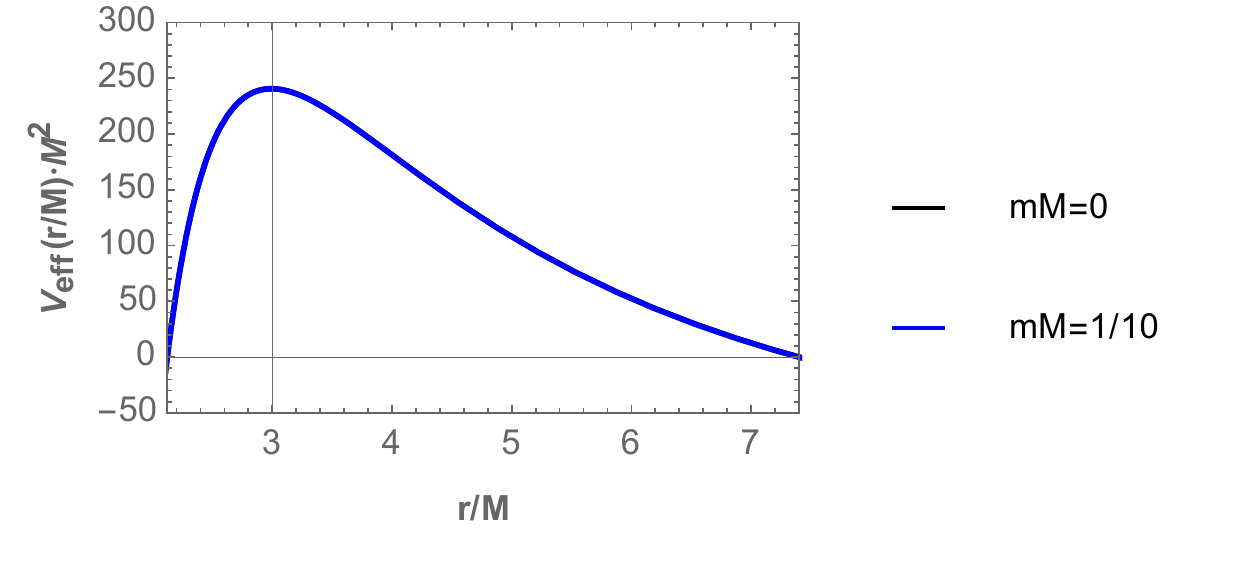}
\end{center}
\caption{The effective potential $V_{eff} M^2$ as a function of $r/M$, for uncharged scalar fields with $\ell=0$ (left panel), $\ell=1$ (center panel), and $\ell=100$ (right panel) with $\Lambda M^2=0.04$, and $Q/M =1/10$. Here $r_H/M\approx 2.123$, and $r_{\Lambda}/M\approx 7.399$.}
\label{Potell004}
\end{figure}

So, in order to visualize the potential, we plot the effective potential for uncharged scalar field with $\ell=0$, $\ell=1$ and $\ell=100$, in Fig. \ref{Potell001} for $\Lambda M^2=0.01$, and in Fig. \ref{Potell004} for $\Lambda M^2=0.04$. Note that a barrier of potential occurs for each case, with it maximum value increasing when the angular number increases, and decreasing when the cosmological constant increases. Also, notice that for a massless scalar field the existence of bound states for $\ell=0$ (left panel), while that when scalar field mass increases the bound states disappear, and there are not bound states for higher values of the angular number $\ell$ (center and right panel). Note the strong correlation between the dominant family and the existence of bound states, the dS family is dominant when there are bound states, otherwise the PS family is dominant.

\subsubsection{Charged scalar field}

\begin{figure}[H]
\begin{center}
\includegraphics[width=0.42\textwidth]{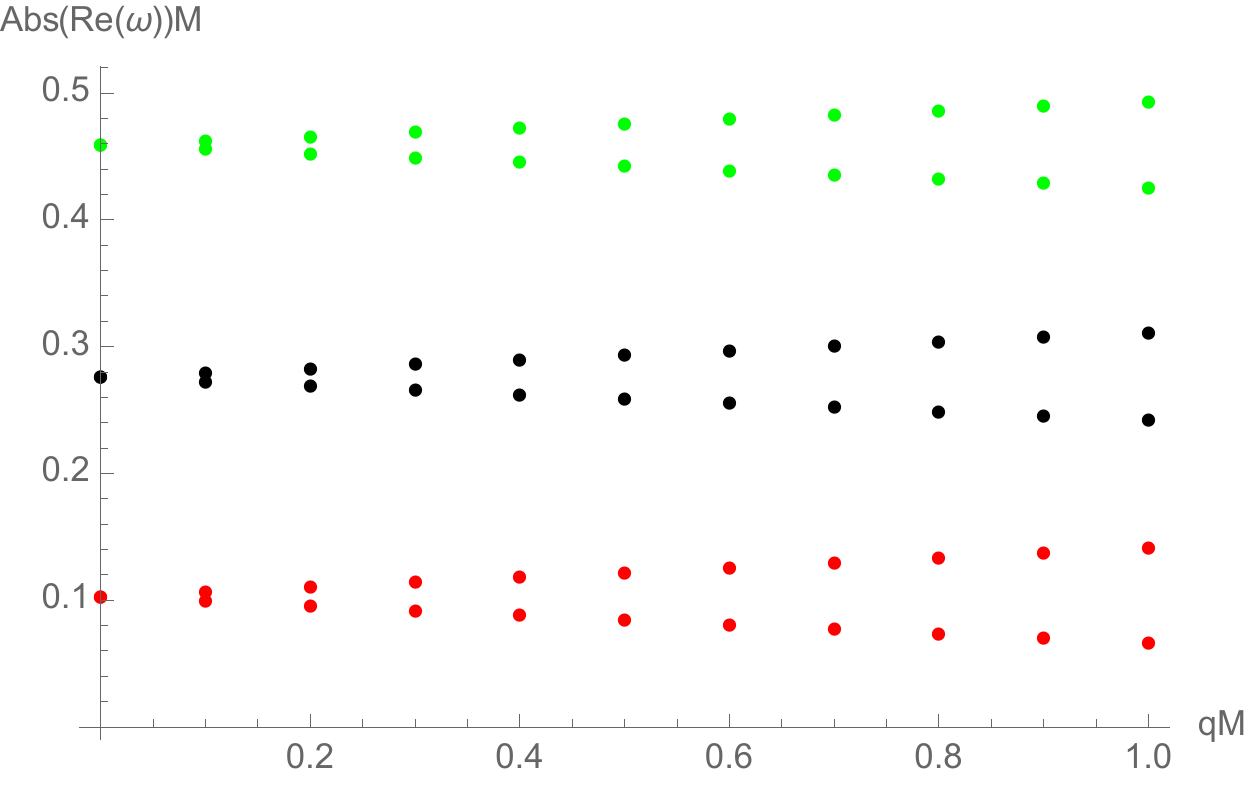}
\includegraphics[width=0.42\textwidth]{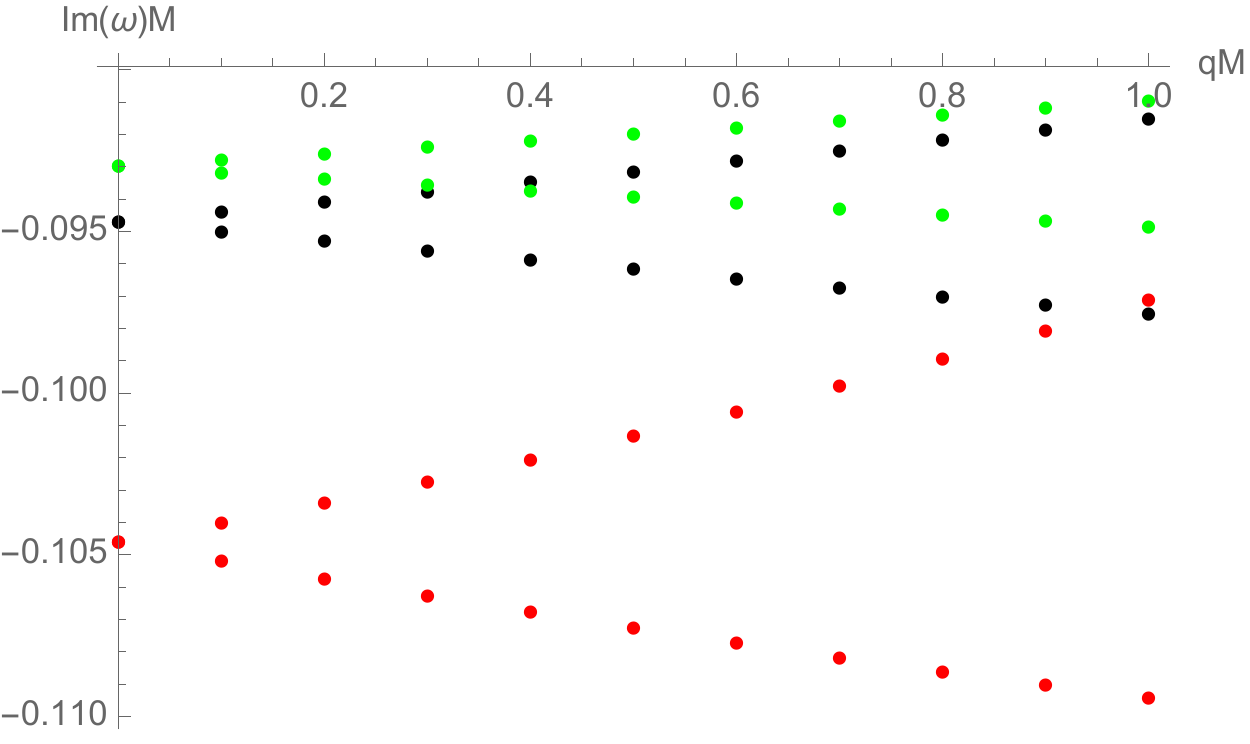}
\end{center}
\caption{
The two branches of the photon sphere modes for massless scalar fields. Left panel for the behaviour of $\abs{Re(\omega)}M$, and
right panels for $Im(\omega)M$,
as a function of $qM$ for $\ell=0$ (red points), $\ell=1$ (black points), and $\ell=2$ (green points) with
$\Lambda M^2=0.01$, and  $Q/M=0.1$.
}

\label{FAY1}
\end{figure}

When the scalar field acquires electric charge, the spectrum showed in Fig. \ref{PotL004} for small values of $qQ$ does not show  a major change; however, the dS family, characterized by purely imaginary modes when $qQ$ vanishes, presents a frequency of oscillation which increases when $qQ$ increases, and the PS family characterized by modes of the form $\pm Re(\omega)-Im(\omega)$ when $qQ$ vanishes, is divided in two different  branches of modes with a real and an imaginary part different, see Table \ref{CH1} for 
$\Lambda M^2 = 0.01$ and Table \ref{CH2} for  
$\Lambda M^2 = 0.04$.  We can observe that the real part of the PS modes is negative/positive which we refer as negative/positive branch. The above behaviour also was observed in Ref.  \cite{Gonzalez:2021vwp}. Also, notice the existence of unstable modes for small values of $qQ$, $\ell=0$ and 
$\Lambda M^2=0.01$ in Table \ref{CH1}, which will be discussed in detail in Section \ref{SM}. Also, notice that for the PS family for massless scalar field  the longest-lived QNMs are those with a high angular number $\ell$. Massive scalar field will be analyzed in the next Section. In Fig. \ref{FAY1}, we plot the PS modes, the positive/negative branch is described by an  absolute value of the frequency of oscillation and of decay rate that increases/decreases when  $qM$ increases.

\begin{table}[H]
\caption {The QNFs $\omega M$ for massless charged scalar fields in the background of RNdS black holes with $\Lambda M^2= 0.01 $, $Q/M =1/10$, and different values of $\ell$.}
\label {CH1}\centering
\scalebox{0.65} {
\begin {tabular} { | c | c | c | c | c | c | c |}
\hline
$\ell$ & $q M = 0$ & $q M=1/100$ & $q M=1/10$ & $q M=1$ &  $q M=5$ & $q M=10$   \\\hline
$0$ &
$0$ &
$0.000068288944 + 1.5960*10^{-8} I$ &
$0.00068292882 + 1.59543*10^{-6} I$ &
$0.0068680433 + 0.0001537376 I$ &
$0.037263872 + 0.001162657 I$ &
$0.075618666 - 0.004457386 I$ \\
${}$ &
$-0.10447782 - 0.10454954 I$ &
$-0.10410424 - 0.10449081 I$ &
$-0.10075144 - 0.10395079 I$ &
$-0.068262518 - 0.097055610 I$ &
$0.008153556 - 0.061892090 I$ &
$0.055746508 - 0.058623272 I$ \\
${}$ &
$0.10447782 - 0.10454954 i$ &
$0.10485160 - 0.10460803 I$ &
$0.10822484 - 0.10512355 I$ &
$0.14280836 - 0.10937694 I$ &
$0.31019996 - 0.11765442 I$ &
$0.048380767 - 0.114955318 I$ \\
${}$ &
$-0.119131058 I$ &
$-0.000070968 - 0.119131074 I$ &
$-0.000709749 - 0.119132655 I$ &
$-0.007121532 - 0.119572504 I$ &
$0.007383911 - 0.123318076 I$ &
$0.53728349 - 0.11955153 I$  \\\hline

$1$ &
$-0.057704317 I$ &
$0.000039041 - 0.057704310 I$ &
$0.000390410 - 0.057703656 I$ &
$0.003903586 - 0.057638138 I$ &
$0.019442457 - 0.056024388 I$ &
$0.037474504 - 0.050704906 I$ \\
${}$ &
$-0.27747429 - 0.09463684 I$ &
$-0.27713063 - 0.09460663 I$ &
$-0.27404276 - 0.09433325 I$ &
$-0.24367994 - 0.09144648 I$ &
$-0.12188700 - 0.07509154 I$ &
$-0.008009442 - 0.052556715 I$ \\
${}$ &
$0.27747429 - 0.09463684 I$ &
$0.27781806 - 0.09466701 I$ &
$0.28091704 - 0.09493706 I$ &
$0.31239189 - 0.09749179 I$ &
$0.46162862 - 0.10604435 I$ &
$0.66474833 - 0.11208790 I$ \\
${}$ &
$-0.17411414 I$ &
$-0.00003589 - 0.17411410 I$ &
$-0.00035884 - 0.17410950 I$ &
$-0.00356766 - 0.17365266 I$ &
$-0.01622963 - 0.16307184 I$ &
$0.001893817 - 0.130350119 I$ \\\hline

$2$ &
$-0.46049358 - 0.09291697 I$ &
$-0.46015575 - 0.09289753 I$ &
$-0.45711846 - 0.09272200 I$ &
$-0.427068283 - 0.090908339 i$ &
$-0.30117685 - 0.08150837 I$ &
$-0.16417351 - 0.06672915 I$ \\
${}$ &
$0.46049358 - 0.09291697 I$ &
$0.46083149 - 0.09293639 I$ &
$0.46387583 - 0.09311064 I$ &
$0.494631199 - 0.094796321 i$ &
$0.63767763 - 0.10112513 I$ &
$0.82898954 - 0.10679059 I$ \\
${}$ &
$-0.115457458 I$ &
$0.000031078 - 0.115457454 I$ &
$0.000310781 - 0.115457087 I$ &
$0.0031073852 - 0.1154203741 i$ &
$0.015484776 - 0.114528673 I$ &
$0.030616572 - 0.111727404 I$ \\
${}$ &
$-0.23149589 I$ &
$-0.00002547 - 0.23149586 I$ &
$-0.00025471 - 0.23149304 I$ &
$-0.002535609 - 0.231212385 i$ &
$-0.01157802 - 0.22493921 I$ &
$-0.12926060 - 0.20298056 I$
\\\hline

$10$ &
$-1.9313201 - 0.0919205 I$ &
$-1.9309858 - 0.0919157 I$ &
$-1.9279780 - 0.0918724 I$ &
$-1.8979778 - 0.0914361 I$ &
$-1.7663979 - 0.0894218 I$ &
$-1.6060898 - 0.0867271 I$ \\
${}$ &
$1.9313201 - 0.0919205 I$ &
$1.9316544 - 0.0919253 I$ &
$1.9346640 - 0.0919685 I$ &
$1.9648376 - 0.0923973 I$ &
$2.1006231 - 0.0942298 I$ &
$2.2740799 - 0.0963579 I$ \\
${}$ &
$-1.9269343 - 0.2760096 I$ &
$-1.9265991 - 0.2759953 I$ &
$-1.9235829 - 0.2758658 I$ &
$-1.8934989 - 0.2745607 I$ &
$-1.7615362 - 0.2685330 I$ &
$-1.6007294 - 0.2604647 I$ \\
${}$ &
$1.9269343 - 0.2760096 I$ &
$1.9272695 - 0.2760240 I$ &
$1.9302874 - 0.2761533 I$ &
$1.9605438 - 0.2774358 I$ &
$2.0966855 - 0.2829163 I$ &
$2.2705598 - 0.2892786 I$  \\\hline

$100$ &
$-18.484286 - 0.091860 I$ &
$-18.483952 - 0.091859 I$ &
$-18.480945 - 0.091855 I$ &
$-18.450887 - 0.091809 I$ &
$-18.317476 - 0.091607 I$ &
$-18.151127 - 0.091352 I$ \\
${}$ &
$18.484286 - 0.091860 I$ &
$18.484620 - 0.091860 I$ &
$18.487627 - 0.091865 I$ &
$18.517703 - 0.091910 I$ &
$18.651555 - 0.092110 I$ &
$18.819280 - 0.092359 $ \\
${}$ &
$-18.483825 - 0.275581 I$ &
$-18.483491 - 0.275580 I$ &
$-18.480484 - 0.275566 I$ &
$-18.450425 - 0.275430 I$ &
$-18.317010 - 0.274824 I$ &
$-18.150655 - 0.274059 I$ \\
${}$ &
$18.483825 - 0.275581 I$ &
$18.484159 - 0.275583 I$ &
$18.487166 - 0.275596 I$ &
$18.517243 - 0.275732 I$ &
$18.651099 - 0.276333 I$ &
$18.818829 - 0.277078 I$ \\\hline
\end {tabular}}
\end{table}

\begin{table}[H]
\caption {The QNFs $\omega M$ for massless charged scalar fields in the background of RNdS black holes with $\Lambda M^2= 0.04 $, $Q/M=1/10$, and diferent values of $\ell$.}
\label {CH2}\centering
\scalebox{0.65} {
\begin {tabular} { | c | c | c | c | c | c |c |}
\hline
$\ell$ & $qM= 0$  & $qM=1/100$ & $qM=1/10$ & $qM=1$ &  $qM=5$ & $qM=10$   \\\hline
$0$ &
$0$ &
$0.000160710416 - 5.957*10^-9 i$ &
$0.00160712286 - 5.9621*10^-7 i$ &
$0.0160888408 - 0.0000648039 i$ &
$0.0808951143 - 0.0030966209 i$ &
$0.156989928 - 0.013457290 i$ \\
${}$ &
$-0.082248297 - 0.101449188 I$ &
$-0.0819134372 - 0.1014350506 i$ &
$-0.0789157290 - 0.1013047585 i$ &
$-0.0506096677 - 0.0997587197 i$ &
$0.0434443000 - 0.0945027927 i$ &
$0.129734857 - 0.095349783 i$ \\
${}$ &
$0.082248297 - 0.101449188 I$ &
$0.0825835108 - 0.1014632555 i$ &
$0.0856162186 - 0.1015867226 i$ &
$0.117382003 - 0.102494044 i$ &
$0.278446867 - 0.102026003 i$ &
$0.499072078 - 0.099936486 i$ \\
${}$ &
$-0.24325499 I$ &
$-0.000228559 - 0.243254121 i$ &
$-0.002290773 - 0.243167400 i$ &
$-0.019134099 - 0.228231434 i$ &
$0.035104028 - 0.189997802 i$ &
$0.118368800 - 0.181492484 i$ \\\hline

$1$ &
$-0.22532155 - 0.08215986 I$ &
$-0.224975949 - 0.082141416 i$ &
$-0.221869566 - 0.081974673 i$ &
$-0.191215562 - 0.080231980 i$ &
$-0.0644748822 - 0.0709436694 i$ &
$0.0685228648 - 0.0513794785 i$ \\
${}$ &
$0.22532155 - 0.08215986 I$ &
$0.225667251 - 0.082178283 i$ &
$0.228782562 - 0.082343335 i$ &
$0.260328858 - 0.083917403 i$ &
$0.408296423 - 0.089300716 i$ &
$0.607288942 - 0.093097369 i$ \\
${}$ &
$-0.115250221 I$ &
$0.0000789890 - 0.1152502063 i$ &
$0.0007898884 - 0.1152487639 i$ &
$0.0078970395 - 0.1151046945 i$ &
$0.0393238301 - 0.1118195972 i$ &
$0.086146945 - 0.110340417 I$ \\
${}$ &
$-0.21593099 - 0.24781626 I$ &
$-0.215571670 - 0.247761346 i$ &
$-0.212341954 - 0.247264066 i$ &
$-0.180483436 - 0.241978499 i$ &
$-0.054437779 - 0.212870158 i$ &
$0.06013021 - 0.19284279 I$
\\\hline

$2$ &
$-0.38180841 - 0.07887236 I$ &
$-0.381470092 - 0.078859047 i$ &
$-0.378427903 - 0.078738920 i$ &
$-0.348275584 - 0.077502899 i$ &
$-0.220569049 - 0.071271211 i$ &
$-0.0769154533 - 0.0621653378 i$ \\
${}$ &
$0.38180841 - 0.07887236 I$ &
$0.382146790 - 0.078885656 i$ &
$0.385194883 - 0.079005010 i$ &
$0.415937640 - 0.080163837 i$ &
$0.557952243 - 0.084571487 i$ &
$0.746167517 - 0.088561966 i$ \\
${}$ &
$-0.23084162 I$ &
$0.000062651 - 0.230841611 i$ &
$0.000626512 - 0.230840816 i$ &
$0.006263889 - 0.230761345 i$ &
$-0.207478680 - 0.215808557 i$ &
$-0.060802232 - 0.188576710 i$ \\
${}$ &
$-0.37266985 - 0.23829238 I$ &
$-0.372323697 - 0.238253551 i$ &
$-0.369210910 - 0.237902937 i$ &
$-0.338343509 - 0.234286458 i$ &
$0.031169970 - 0.228841188 i$ &
$0.061478108 - 0.222998822 i$
\\\hline

$10$ &
$-1.6197944 - 0.0771937 I$ &
$-1.61946013 - 0.07719035 i$ &
$-1.61645200 - 0.07715990 i$ &
$-1.58643651 - 0.07685335 i$ &
$-1.45450164 - 0.07544731 i$ &
$-1.293046403 - 0.073592022 i$ \\
${}$ &
$1.6197944 - 0.0771937 I$ &
$1.62012876 - 0.07719711 i$ &
$1.62313834 - 0.07722752 i$ &
$1.65329947 - 0.07752962 i$ &
$1.78876440 - 0.07882806 i$ &
$1.96124764 - 0.08034961 i$ \\
${}$ &
$-1.6172051 - 0.2316855 I$ &
$-1.61687036 - 0.23167535 i$ &
$-1.61385807 - 0.23158402 i$ &
$-1.58380093 - 0.23066468 i$ &
$-1.45168211 - 0.22644644 i$ &
$-1.290003983 - 0.220876636 i$ \\
${}$ &
$1.6172051 - 0.2316855 I$ &
$1.61753992 - 0.23169563 i$ &
$1.62055366 - 0.23178682 i$ &
$1.65075642 - 0.23269269 i$ &
$1.78640548 - 0.23658506 i$ &
$1.95911371 - 0.24114406 i$  \\\hline

$100$ &
$-15.513303 - 0.077096 I$ &
$-15.5129689 - 0.0770957 i$ &
$-15.5099623 - 0.0770925 i$ &
$-15.4799027 - 0.0770606 i$ &
$-15.3464557 - 0.0769182 i$ &
$-15.1799951 - 0.0767391 i$ \\
${}$ &
$15.513303 - 0.077096 I$ &
$15.5136371 - 0.0770964 i$ &
$15.5166439 - 0.0770996 i$ &
$15.5467187 - 0.0771314 i$ &
$15.6805351 - 0.0772726 i$ &
$15.8481503 - 0.0774480 i$ \\
${}$ &
$-15.513030 - 0.231289 I$ &
$-15.5126964 - 0.2312882 i$ &
$-15.5096897 - 0.2312786 i$ &
$-15.4796296 - 0.2311828 i$ &
$-15.3461807 - 0.2307557 i$ &
$-15.1797175 - 0.2302186 i$ \\
${}$ &
$15.513030 - 0.231289 I$ &
$15.5133645 - 0.2312903 i$ &
$15.5163714 - 0.2312998 i$ &
$15.5464466 - 0.2313955 i$ &
$15.6802651 - 0.2318190 i$ &
$15.8478828 - 0.2323452 i$

\\\hline

\end {tabular}
}
\end{table}

\subsection{Superradiant modes}
\label{SM}

\begin{figure}[H]
\begin{center}
\includegraphics[width=0.5\textwidth]{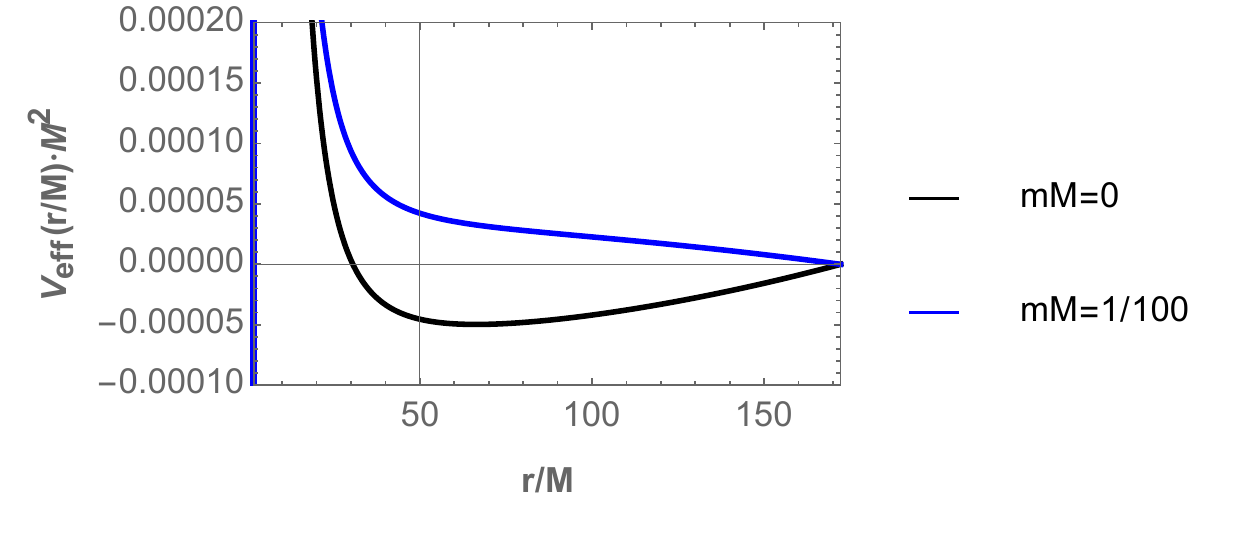}
\end{center}
\caption{The effective potential $V_{eff} (r/M) M^2$ as a function of $r/M$, with $\Lambda M^2=1/10000$, $Q/M=0.50$, $qM =0.1$, and $\ell=0$. Here, $\omega^2 M^2=9*10^{-8}$, and $V_{eff}(r_{\Lambda} /M) M^2\approx 9*10^{-8}$. Here $r_H/M\approx 1.866$, and $r_{\Lambda}/M\approx 172.197$. }
\label{qcritical}
\end{figure}

As was pointed out in  \cite{Zhu:2014sya}, four-dimensional RNdS black holes are unstable against charged scalar perturbations with vanishing angular momentum, $\ell=0$, which is caused by superradiance, and such instability does not occur for a larger angular index. Also, it was shown that the reflected wave has larger amplitude than the incident one when
\begin{equation}
\label{sc}
    \frac{qQ}{r_{\Lambda}}<\omega<\frac{qQ}{r_H}\,,
\end{equation}
knowing as superradiant condition. Note that, when the cosmological constant $\Lambda$ approaches  zero, $r_{\Lambda}$ becomes infinite, and the superradiant condition is equivalent to the one for the RN black hole in the presence of charged scalar perturbations, $0<\omega <\frac{qQ}{r_+}$  \cite{Bekenstein}, where $\omega$ is the real oscillation frequency of the perturbation. Here, we consider massless, see Table \ref{S2}, and massive charged scalar fields, see Table \ref{S2M}, and we observe that superradiant modes are present for massless charged scalar field and for $\ell=0$, as in Ref.  \cite{Zhu:2014sya}. However, also we observe that for small values of the charged scalar field mass, and $\ell=0$, the superradiant modes are present, see Table \ref{smass}. By analyzing the superradiance condition we can see the existence of unstable and also stable superradiant modes, so we can say that the stability of the superradiant modes depend on the scalar field mass. In Fig. \ref{qcritical}, we show the behaviour of the effective potential in order to visualize how is the effect of the scalar field mass when the superradiance condition (\ref{sc}) is satisfied. Note, for instance, that for a certain interval of scalar field mass a potential well is possible, and there are bound states for charged scalar fields which allows to accumulate the energy to trigger the instability. However, for $m>m_c$ there are not bound states and the perturbation wave can be easily absorbed by the black hole and the corresponding background becomes stable under charged scalar perturbations.
 The potential well exists when $V_{eff}'(r_{\Lambda}) >0$, which yields
\begin{equation}
\ell (\ell+1) + m^2 r_{\Lambda}^2 < - f'(r_{\Lambda}) r_{\Lambda} + \frac{2 qQ}{f'(r_{\Lambda})} \left(\omega - \frac{qQ}{r_{\Lambda}} \right) \,.
\end{equation}
 For the parameters satisfying this condition, the superradiant modes are unstable. It is worth mentioning that the superradiant modes has been studied in detail in Ref. \cite{Destounis:2019hca}, and a similar behaviour was observed.

\begin{table}[H]
\caption {The QNFs $\omega M$ for massless scalar fields in the background of RNdS black holes with $\Lambda M^2= 10^{-4} $, $Q/M =0.5$, and different values of $\ell$,  and $q M$.}
\label {S2}\centering
\scalebox{0.7} {

\begin {tabular} { | c | c | c | c | c | c |}
\hline
${}$ & $qM= 0$ & $qM= 0.1$  & $qM= 0.2$ & $qM= 0.3$ & $qM= 0.4$ \\\hline
$\omega M(\ell=0)$ &
$0$ &
$0.00029375585 + 5.90732*10^{-6} i$ &
$0.00058922420 + 0.00002374651 i$ &
$0.00088833967 + 0.00005385197 i$ &
$0.00119351969 + 0.00009670646 i$   \\\hline
$\omega M(\ell=0)$ &
$-0.0115497846 i$ &
$-0.0002116813 - 0.0115187405 i$ &
$-0.0004203766 - 0.0114251387 i$ &
$-0.0006236717 - 0.0112669930 i$ &
$-0.0008196092 - 0.0110390365 i$   \\\hline
$\omega M(\ell=0)$ &
$-0.017453649 i$ &
$-0.000212916 - 0.017314311 i$ &
$-0.000263347 - 0.017138539 i$ &
$-0.000260873 - 0.016982739 i$ &
$-0.000241439 - 0.016836033 i$   \\\hline
$\omega M(\ell=0)$ &
$-0.019051426 i$ &
$0.001150771 - 0.018832589 i$ &
$0.001685334 - 0.018593204 i$ &
$0.002027273 - 0.018431775 i$ &
$-0.003094082 - 0.018255268 i$   \\\hline
$\omega M(\ell=0)$ &
$-0.004290674 - 0.020103073 i$ &
$-0.003716691 - 0.019487876 i$ &
$-0.003477389 - 0.018995885 i$ &
$-0.003288733 - 0.018606549 i$ &
$0.002298702 - 0.018311279 i$   \\\hline
$\omega M(\ell=0)$ &
$0.004290674 - 0.020103073 i$ &
$0.005195244 - 0.020134942 i$ &
$0.005689605 - 0.019889144 i$ &
$-0.007436085 - 0.019663868 i$ &
$-0.007215316 - 0.019308781 i$   \\\hline\hline

$\omega M(\ell=1)$ &
$-0.0057734786 i$ &
$0.0001927509 - 0.0057720234 i$ &
$0.0003854998 - 0.0057676545 i$ &
$0.0005782445 - 0.0057603622 i$ &
$0.0007709822 - 0.0057501298 i$   \\\hline
$\omega M(\ell=1)$ &
$-0.017321716 i$ &
$-0.000122282 - 0.017310465 i$ &
$-0.000242806 - 0.017276936 i$ &
$-0.000360010 - 0.017221745 i$ &
$-0.000472659 - 0.017145763 i$   \\\hline

$\omega M(\ell=1)$ &
$-0.022286414 i$ &
$0.000447952 - 0.022157367 i$ &
$0.000723994 - 0.021971393 i$ &
$0.000923035 - 0.021819014 i$ &
$0.001090578 - 0.021693638 i$   \\\hline

$\omega M(\ell=1)$ &
$-0.003106164 - 0.023416978 i$ &
$-0.002617271 - 0.022894427 i$ &
$-0.002448842 - 0.022471247 i$ &
$-0.002344076 - 0.022161069 i$ &
$-0.002250014 - 0.021907289 i$   \\\hline

$\omega M(\ell=1)$ &
$0.003106164 - 0.023416978 i$ &
$0.003866336 - 0.023455011 i$ &
$0.004310642 - 0.023241390 i$ &
$0.004590184 - 0.023058403 i$ &
$0.004805912 - 0.022911344 i$   \\\hline

$\omega M(\ell=1)$ &
$-0.007211353 - 0.024617310 i$ &
$-0.006739418 - 0.024096567 i$ &
$-0.006545579 - 0.023665653 i$ &
$-0.006419197 - 0.023330274 i$ &
$-0.006309101 - 0.023045070 i$   \\\hline\hline

$\omega M(\ell=30)$ &
$6.1510961 - 0.0189954 i$ &
$6.1634991 - 0.0188729 i$ &
$6.1759151 - 0.0187616 i$ &
$6.1883439 - 0.0186612 i$ &
$6.2007853 - 0.0185716 i$   \\\hline
$\omega M(\ell=30)$ &
$-6.1510961 - 0.0189954 i$ &
$-6.1387062 - 0.0191292 i$ &
$-6.1263296 - 0.0192746 i$ &
$-6.1139664 - 0.0194317 i$ &
$-6.1016168 - 0.0196007 i$   \\\hline

$\omega M(\ell=30)$ &
$-5.7102395 - 0.0221797 i$ &
$-5.6989495 - 0.0219258 i$ &
$-5.6876736 - 0.0216726 i$ &
$-5.6764118 - 0.0214202 i$ &
$-5.6651640 - 0.0211685 i$   \\\hline

$\omega M(\ell=30)$ &
$5.7102395 - 0.0221797 i$ &
$5.7215437 - 0.0224343 i$ &
$5.7328622 - 0.0226895 i$ &
$5.7441948 - 0.0229452 i$ &
$5.7555417 - 0.0232015 i$   \\\hline

$\omega M(\ell=30)$ &
$-5.2723392 - 0.0332193 i$ &
$-5.2617881 - 0.0329601 i$ &
$-5.2512526 - 0.0327006 i$ &
$-5.2407329 - 0.0324408 i$ &
$-5.2302289 - 0.0321808 i$   \\\hline

$\omega M(\ell=30)$ &
$5.2723392 - 0.0332193 i$ &
$5.2829060 - 0.0334782 i$ &
$5.2934885 - 0.0337369 i$ &
$5.3040867 - 0.0339952 i$ &
$5.3147006 - 0.0342533 i$   \\\hline

\end {tabular}
}
\end{table}\leavevmode\newline

\begin{table}[H]
\caption {The QNFs $\omega M$ for massive ($m M=0.20$) scalar fields in the background of RNdS black holes with $\Lambda M^2 = 10^{-4} $, $Q/M=0.50$, and different values of $\ell$, and $qM$.}
\label {S2M}\centering
\scalebox{0.7} {

\begin {tabular} { | c | c | c | c | c | c |}
\hline
${}$ & $qM = 0$ & $qM = 0.1$  & $qM = 0.2$ & $qM = 0.3$ & $qM = 0.4$ \\\hline
$\omega M(\ell=0)$ &
$-0.19132860 - 0.00315499 i$ &
$-0.18867665 - 0.00305974 i$ &
$-0.18626080 - 0.00356834 i$ &
$-0.18443675 - 0.00462611 i$ &
$-0.18379078 - 0.00561202 i$   \\\hline
$\omega M(\ell=0)$ &
$0.19132860 - 0.00315499 i$ &
$0.19396839 - 0.00383565 i$ &
$0.19638169 - 0.00501415 i$ &
$0.19841744 - 0.00651629 i$ &
$0.20004933 - 0.00812810 i$   \\\hline
$\omega M(\ell=0)$ &
$-0.19373987 - 0.01353939 i$ &
$-0.19152607 - 0.01254376 i$ &
$-0.18912372 - 0.01171669 i$ &
$-0.18636279 - 0.01109345 i$ &
$-0.18284215 - 0.01128977 i$   \\\hline
$\omega M(\ell=0)$ &
$0.19373987 - 0.01353939 i$ &
$0.19579903 - 0.01471733 i$ &
$0.19767161 - 0.01608517 i$ &
$0.19928772 - 0.01760208 i$ &
$0.20061019 - 0.01915549 i$   \\\hline
$\omega M(\ell=0)$ &
$-0.19505985 - 0.02462279 i$ &
$-0.19317932 - 0.02357447 i$ &
$-0.19124917 - 0.02263217 i$ &
$-0.18924279 - 0.02176383 i$ &
$-0.18710423 - 0.02092316 i$   \\\hline
$\omega M(\ell=0)$ &
$0.19505985 - 0.02462279 i$ &
$0.19689063 - 0.02580918 i$ &
$0.19862582 - 0.02716048 i$ &
$0.20015996 - 0.02865497 i$ &
$0.20140724 - 0.03017056 i$   \\\hline\hline

$\omega M(\ell=1)$ &
$-0.19608114 - 0.01037118 i$ &
$-0.19497988 - 0.00939555 i$ &
$-0.19381800 - 0.00841726 i$ &
$-0.19258731 - 0.00743797 i$ &
$-0.19127666 - 0.00646013 i$   \\\hline
$\omega M(\ell=1)$ &
$0.19608114 - 0.01037118 i$ &
$0.19712806 - 0.01134282 i$ &
$0.19812572 - 0.01230928 i$ &
$0.19907835 - 0.01326949 i$ &
$0.19998962 - 0.01422247 i$   \\\hline

$\omega M(\ell=1)$ &
$-0.19703829 - 0.02183067 i$ &
$-0.19615203 - 0.02087952 i$ &
$-0.19523592 - 0.01990747 i$ &
$-0.19428639 - 0.01890898 i$ &
$-0.19329748 - 0.01787596 i$   \\\hline

$\omega M(\ell=1)$ &
$0.19703829 - 0.02183067 i$ &
$0.19789713 - 0.02276473 i$ &
$0.19873039 - 0.02368430 i$ &
$0.19953959 - 0.02459121 i$ &
$0.20032603 - 0.02548680 i$   \\\hline

$\omega M(\ell=1)$ &
$-0.19794190 - 0.03318339 i$ &
$-0.19718303 - 0.03230904 i$ &
$-0.19639899 - 0.03140143 i$ &
$-0.19559464 - 0.03044962 i$ &
$-0.19477477 - 0.02944099 i$   \\\hline

$\omega M(\ell=1)$ &
$0.19794190 - 0.03318339 i$ &
$0.19867304 - 0.03403459 i$ &
$0.19937697 - 0.03487166 i$ &
$0.20005735 - 0.03570200 i$ &
$-0.19459457 - 0.03595925 i$   \\\hline\hline

$\omega M(\ell=30)$ &
$6.1525755 - 0.0190319 i$ &
$6.1649792 - 0.0189084 i$ &
$6.1773958 - 0.0187960 i$ &
$6.1898253 - 0.0186946 i$ &
$6.2022674 - 0.0186040 i$   \\\hline
$\omega M(\ell=30)$ &
$-6.1525755 - 0.0190319 i$ &
$-6.1401850 - 0.0191669 i$ &
$-6.1278077 - 0.0193134 i$ &
$-6.1154439 - 0.0194716 i$ &
$-6.1030937 - 0.0196419 i$   \\\hline

$\omega M(\ell=30)$ &
$-5.7120655 - 0.0221867 i$ &
$-6.1401850 - 0.0191669 i$ &
$-6.1278077 - 0.0193134 i$ &
$-5.6782358 - 0.0214276 i$ &
$-5.6669874 - 0.0211761 i$   \\\hline

$\omega M(\ell=30)$ &
$5.7120655 - 0.0221867 i$ &
$5.7233703 - 0.0224411 i$ &
$5.7346893 - 0.0226962 i$ &
$5.7460225 - 0.0229518 i$ &
$5.7573699 - 0.0232080 i$   \\\hline

$\omega M(\ell=30)$ &
$-5.2745392 - 0.0332292 i$ &
$-5.2639876 - 0.0329701 i$ &
$-5.2534517 - 0.0327107 i$ &
$-5.2429314 - 0.0324511 i$ &
$-5.2324269 - 0.0321912 i$   \\\hline

$\omega M(\ell=30)$ &
$5.2745392 - 0.0332292 i$ &
$5.2851065 - 0.0334880 i$ &
$5.2956895 - 0.0337465 i$ &
$5.3062881 - 0.0340047 i$ &
$5.3169025 - 0.0342626 i$   \\\hline

\end {tabular}
}
\end{table}\leavevmode\newline

\begin{table}[H]
\caption {The QNFs $\omega M$ for massive  scalar fields in the background of RNdS black holes with $\Lambda M^2= 10^{-4} $, $Q/M=0.50$, $\ell=0$, and different values of $qM$.}
\label {smass}\centering
\scalebox{0.7} {

\begin {tabular} { | c | c | c | c | c | c | c |}
\hline
$q M$ & $m M= 0$  & $mM =1/5000$ &  $mM =1/2500$ & $mM = 1/1000$ & $mM = 1/100$   \\\hline
$0.0$ &
$0$ &
$-2.2959973*10^{-6} i$ &
$-9.1877087*10^{-6} i$ &
$-0.000057586955 i$ &
$-0.0050437442 - 0.0086028401 i$ \\\hline
$0.1$ &
$0.00029375585 + 5.90732*10^{-6} i$ &
$0.00029368229 + 3.61271*10^{-6} i$ &
$0.00029346139 - 3.27482*10^{-6} i$ &
$0.00029190610 - 0.00005164471 i$  &
$-0.0049041729 - 0.0082629975 i$ \\\hline
$0.2$ &
$0.00058922420 + 0.00002374651 i$ &
$0.00058907652 + 0.00002145619 i$ &
$0.00058863305 + 0.00001458151 i$ &
$0.00058551056 - 0.00003369793 i$  &
$-0.0047602485 - 0.0079008639 i$ \\\hline
$0.3$ &
$0.00088833967 + 0.00005385197 i$ &
$0.00088811674 + 0.00005156919 i$ &
$0.00088744730 + 0.00004471714 i$ &
$0.00088273365 - 3.40310*10^{-6} i$ &
$-0.0046095965 - 0.0075113920 i$ \\\hline
$0.4$ &
$0.00119351969 + 0.00009670646 i$ &
$0.00119321983 + 0.00009443516 i$ &
$0.00119231935 + 0.00008761761 i$ &
$0.00118597879 + 0.00003974005 i$  &
$-0.0044475144 - 0.0070875010 i$  \\\hline

\end {tabular}
}
\end{table}

\subsection{Anomalous decay rate}

 Using the WKB method and the pseudospectral Chebyshev method in this subsection we will look for possible anomalous decay rate
for quasinormal modes.

\subsubsection{Analysis using the WKB method}
\label{WKBJ}

In order to get some analytical insight of the behaviour
of the QNFs, we use the WKB method \cite{Mashhoon, Schutz:1985zz, Iyer:1986np, Konoplya:2003ii, Matyjasek:2017psv, Konoplya:2019hlu} that can be used for effective potentials which have the form of a barrier potential, approaching to a constant value at the event horizon and cosmological horizon or spatial infinity \cite{Konoplya:2011qq}. Here, we consider the eikonal limit $\ell \rightarrow \infty$   to estimate the critical scalar field mass, by
considering $\omega_I^\ell=\omega_I^{\ell+1}$ as a proxy for where the transition or critical behaviour occurs \cite{Lagos:2020oek}.
Here, we obtain the following analytical QNFs valid for small values of $qM$, $Q/M$ and large values of $L=\sqrt{\ell(\ell+1)}$ by using  the WKB method (see appendix \ref{AWKB}):

\begin{equation}
\label{omegarelation}
\omega \approx \omega_1 L+ \omega_0 +\omega_{-1} L^{-1} + \omega_{-2} L^{-2} + \mathcal{O} (L^{-3})\,,
\end{equation}
where

\begin{equation}
\nonumber
  \omega_1 M \approx  \pm \frac{\sqrt{1-9 \Lambda M^2}}{3 \sqrt{3}} \pm \frac{(Q/M)^2}{18 \sqrt{3(1-9 \Lambda M^2)}} \pm \frac{(13-144 \Lambda M^2)(Q/M)^4}{648 \sqrt{3} (1- 9 \Lambda M^2)^{3/2}} + \dots  \,,  \\
\end{equation}
 
 \begin{equation}
     \nonumber
     \omega_0 M \approx \frac{qQ}{3} +\frac{2 q M (Q/M)^3}{27} -i \frac{\sqrt{1-9 \Lambda M^2}}{6 \sqrt{3}}-i \frac{(1+18 \Lambda M^2)(Q/M)^2}{108\sqrt{3(1-9 \Lambda M^2)}}+i \frac{(1-18 \Lambda M^2)^2 (Q/M)^4}{432 \sqrt{3(1-9 \Lambda M^2)}(1- 9 \Lambda M^2)}+ \dots  \,, \\
 \end{equation}
 
 \begin{eqnarray}
\nonumber
 \omega_{-1} M &\approx&  \pm \frac{(34 + 9 (108 (\mu M)^2 -61 \Lambda M^2)) \sqrt{ 1-9 \Lambda M^2}}{648  \sqrt{3}} \\
&& \nonumber
  \pm \frac{(122+  9 (\Lambda M^2 (251-4086 \Lambda M^2) -108 (\mu M)^2 (5-72 \Lambda M^2)))(Q/M)^2}{11664 \sqrt{3 (1-9\Lambda M^2 )}}  \\
&& \nonumber
 \pm \frac{ (1550-27  (\Lambda M^2 (-139+36 \Lambda  M^2(313-1524 \Lambda M^2)))) (Q/M)^4}{419904 \sqrt{3 (1- 9\Lambda M^2)} (1-9 \Lambda M^2)} \\
&& \nonumber \mp \frac{ (27 (-36 (\mu M)^2 (41+288 \Lambda M^2 (-2+9 \Lambda M^2)))) (Q/M)^4}{419904 \sqrt{3 (1- 9\Lambda M^2)} (1-9 \Lambda M^2)} 
 \mp i \frac{q M (Q/M)^3 \Lambda M^2}{6} \mp i \frac{q Q (1- 9 \Lambda M^2)}{18} + \dots\,, \\
 \end{eqnarray}

 \begin{eqnarray}
\nonumber
 \omega_{-2} M  &\approx& -i \frac{(137 -45 (648 (\mu M)^2-401 \Lambda M^2)) (1- 9 \Lambda M^2 )^{3/2}}{ 23328 \sqrt{3}} \\
 && \nonumber +i \frac{(145 + 27  \Lambda M^2 (1265 -31098 \Lambda M^2))(1- 9\Lambda M^2)^{1/2} (Q/M)^2}{139968 \sqrt{3}} \\
&&  \nonumber
 -i \frac{5832 (\mu M)^2 (11-234 \Lambda M^2)(1- 9\Lambda M^2)^{1/2} (Q/M)^2}{139968 \sqrt{3}} -i \frac{3888 (q M)^2 (2 +9\Lambda M^2)(1- 9\Lambda M^2)^{1/2} (Q/M)^2}{139968 \sqrt{3}}  \\
&&  \nonumber
  +i\frac{(1505 + 3 (5832 (\mu M)^2 (1+6 \Lambda M^2)(-11+ 126 \Lambda M^2))) (Q/M)^4}{1679616  \sqrt{3(1- 9\Lambda M^2)}} \\
&&  \nonumber  +i\frac{3 (31283 \Lambda M^2 - 36 (\Lambda M^2)^2  (5530 + 74151 \Lambda M^2)) (Q/M)^4}{1679616 \sqrt{3(1- 9\Lambda M^2)}} \\
&&  \nonumber  - i\frac{2592 (q M)^2 (16 -9 \Lambda M^2 (17 - 90\Lambda M^2)) (Q/M)^4}{1679616  \sqrt{3(1- 9\Lambda M^2)}}+ \dots  \,. 
\end{eqnarray}

\begin{figure}[h]
\begin{center}
\includegraphics[width=0.48\textwidth]{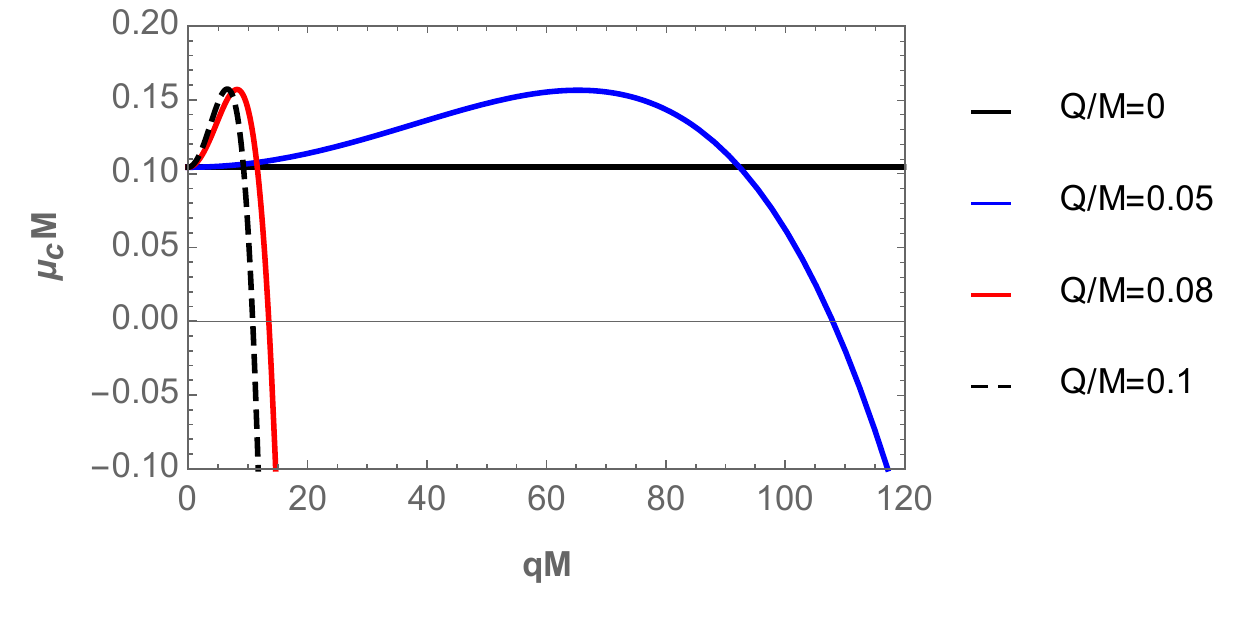}
\includegraphics[width=0.48\textwidth]{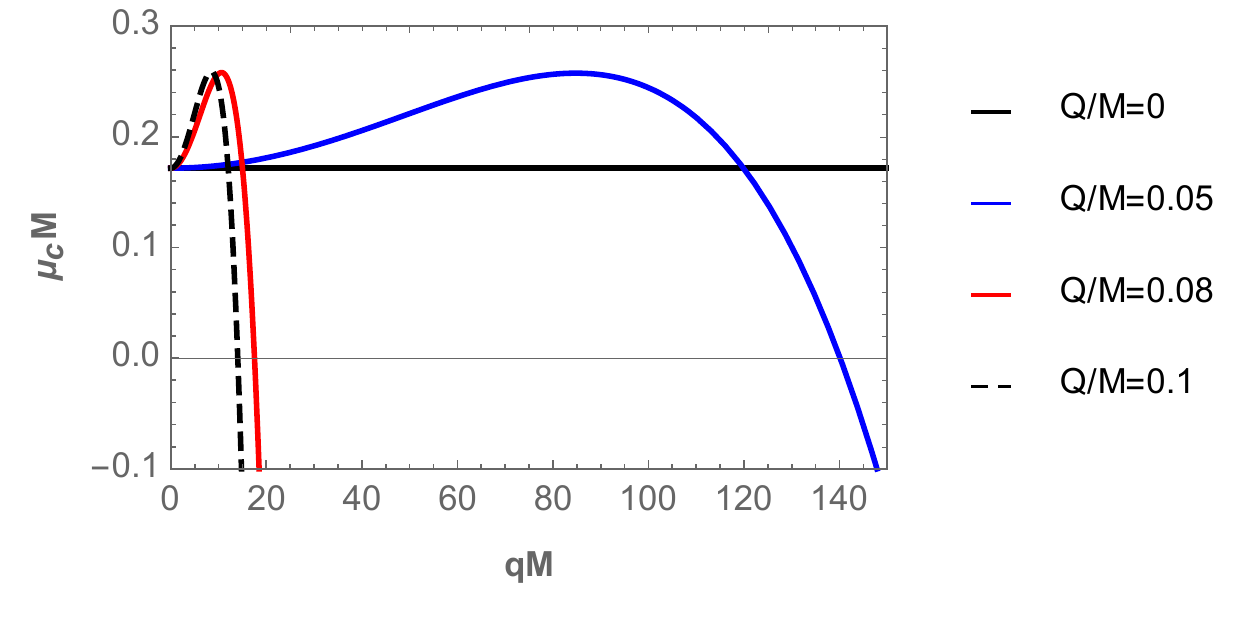}
\end{center}
\caption{The behaviour of the critical scalar field mass $\mu_c M$ as a function of $qM$, with $\Lambda M^2=0.01$ (left panel), $\Lambda M^2=0.04$ (right panel).}
\label{WKB}
\end{figure}

 The dots $\dots$ represents terms of higher order in $qM$ and $Q/M$. The upper/lower sign corresponds to the positive/negative branch.
The imaginary part of $\omega_{-2}$ is zero at the value of the critical mass $\mu_c$, which is given by \\

\begin{eqnarray}
\label{mc}
\nonumber
\mu_c M  &\approx&  \frac{\sqrt{137+18045 \Lambda M^2 }}{54 \sqrt{10}} \Big(1+\frac{391 + 9720(qM)^2 (2+ 9\Lambda M^2) -9 \Lambda M^2 (241 + 1350 \Lambda M^2)}{ 30 (1- 9 \Lambda M^2) (137+18045\Lambda M^2)}(\frac{Q}{M})^2 \\
\nonumber
&&- \frac{3277- 60 \Lambda M^2 (408779+12871800 \Lambda M^2)}{1800 (137+18045 \Lambda M^2)^2} (\frac{Q}{M})^4 \\
\nonumber
&& + \frac{18 (qM)^2 (17656 -27 \Lambda M^2 (-88169 + 9 \Lambda M^2 (139751 +167070 \Lambda M^2)))}{5 (1- 9 \Lambda M^2)^2 (137+18045 \Lambda M^2)^2} (\frac{Q}{M})^4  \\  
&& - \frac{52488 (qM)^4 (2 + 9 \Lambda M^2)^2}{(1-9\Lambda M^2)^2 (137 + 18045 \Lambda M^2)^2} (\frac{Q}{M})^4 \Big) + \mathcal{O}(Q^5)\,, 
\end{eqnarray}
and it is valid for small values of $qM$, $Q/M$ and $n_{PS}=0$. For $Q=0$ we recover the result of critical mass for Schwarzschild-dS black holes \cite{Aragon:2020tvq},  while for $q=0$ we recover the Reissner-Nordstr\"om black holes \cite{Fontana:2020syy}. In Fig. \ref{WKB} we show the behaviour of the critical mass $\mu_c M$ as a function of the scalar field charge. The curves are valid in the regime $Q/M$, $qM$ small and $n_{PS}=0$.

Also, we can observe that the value of the critical mass increases when scalar charge increases until
\begin{equation}
    q_0= \frac{1}{54\sqrt{10}}\sqrt{\frac{9 M^2 \left(3 \Lambda  \left(88169 Q^2-9 M^2 \left(\Lambda  \left(\eta+139751 Q^2\right)-12910\right)\right)+2740\right)+17656 Q^2}{M^2 Q^2 \left(9 \Lambda  M^2 \left(9 \Lambda  M^2+4\right)+8\right)}}\,,
\end{equation}
where the critical mass reaches a maximum value, with $\eta=30 M^2 \left(18045 \Lambda  M^2+5569 \Lambda  Q^2+2142\right)$, certainly, beyond $q_0$ the conditions of small $qM$ and $Q/M$ are not satisfied. Despite it,  then, the critical scalar field mass begin to decreases when $q$ increases. Also, notice that there is a positive range of $q$ for which the critical mass is positive, then it take negative values. Also, the critical scalar field mass increases when the cosmological constant increases.

\begin{figure}[H]
\begin{center}
\includegraphics[width=0.5\textwidth]{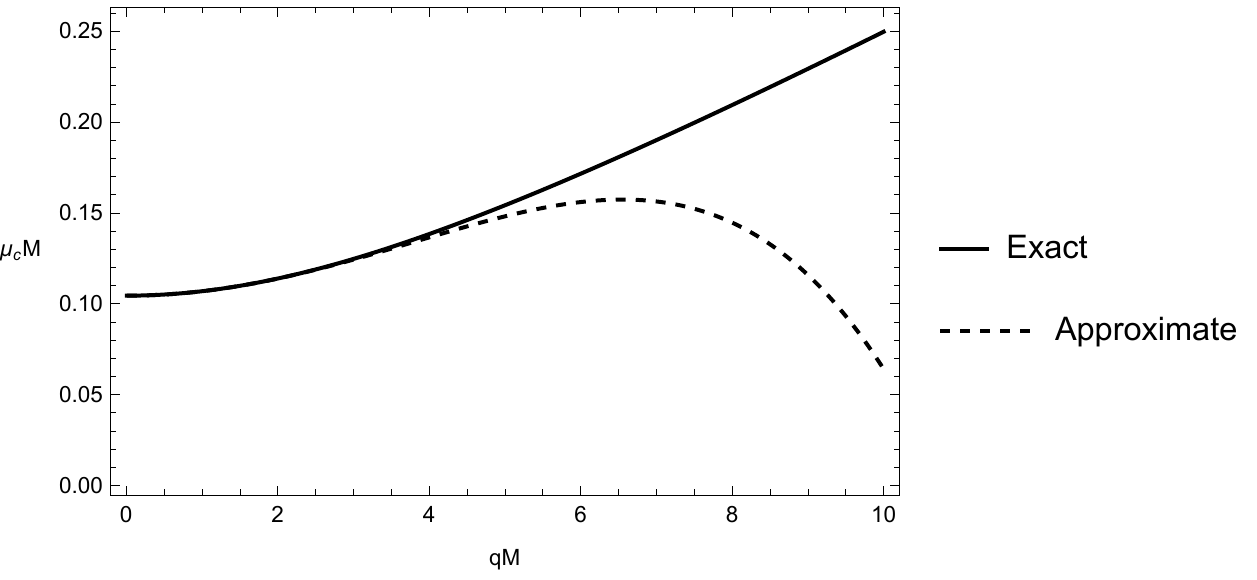}
\end{center}
\caption{The behaviour of the critical scalar field mass $\mu_c M$ as a function of $qM$, with $Q/M =0.1$ and $\Lambda M^2 =0.01$. The continuous line corresponds to the exact results, and the dashed line corresponds to the approximate results given by (\ref{mc}). }
\label{result}
\end{figure}

\begin{figure}[H]
\begin{center}
\includegraphics[width=0.4\textwidth]{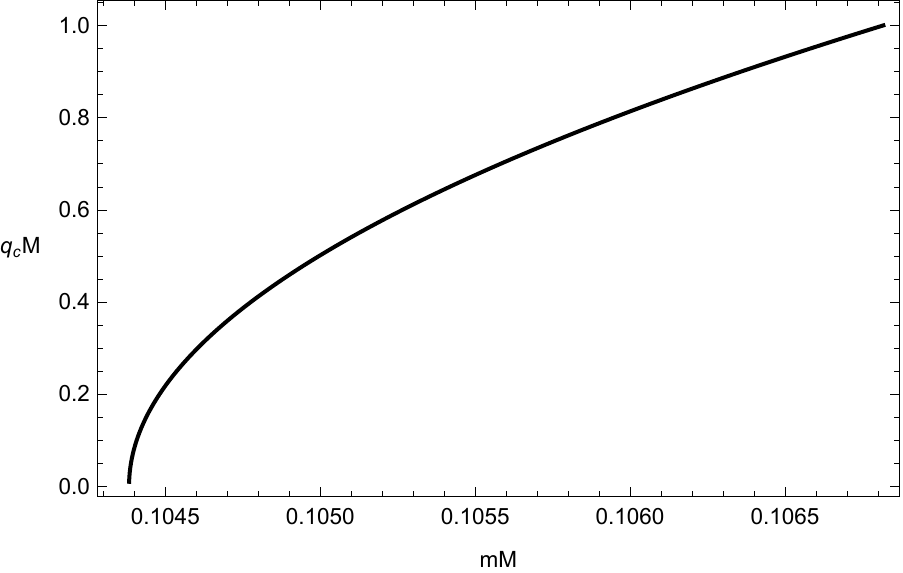}
\end{center}
\caption{The behaviour of the critical scalar field charge $q_c M$ as a function of $mM$, with $Q/M =0.1$ and $\Lambda M^2 =0.01$.}
\label{critica}
\end{figure}

The analytical expression for the critical mass without approximations is very lengthy and we do not write it here. However, in Fig. (\ref{result}) we plot the exact values of the critical mass and the approximate values given by ($\ref{mc}$) as a function of the scalar field charge. Notice that the approximate results are in very good agreement with the exact results for $qM \lesssim 5$. Also, in Fig. (\ref{critica}) we observe that for a fixed value of the scalar field mass, there exist a critical value of the scalar field charge where beyond this value the behavior of the decay rate is inverted. However for $mM \lesssim 0.1044$ there is not a critical charge.

\subsubsection{Analysis using the  pseudospectral Chebyshev method}

\begin{figure}[H]
\begin{center}
\includegraphics[width=0.42\textwidth]{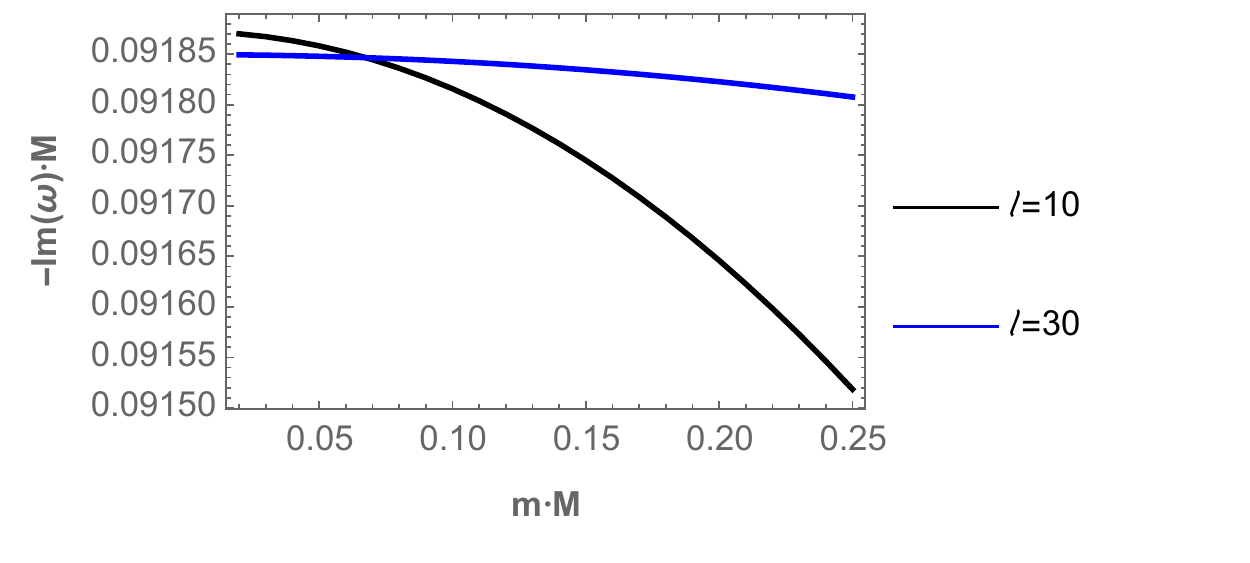}
\includegraphics[width=0.42\textwidth]{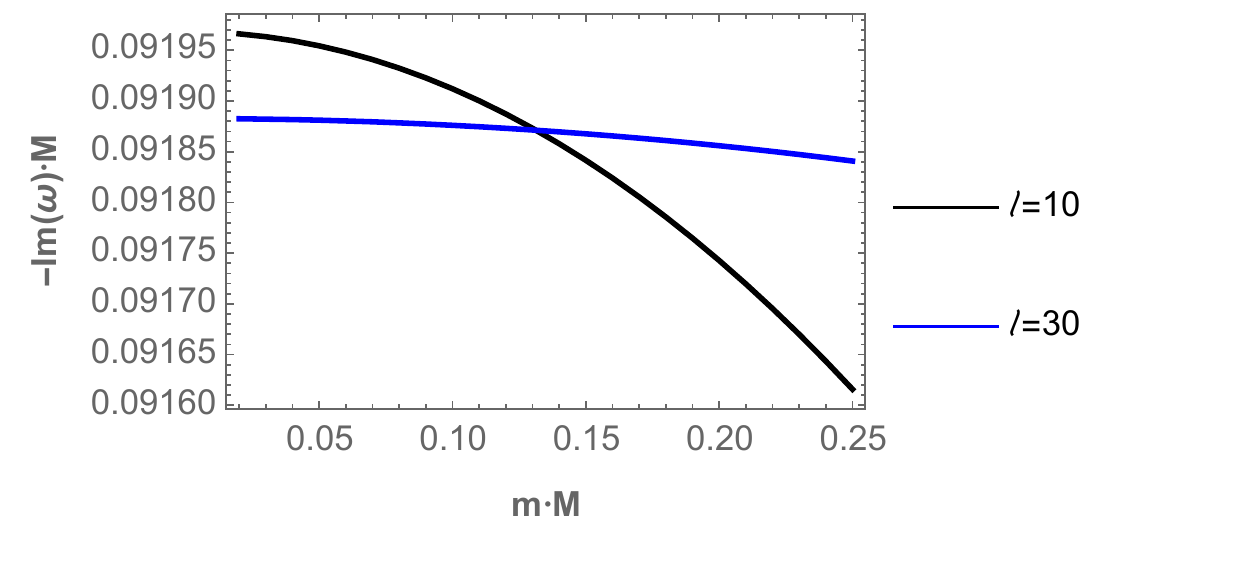}
\includegraphics[width=0.42\textwidth]{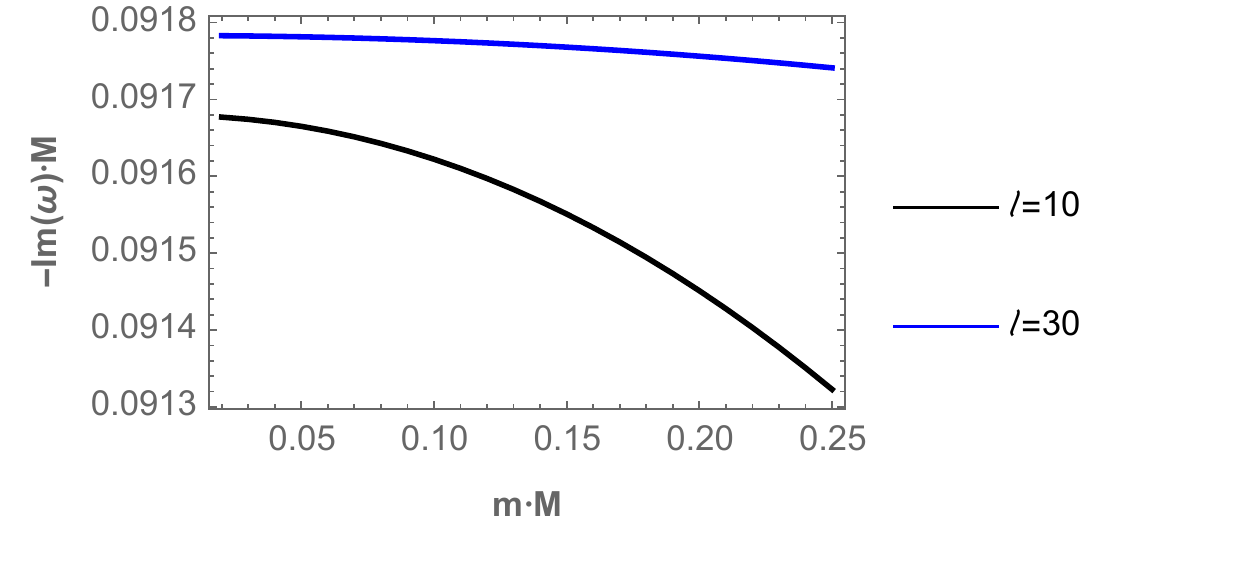}
\includegraphics[width=0.42\textwidth]{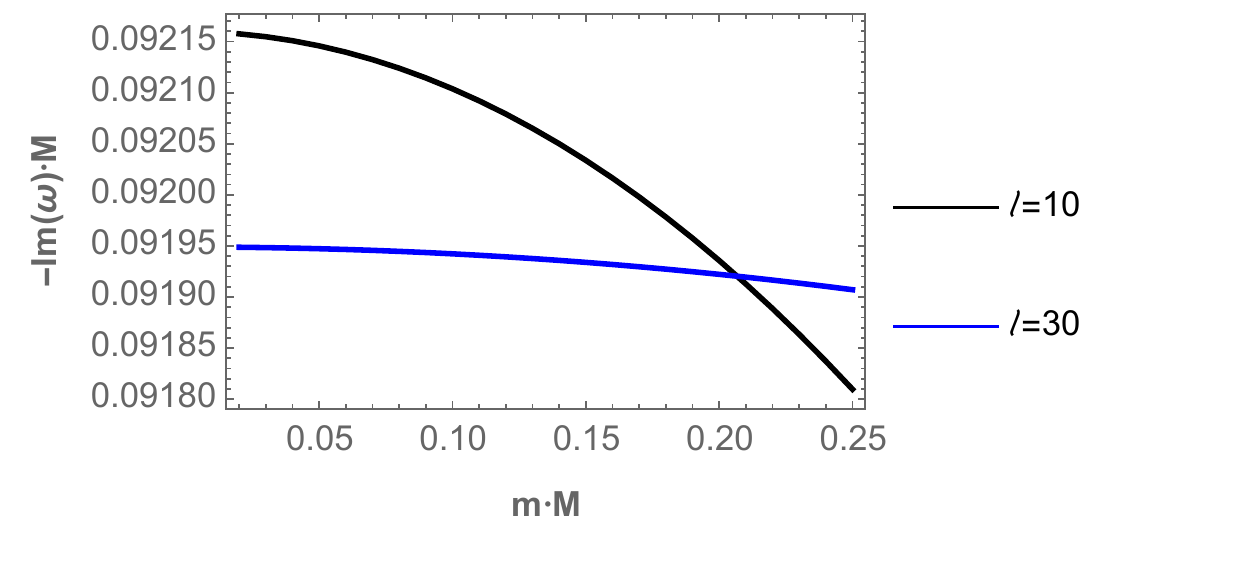}
\end{center}
\caption{
Top panel for the behaviour of
$Im(\omega) M$ of the photon sphere modes,
left panels for the dominant mode $n_{PS}=0$, and right panels for the first overtone $n_{PS}=1$,
as a function of $mM$ for $\ell=10,30$, with
$\Lambda M^2=0.01$,  $qQ=0.01$ (top panels), and $qQ=0.05$ (bottom panel).
Here, the WKB method gives via Eq. (\ref{mc}) $\mu_c M\approx 0.109$, for $qQ=0.01$, and  $\mu_c M\approx 0.110$, for $qQ=0.05$.
}
\label{FA}
\end{figure}

\begin{figure}[H]
\begin{center}
\includegraphics[width=0.42\textwidth]{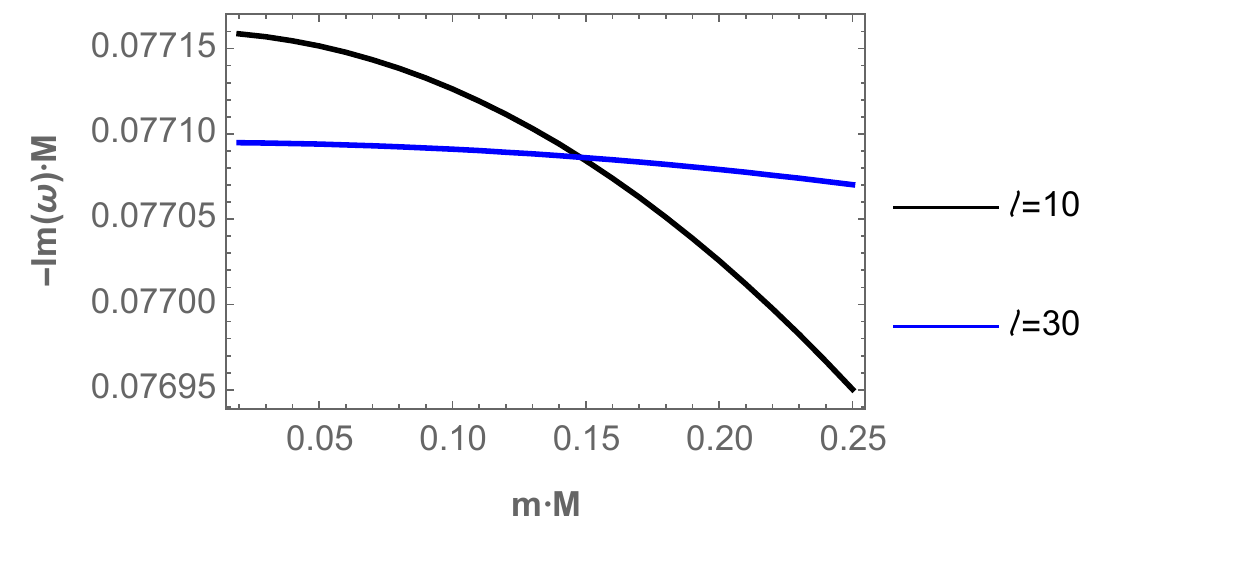}
\includegraphics[width=0.42\textwidth]{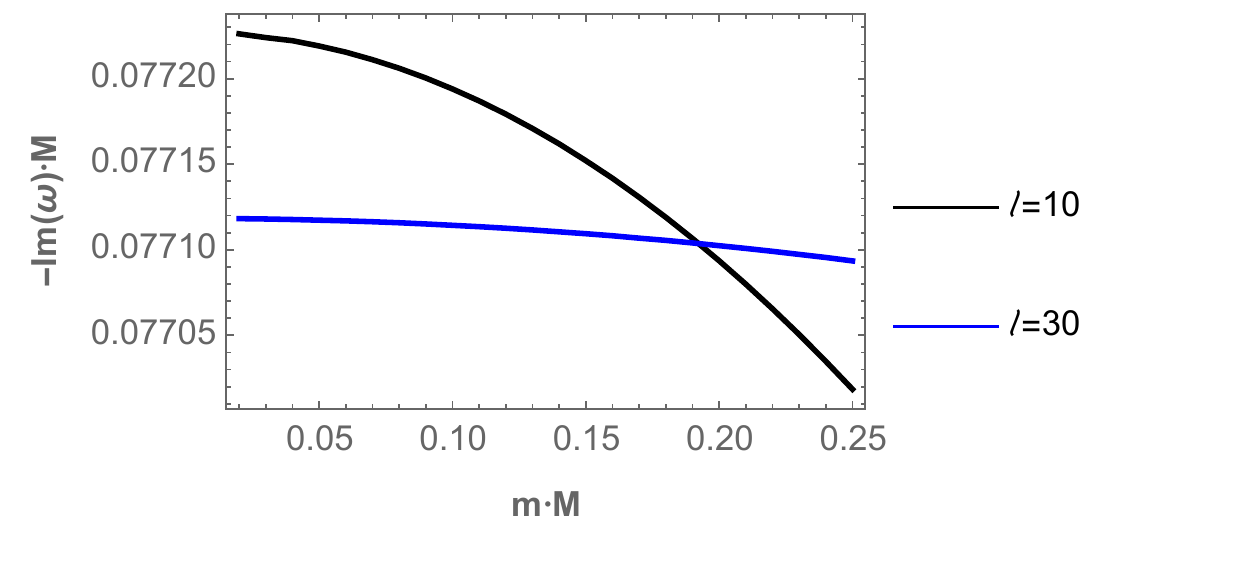}
\includegraphics[width=0.42\textwidth]{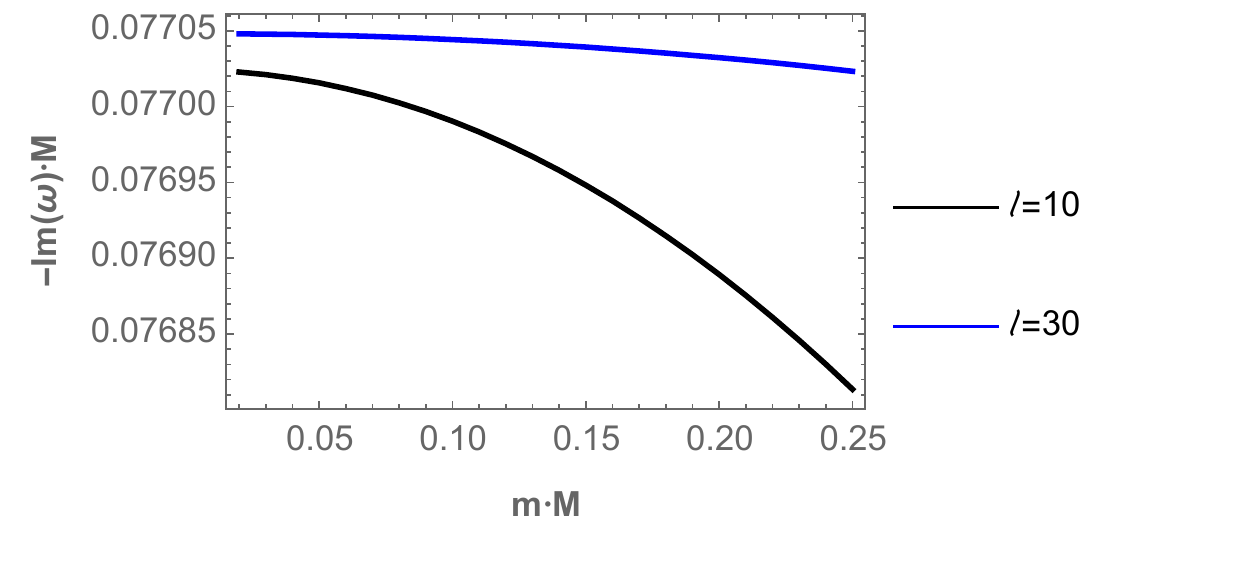}
\includegraphics[width=0.42\textwidth]{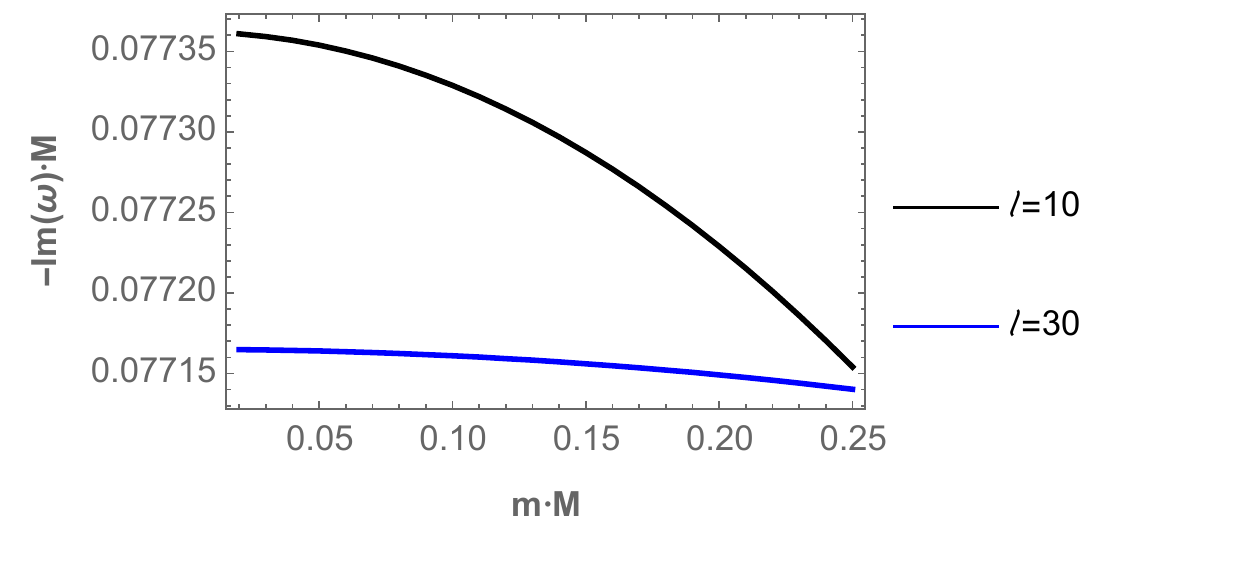}
\end{center}
\caption{
Top panel for the behaviour of
$-Im(\omega) M$ of the photon sphere modes,
left panels for the mode $n_{PS}=0$, and right panels for the mode  $n_{PS}=1$,
as a function of $mM$ for $\ell=10,30$, with
$\Lambda M^2=0.04$, $qQ=0.01$ (top panels), and $qQ=0.05$ (bottom panel).
Here, the WKB method gives via Eq. (\ref{mc}) $\mu_c M\approx 0.175$, for $qQ=0.01$, and  $\mu_c M\approx 0.176$, for $qQ=0.05$.}
\label{FB}
\end{figure}

In order to study the effect of the scalar field charge in the anomalous behaviour of the decay rate we plot  $-Im(\omega)M$ as a function of $mM$ for $\Lambda M^2=0.01$ in Fig. \ref{FA}, and $\Lambda M^2=0.04$ in Fig. \ref{FB} with $qQ=0.01$ (top panels) and $qQ=0.05$ (bottom panels).
We can observe the existence of an anomalous decay rate, that is, when the longest-lived modes are the ones with higher angular number, and the existence of a critical scalar field mass, where beyond of this critical mass, the behaviour of the longest-lived modes is inverted.  This can be seen in Fig. \ref{FA}, and Fig. \ref{FB} (Top panels) for different values of $\Lambda M^2$, $qQ=0.01$ and $n_{PS}=0,1$. However, when the parameter $qQ$ increases, the anomalous decay rate and a critical scalar field mass is clear for $n_{PS}=1$, see bottom panels Fig. \ref{FA}, and  Fig. \ref{FB}. The numerical values are in appendix \ref{NV}. This suggests the existence of a critical value of $qQ$ (previously predicted  by the WKB analysis) and a critical scalar field mass for $n_{PS}=0$, where beyond this value the anomalous decay rate disappears. It occurs at $qQ\approx 0.02$ for $\Lambda M^2= 0.01$, and at $qQ\approx 0.04 $ for $\Lambda M^2= 0.04$, see Table \ref{TqQ}.
Thus, the critical value of  $qQ$ increases when the cosmological constant increases, and the anomalous behaviour of the decay rate could be avoided beyond this value for the fundamental mode.

\begin {table}[h]
\caption {Photon sphere modes $\omega M$ ($n_{PS}=0,1$) for massive scalar fields $mM=0.02$ with $\ell = 10$, and $30$ in the background of a RNdS black hole with different values of $qQ$.}
\label {TqQ}\centering
\scalebox{0.7}{
\begin {tabular} { | c | c | c | c | c | c | c |}
\hline
$\Lambda M^2$ & $\ell$  & $qQ = 0.01$ & $qQ = 0.02$ & $qQ = 0.03$ & $qQ = 0.04$ & $qQ = 0.05$ \\\hline
$0.01 $ & $10$ &
$-1.9280093 - 0.0918701 I$ &
$-1.9246689 - 0.0918220 I$ &
$-1.9213303 - 0.0917737 I$ &
$-1.9179935 - 0.0917254 I$ &
$-1.9146584 - 0.0916770 I$  \\
\hline
$0.01 $ & $10$ &
$1.9346952 - 0.0919662 I$ &
$1.9380408 - 0.0920142 I$ &
$1.9413882 - 0.0920621 I$ &
$1.9447373 - 0.0921099 I$ &
$1.9480882 - 0.0921576 I$  \\
\hline
$0.01 $ & $30$ &
$-5.6063683 - 0.0918493 I$ &
$-5.6030281 - 0.0918327 I$ &
$-5.5996886 - 0.0918161 I$ &
$-5.5963497 - 0.0917995 I$ &
$-5.5930114 - 0.0917829 I$  \\
\hline
$0.01 $ & $30$ &
$5.6130503 - 0.0918825 I$ &
$5.6163923 - 0.0918990 I$ &
$5.6197348 - 0.0919156 I$ &
$5.6230780 - 0.0919321 I$ &
$5.6264217 - 0.0919487 I$  \\
\hline\hline
$0.04 $ & $10$ &
$-1.6164783 - 0.0771586 I$ &
$-1.6131374 - 0.0771247 I$ &
$-1.6097979 - 0.0770907 I$ &
$-1.6064599 - 0.0770568 I$ &
$-1.6031234 - 0.0770228 I$  \\
\hline
$0.04 $ & $10$ &
$1.6231646 - 0.0772262 I$ &
$1.6265100 - 0.0772599 I$ &
$1.6298568 - 0.0772936 I$ &
$1.6332051 - 0.0773273 I$ &
$1.6365548 - 0.0773609 I$  \\
\hline
$0.04 $ & $30$ &
$-4.7043696 - 0.0770948 I$ &
$-4.7010293 - 0.0770831 I$ &
$-4.6976895 - 0.0770714 I$ &
$-4.6943502 - 0.0770597 I$ &
$-4.6910115 - 0.0770480 I$  \\
\hline
$0.04 $ & $30$ &
$4.7110517 - 0.0771182 I$ &
$4.7143935 - 0.0771298 I$ &
$4.7177359 - 0.0771415 I$ &
$4.7210787 - 0.0771531 I$ &
$4.7244220 - 0.0771648 I$  \\
\hline
\end {tabular} }
\end {table}

\begin{figure}[H]
\begin{center}
\includegraphics[width=0.42\textwidth]{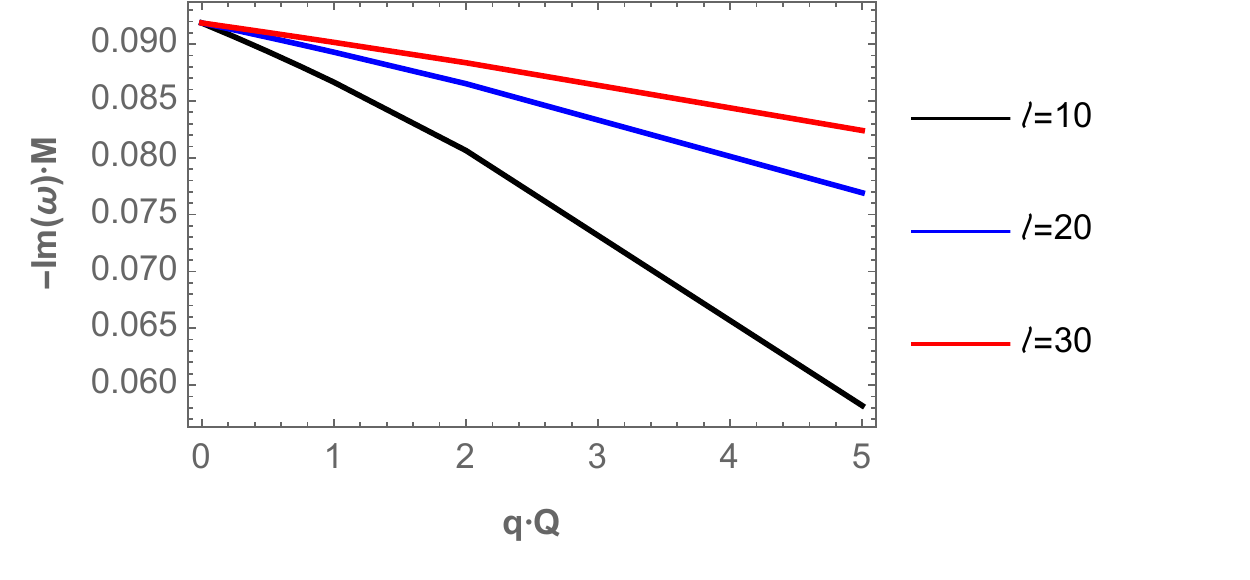}
\includegraphics[width=0.42\textwidth]{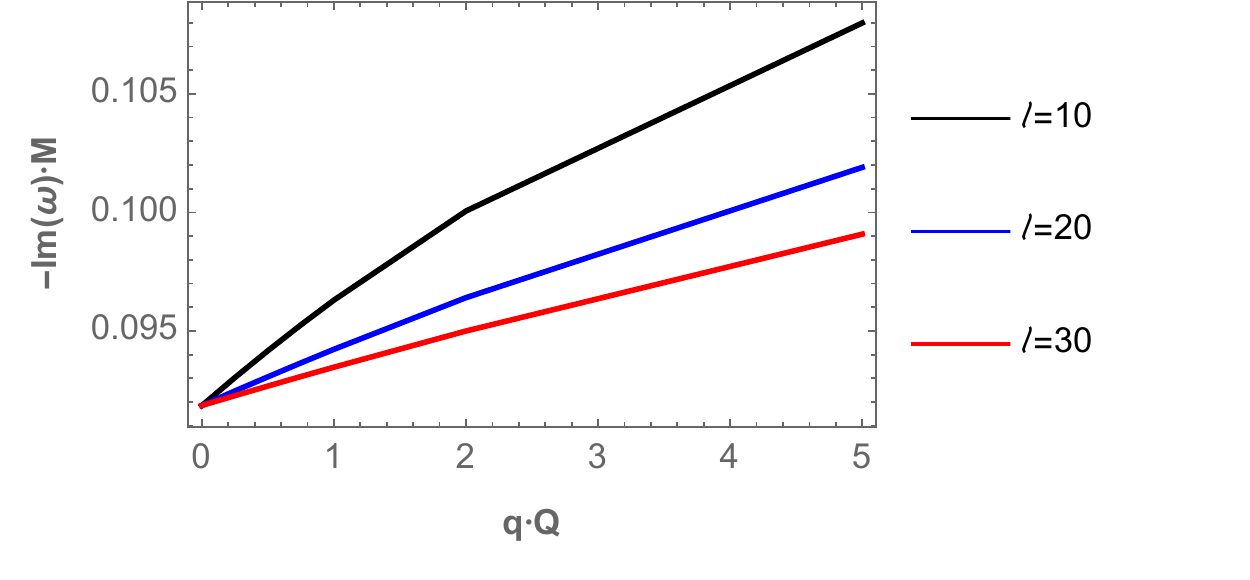}
\includegraphics[width=0.42\textwidth]{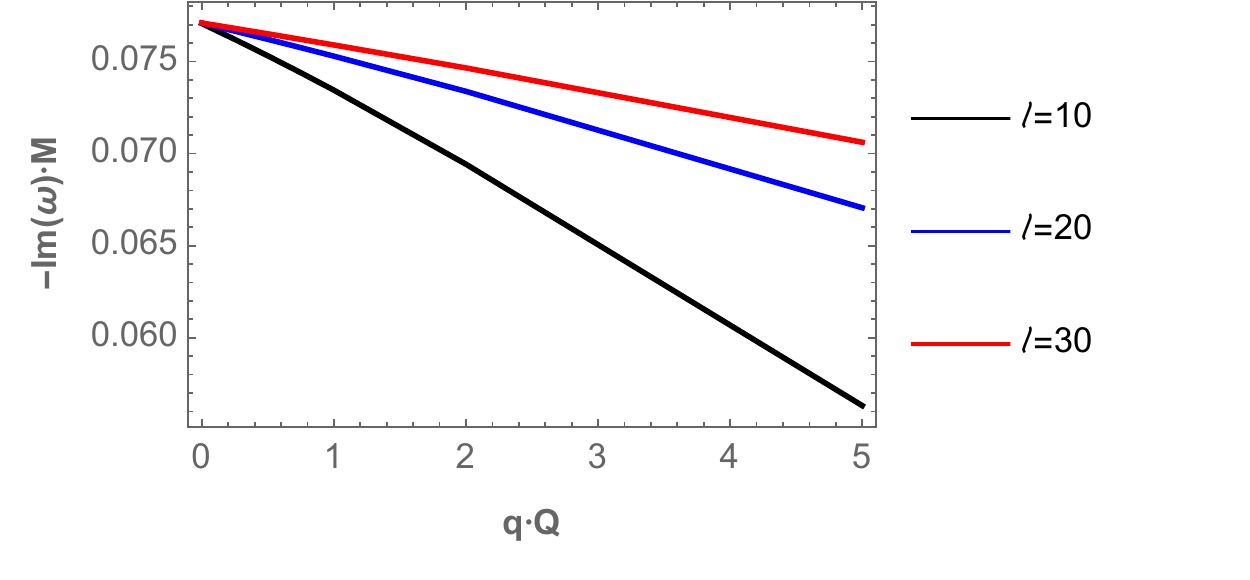}
\includegraphics[width=0.42\textwidth]{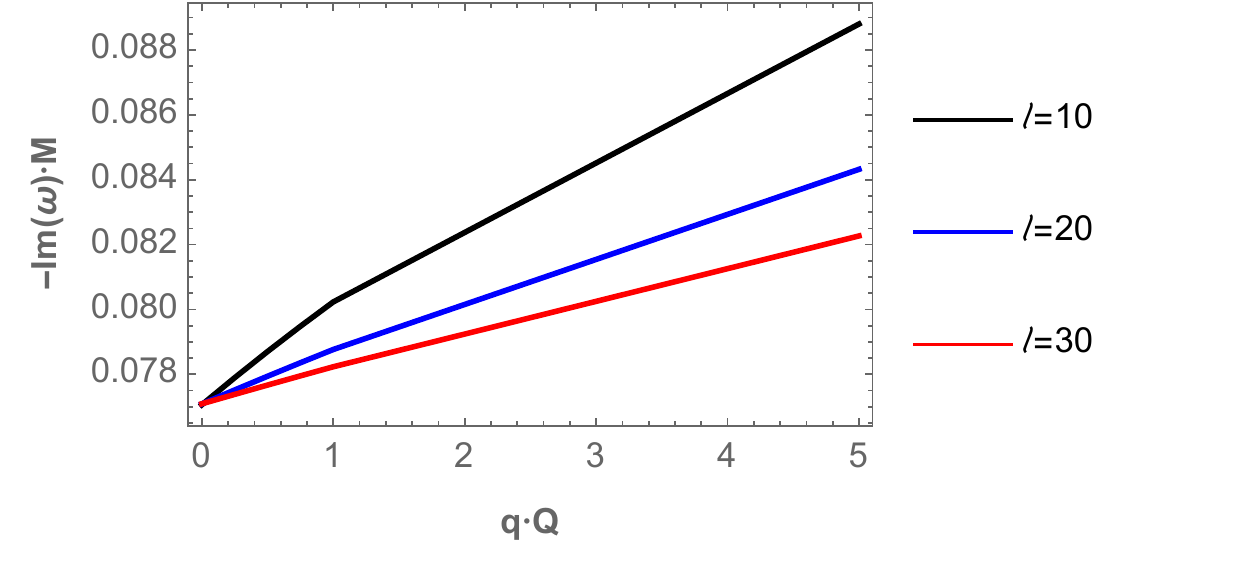}
\end{center}
\caption{Behaviour of
$-Im(\omega) M$ as a function of $qQ$,
left panels for the dominant mode $n_{PS}=0$ (negative branch), and right panels for the first overtone $n_{PS}=1$ (positive branch) for different values of the parameter $\ell=10,20,30$, with
$\Lambda M^2=0.01$, $mM=0.11$ (top panels), and $\Lambda M^2=0.04$, $mM=0.20$ (bottom panels).
}
\label{FA2}
\end{figure}

Now, in order to see the effect of the parameter $qQ$ on the behaviour of $-Im(\omega)M$, for bigger scalar field mass we set it 
, see Fig. \ref{FA2}. For $n_{PS}=0$ (left panels), the longest live modes are the ones with smallest angular number and the decay rate decreases when the parameter $qQ$ increases, while that for $n_{PS}=1$ (right panels), the longest live modes are the ones with higher angular number, and the decay rate increases when the parameter $qQ$ increases.

\subsection{Near extremal modes}

 The near extremal (NE) modes are generated  in the limit when the Cauchy  and  event  horizon  radius  approach  each  other.
Here, we show numerically that the "near  extremal" family, characterized by purely imaginary QNFs when the field is uncharged, acquires a real part when the  scalar field is charged.
The transition from complex PS modes to NE dominance modes,  when the inner and outer horizon get closer
has been found in various charged spacetimes \cite{Richartz:2014jla, Richartz:2015saa, Panotopoulos:2019tyg, Cardoso:2017soq, Berti:2003ud, Cardoso:2018nvb, Destounis:2018qnb, Liu:2019lon, Destounis:2019hca, Destounis:2019omd, Destounis:2020pjk, Destounis:2020yav,Aragon:2021ogo}. Here,
by excluding in the analysis the $\ell=0$ mode, as in  \cite{Cardoso:2017soq}, for $\Lambda M^2 =0.01$, see appendix \ref{NE} Table \ref{NE1}, the longest live NE modes are the ones with smallest angular number and the PS modes become dominant when $\ell$ increases. While that for $\Lambda M^2 = 0.04$, see appendix \ref{NE} Table \ref{NE2}, the PS modes are dominant. So, the near extremal modes does not present an anomalous decay rate. The same behavior was observed for RN black holes \cite{Fontana:2020syy}. It is worth mentioning that for a bigger value of the scalar field mass the dominant family is the PS family for the cases analyzed, see appendix \ref{NE}, Table \ref{TNE1} and \ref{TNE2}.

\section{Conclusions}
\label{conclusion}

In this work, we studied the propagation of charged and uncharged massive scalar fields in the Reissner-Nordstr\"om-dS black hole as a background, and we analyzed their QNFs.  For uncharged scalar field we  showed that 
more than one family of QNMs modes  are present,
and one of them is a family of complex QNFs, and the other one is a family of purely imaginary modes. The purely imaginary modes belong to the family of de Sitter modes, and they continuously
approach those of pure de Sitter space in the limit that the black hole vanishes.
The complex ones are those of the Reissner-Nordstr\"om black hole, and this family corresponds to the photon sphere modes. As in Ref. \cite{Cardoso:2017soq}, by excluding in the analysis the $\ell=0$ mode, we have shown that the dS family is dominant  for small values of the cosmological constant $\Lambda$, 
and in the opposite regime the PS modes are dominant. When the scalar field  acquires mass, the anomalous behaviour of the decay rate is present for the PS family \cite{Fontana:2020syy}. Also, we showed that for small values of the scalar field mass the behaviour in the dominant family is the same. However, for a bigger value of the scalar field mass the dominant family is the PS family.
.

When the massive scalar fields acquires charge, we  showed that de Sitter modes  acquire a real part. Also, we  showed that when the superradiance condition  is  fulfilled, a potential well is generated, and if the mass of the scalar field is smaller than a critical value, there are bound states which allow the accumulation of  energy to trigger the instability for $\ell=0$. However, for larger values that the critical one, there are not bound states and the perturbation wave can be easily absorbed by the black hole and the corresponding background becomes stable under charged scalar perturbations.

Using the WKB method we were able to estimate the value of the critical scalar field mass, and to found their dependence on the the scalar field charge in the eikonal limit for small values of $qM$, $Q/M$ and $n_{PS}=0$. The  value of the critical scalar field mass increases when the scalar charge increases until a certain value of $q$, where the critical mass takes a maximum value, and also  the cosmological constant increases. Using the pseudospectral Chebyshev method allowed us to perform an analysis of the QNFs in a space of parameter more larger than the WKB method and we showed  that the anomalous behaviour is present for massive charged scalar field, up to a critical scalar field mass, which  beyond  this value the behaviour is inverted.

We also showed that both branches of modes (negative and positive)  present an  anomalous decay rate behaviour for small values of the parameters $qQ=0.01$. %and $n_{PS}=0,1$.
However, when the parameter $qQ$ is increased to $qQ=0.05$, we found an anomalous decay rate behaviour for $n_{PS}=1$ (positive branch), which  it was not present for $n_{PS}=0$ (negative branch). This behaviour allowed us to determine and establish a critical value of the parameter $qQ$ that increases when the cosmological constant increases, and beyond this value the anomalous behaviour of the decay rate could be avoided for the fundamental mode (negative branch). Also, the positive/negative branch is described by an  absolute value of the frequency of oscillation and a decay rate  that increases/decreases when the parameter $qM$ increases. Also, we showed that the near extremal modes does not present an anomalous decay rate.

It would be interesting to extent this work to the case in which the background black hole is a hairy black hole \cite{Kolyvaris:2010yyf,Kolyvaris:2011fk}. In the case that the  charge of the scalar field backreacting to the background appears in the hairy black hole metric, i.e it is a primary scalar charge \cite{Rinaldi:2012vy,Gonzalez:2013aca,Charmousis:2014zaa,Gonzalez:2014tga}, it could influence the decay rate of the quasinormal modes generated by a test scalar field perturbing the hairy black hole. We expect that the matter distribution provided by the scalar field outside the horizon of the hairy black hole it would be influenced by the perturbations of the test scalar field and it will leave a trace on the quasinormal spectrum.

\acknowledgments
 We thank Kyriakos Destounis and Almendra Arag\'{o}n for his/her valuable comments. This work is partially supported by ANID Chile through FONDECYT Grant No 1210635 (J. S.).\\

\appendix{}

\section{WKB method}
\label{AWKB}

The QNMs are determined by the behaviour of the effective potential near its maximum value $r^*_{max}$. The Taylor series expansion of the potential around its maximum is given by
\begin{equation}
V(r^*)= V(r^*_{max})+ \sum_{i=2}^{i=\infty} \frac{V^{(i)}}{i!} (r^*-r^*_{max})^{i}   \,,
\end{equation}
where
\begin{equation}
V^{(i)}= \frac{d^{i}}{d r^{*i}}V(r^*)|_{r^*=r^*_{max}}
\end{equation}
corresponds to the $i$-th derivative of the potential with respect to $r^*$ evaluated at the maximum of the potential $r^*_{max}$. Using the WKB approximation up to 6th order the QNFs are given by the following expression \cite{Hatsuda:2019eoj}

\begin{equation} \label{omega}
\omega^2 = V(r^*_{max}) -2 i U \,,
\end{equation}
where

\begin{multline}
 U = N\sqrt{-V^{(2)}/2}
+\frac{i}{64} \left( -\frac{1}{9}\frac{V^{(3)2}}{V^{(2)2}} (7+60N^2)+\frac{V^{(4)}}{V^{(2)}}(1+4 N^2) \right)
+\frac{N}{2^{3/2} 288} \Bigg( \frac{5}{24} \frac{V^{(3)4}}{(-V^{(2)})^{9/2}} (77+188N^2) \\
+\frac{3}{4} \frac{V^{(3)2} V^{(4)}}{(-V^{(2)})^{7/2}}(51+100N^2)  + \frac{1}{8} \frac{V^{(4)2}}{(-V^{(2)})^{5/2}}(67+68 N^2)
+\frac{V^{(3)}V^{(5)}}{(-V^{(2)})^{5/2}}(19+28N^2)+\frac{V^{(6)}}{(-V^{(2)})^{3/2}} (5+4N^2)  \Bigg)\,,
\end{multline}

and $N=n_{PS}+1/2$, with $n_{PS}=0,1,2,\dots$, the overtone number.
\newline

Defining $L^2= \ell (\ell+1)$, we find that for large values of $L$ and low values of $Q$, the maximum of the potential is approximately at
\begin{equation}
\notag r_{max} \approx r_0+ \frac{1}{L^2} r_1 + \mathcal{O}(L^{-4})~,
\end{equation}
where

\begin{eqnarray}
 r_0 &\approx&  \frac{1}{2} \left( 3M+\sqrt{9M^2-8Q^2} \right) \approx 3M-\frac{2Q^2}{3M} - \frac{4 Q^4}{27 M^3} + \mathcal{O}(Q^6)\,, \\
 \nonumber r_1 &\approx& -\frac{M}{3} \left(1-27 \mu^2 M^2+18 \Lambda M^2 \right) \left(1- 9 \Lambda M^2 \right) - 9 \omega M^2 q Q
+ \left(  \frac{5}{27 M} + M \left( \frac{2 \Lambda}{3} -5 \mu^2 + 3 q^2 \right) +24 M^3 (3 \mu^2 -2\Lambda) \Lambda \right) Q^2  \\
 && + 2 \omega q Q^3  + \left(  4 M \Lambda \left( -3 \mu^2 +2 \Lambda \right) +  \frac{4}{81 M^3}\right) Q^4 +\mathcal{O}(Q^6)\,,
\end{eqnarray}
and
\begin{equation} \label{coa}
V(r^*_{max}) \approx V_0 L^2+ V_1 +\mathcal{O}(L^{-2}) \,,
\end{equation}
where
\begin{eqnarray}
\nonumber
 V_0 &\approx&  \frac{1}{27 M^2} \left(1-9 \Lambda M^2 \right) +\frac{Q^2}{81 M^4} +\frac{4 Q^4}{729 M^6} + \mathcal{O} (Q^6)~, \\
\nonumber
 V_1 &\approx&  \frac{\left(2+27\mu^2 M^2-18 \Lambda M^2 \right) \left(1-9 \Lambda M^2 \right)}{81 M^2}+\frac{2 \omega q Q}{3 M}  + \frac{4+36 \Lambda M^2-648 \Lambda^2 M^4-81 M^2 q^2 -27 \mu^2M^2 \left(1-36 \Lambda M^2 \right)}{729 M^4} Q^2  \\
\nonumber
&& \nonumber +  \frac{4 \omega q Q^3}{27 M^3} + \frac{2 \left( 5 +72\Lambda M^2 -324 \Lambda^2 M^4 - 162 M^2 q^2 -54 \mu^2 M^2 \left( 1- 9 \Lambda M^2 \right) \right)}{6561 M^6} Q^4 + \frac{16 \omega q Q^5}{243 M^5} + \mathcal{O} (Q^6)~,\\
\end{eqnarray}
while the second derivative of the potential evaluated at $r^*_{max}$ is given by
\begin{equation}
V^{(2)}(r^*_{max}) \approx V^{(2)}_0 L^2 +V^{(2)}_1 + \mathcal{O}(L^{-2}) \,,
\end{equation}
where\\

\begin{equation}
    \nonumber
    V^{(2)}_0   \approx  -\frac{2 \left(  1- 9 \Lambda M^2\right)^2}{729 M^4} - \frac{4 \left( 1- 9\Lambda M^2 \right) \left( 2 +9\Lambda M^2 \right)}{6561 M^6} Q^2 - \frac{2 \left( 13+36\Lambda M^2 -648 \Lambda^2 M^4 \right)}{59049 M^8} Q^4 +\mathcal{O} (Q^6)~,   \\
\end{equation}

\begin{eqnarray}
\nonumber V^{(2)}_1  &\approx&  -\frac{(16M^{-2}+36 \Lambda  -1620 \Lambda^2 M^2 -54 \mu^2  (2-45 \Lambda M^2) )}{6561 M^2( 1- 9 \Lambda M^2 )^{-2} }    -\frac{4 \omega (1- 9 \Lambda M^2)^2 q Q}{243 M^3}\\
 && \nonumber - \frac{2 (1- 9 \Lambda M^2)}{59049 M^6} \Big(  22+ 270 \Lambda M^2
 + 2916 \Lambda ^2 M^4 -81 M^2 q^2
-64152 \Lambda^3 M^6
 +729 \Lambda M^4 q^2 \\
 && \nonumber -27 \mu^2 M^2 (1+225 \Lambda M^2 -3564 \Lambda^2 M^4) \Big) Q^2 \\
 && \nonumber  - \frac{8 (1 + 9 \Lambda M^2  - 162 \Lambda^2 M^4 )}{2187 M^5} \omega qQ^3  \\
 && \nonumber -  \frac{2}{177147 M^8} \Big(2 (1 -9 \Lambda M^2)(1+18 \Lambda M^2)  (5-27 M^2 q^2 + 126 \Lambda M^2  +648 \Lambda^2 M^4)\\
&& \nonumber + 27 \mu^2 M^2 ( 4 - 279 \Lambda M^2 +1620 \Lambda^2 M^4 +11664 \Lambda^3 M^6)   \Big) Q^4 + \mathcal{O} (Q^5)~,
\end{eqnarray}\\

%\begin{centeralign}
%V^{(2)}_1  \approx  -\frac{(16M^{-2}+36 \Lambda  -1620 \Lambda^2 M^2 -54 \mu^2  (2-45 \Lambda M^2) )}{6561 M^2( 1- 9 \Lambda M^2 )^{-2} }    -\frac{4 \omega (1- 9 \Lambda M^2)^2 q Q}{243 M^3}
%- \frac{2 (1- 9 \Lambda M^2)}{59049 M^6} \Big(  22+ 270 \Lambda M^2
% + 2916 \Lambda ^2 M^4 -81 M^2 q^2\\
%\nonumber -64152 \Lambda^3 M^6
% +729 \Lambda M^4 q^2 -27 \mu^2 M^2 (1+225 \Lambda M^2 -3564 \Lambda^2 M^4) \Big) Q^2  - \frac{8 (1 + 9 \Lambda M^2  - 162 \Lambda^2 M^4 )}{2187 M^5} \omega qQ^3 - \frac{2}{177147 M^8} \Big(2 (1 -9 \Lambda M^2)(1+18 \Lambda M^2)  (5-27 M^2 q^2 + 126 \Lambda M^2  +648 \Lambda^2 M^4)
%+ 27 \mu^2 M^2 ( 4 - 279 \Lambda M^2 +1620 \Lambda^2 M^4 +11664 \Lambda^3 M^6)   \Big) Q^4 + \mathcal{O} (Q^5)~,\hspace{2.0cm}
%\end{centeralign}
%\\

and the higher derivatives of the potential evaluated at $r^*_{max}$ yields the following expressions\\

\begin{eqnarray}
 \nonumber V^{(3)}(r^*_{max}) &\approx&  (1-9\Lambda M^2)^3 \Bigg(\frac{4 }{6561M^5}+ \frac{4 (1+18\Lambda M^2)}{19683 M^7 (1-9\Lambda M^2)} Q^2\\
 && \nonumber
-\frac{4 (1-126 \Lambda M^2 +324 \Lambda^2 M^4)}{177147 M^9(1-9\Lambda M^2)^2}Q^4
 +\mathcal{O}(Q^6)  \Bigg) L^2+ \mathcal{O}(L^0)\,, \\
 \nonumber
 V^{(4)}(r^*_{max})  &\approx&  \Bigg( \frac{16 (1-9 \Lambda M^2 )^3}{19683M^6}
+\frac{8 (1-9 \Lambda M^2)^2 (11+54 \Lambda M^2 +81 \Lambda^2 M^4)}{177147 M^8} Q^2\\
&&\nonumber 
+\frac{16 (7 -27 \Lambda M^2-486 \Lambda^2 M^4+1458 \Lambda^3 M^6)}{531441 M^{10}}  Q^4
+\mathcal{O}(Q^5)  \Bigg)   L^2+ \mathcal{O}(L^0)\,, \\
 \nonumber
 V^{(5)}(r^*_{max}) &\approx&  \Bigg(-\frac{40 (1-9 \Lambda M^2)^4}{59049 M^7}
-\frac{160 (1- 9\Lambda M^2)^3 (1+18 \Lambda M^2)}{531441 M^9} Q^2  \\
&& \nonumber + \frac{80 (1-9 \Lambda M^2)^2 (1-288 \Lambda M^2 +324 \Lambda^2 M^4)}{4782969 M^{11}} Q^4 + \mathcal{O} (Q^6) \Bigg) L^2+ \mathcal{O}(L^0)\,,  \\
 \nonumber
  V^{(6)}(r^*_{max}) &\approx&  \Bigg(  -\frac{16(1-9\Lambda M^2)^4(4+15 \Lambda M^2)}{177147 M^8}-\frac{16 (1-9\Lambda M^2)^3(113+45 \Lambda M^2+6966 \Lambda^2 M^4)}{4782969 M^{10}} Q^2  \\
&&  \nonumber - \frac{32  (367-1701\Lambda M^2+18225\Lambda^2 M^4 +40824 \Lambda^3 M^6)}{43046721 M^{12}(1-9 \Lambda M^2)^{-2}} Q^4+ \mathcal{O}(Q^6) \Bigg) L^2 + \mathcal{O}(L^0)\,.
\end{eqnarray}

%\begin{centeralign}
% V^{(3)}(r^*_{max}) \approx  (1-9\Lambda M^2)^3 \Bigg(\frac{4 }{6561M^5}+ \frac{4 (1+18\Lambda M^2)}{19683 M^7 (1-9\Lambda M^2)} Q^2
%-\frac{4 (1-126 \Lambda M^2 +324 \Lambda^2 M^4)}{177147 M^9(1-9\Lambda M^2)^2}Q^4
% +\mathcal{O}(Q^6)  \Bigg) L^2+ \mathcal{O}(L^0)\,, \\
%\nonumber
% V^{(4)}(r^*_{max})  \approx  \Bigg( \frac{16 (1-9 \Lambda M^2 )^3}{19683M^6}
%+\frac{8 (1-9 \Lambda M^2)^2 (11+54 \Lambda M^2 +81 \Lambda^2 M^4)}{177147 M^8} Q^2+\frac{16 (7 -27 \Lambda M^2-486 \Lambda^2 M^4+1458 \Lambda^3 M^6)}{531441 M^{10}}  Q^4
%+\mathcal{O}(Q^5)  \Bigg)   L^2+ \mathcal{O}(L^0)\,, \\
%\nonumber
% V^{(5)}(r^*_{max}) \approx  \Bigg(-\frac{40 (1-9 \Lambda M^2)^4}{59049 M^7}
%-\frac{160 (1- 9\Lambda M^2)^3 (1+18 \Lambda M^2)}{531441 M^9} Q^2 + \frac{80 (1-9 \Lambda M^2)^2 (1-288 \Lambda M^2 +324 \Lambda^2 M^4)}{4782969 M^{11}} Q^4 + \mathcal{O} (Q^6) \Bigg) L^2+ \mathcal{O}(L^0)\,,  \\
%\nonumber
%  V^{(6)}(r^*_{max}) \approx  \Bigg(  -\frac{16(1-9\Lambda M^2)^4(4+15 \Lambda M^2)}{177147 M^8}-\frac{16 (1-9\Lambda M^2)^3(113+45 \Lambda M^2+6966 \Lambda^2 M^4)}{4782969 M^{10}} Q^2 - \\
%  \nonumber \frac{32  (367-1701\Lambda M^2+18225\Lambda^2 M^4 +40824 \Lambda^3 M^6)}{43046721 M^{12}(1-9 \Lambda M^2)^{-2}} Q^4+ \mathcal{O}(Q^6) \Bigg) L^2 + \mathcal{O}(L^0)\,.
%\end{centeralign}
%\\
Using these results together with Eq. (\ref{omega}) we obtain Eq. (\ref{omegarelation}).

\clearpage

\section{Numerical Values}
\label{NV}

\begin{table}[ht]
\caption {Dominant photon sphere QNFs ($\omega M$) for massive scalar fields with $qM=0.10$, $\ell = 0, 1, 10,$ and $30$ in the background of a RNdS black hole with
$\Lambda M^2= 0.04$, and $Q/M=0.1$.}
\label {T1}\centering
\scalebox{0.7}{
\begin {tabular} { | c | c | c | c | c | c |}
\hline
$\ell$ & $n$ & $mM = 0.02 $ & $mM = 0.03 $ & $mM = 0.04 $ & $mM = 0.05 $ \\
\hline
$0 $ & $0$ &
$-0.079123984 - 0.101178151 i$ &
$-0.079378124 - 0.101018746 i$ &
$-0.079722294 - 0.100793055 i$ &
$-0.080144632 - 0.100497644 i$ \\
$0$ & $1$ &
$0.085822016 - 0.101449542 i$ &
$0.086073719 - 0.101277005 i$ &
$0.086415679 - 0.101033089 i$ &
$0.086837282 - 0.100714528 i$ \\
$1$ & $0$ &
$-0.22203715 - 0.08190343 i$ &
$-0.22224678 - 0.08181441 i$ &
$-0.22254054 - 0.08168981 i$ &
$-0.22291870 - 0.08152969 i$ \\
$1$ & $1$ &
$0.22894817 - 0.08227367 i$ &
$0.22915532 - 0.08218660 i$ &
$0.22944559 - 0.08206476 i$ &
$0.22981922 - 0.08190817 i$ \\
$10$ & $0$ &
$-1.6164783 - 0.0771586 i$ &
$-1.6165113 - 0.0771569 i$ &
$-1.6165574 - 0.0771545 i$ &
$-1.6166166 - 0.0771515 i$ \\
$10$ & $1$ &
$1.6231646 - 0.0772262 i$ &
$1.6231975 - 0.0772245 i$ &
$1.6232435 - 0.0772222 i$ &
$1.6233027 - 0.0772191 i$ \\
$30$ & $0$ &
$-4.7043696 - 0.0770948 i$ &
$-4.7043809 - 0.0770946 i$ &
$-4.7043968 - 0.0770943 i$ &
$-4.7044172 - 0.0770940 i$ \\
$30$ & $1$ &
$4.7110517 - 0.0771182 i$ &
$4.7110630 - 0.0771180 i$ &
$4.7110789 - 0.0771177 i$ &
$4.7110993 - 0.0771173 i$ \\
\hline
$\ell$ & $n$ & $mM = 0.06 $ & $mM = 0.07 $ & $mM = 0.08 $ & $mM = 0.09 $ \\
\hline
$0$ & $0$ &
$-0.080629492 - 0.100126671 i$ &
$-0.081157048 - 0.099670532 i$ &
$-0.081702724 - 0.099113741 i$ &
$-0.082236495 - 0.098431462 i$ \\
$0$ & $1$ &
$0.087324579 - 0.100315723 i$ &
$0.087860020 - 0.099827442 i$ &
$0.088422117 - 0.099234823 i$ &
$0.088985153 - 0.098514205 i$ \\
$1$ & $0$ &
$-0.22338164 - 0.08133412 i$ &
$-0.22392982 - 0.08110322 i$ &
$-0.22456377 - 0.08083711 i$ &
$-0.22528416 - 0.08053600 i$ \\
$1$ & $1$ &
$0.23027657 - 0.08171693 i$ &
$0.23081805 - 0.08149112 i$ &
$0.23144416 - 0.08123088 i$ &
$0.23215551 - 0.08093640 i$ \\
$10$ & $0$ &
$-1.6166890 - 0.0771478 i$ &
$-1.6167746 - 0.0771434 i$ &
$-1.6168734 - 0.0771384 i$ &
$-1.6169853 - 0.0771327 i$ \\
$10$ & $1$ &
$1.6233750 - 0.0772155 i$ &
$1.6234604 - 0.0772111 i$ &
$1.6235590 - 0.0772061 i$ &
$1.6236707 - 0.0772004 i$ \\
$30$ & $0$ &
$-4.7044422 - 0.0770935 i$ &
$-4.7044717 - 0.0770930 i$ &
$-4.7045057 - 0.0770924 i$ &
$-4.7045442 - 0.0770917 i$ \\
$30$ & $1$ &
$4.7111242 - 0.0771169 i$ &
$4.7111537 - 0.0771164 i$ &
$4.7111877 - 0.0771158 i$ &
$4.7112262 - 0.0771151 i$ \\
\hline
$\ell$ & $n$ & $mM = 0.10 $ & $mM = 0.11 $ & $mM = 0.12 $ & $mM = 0.13 $ \\
\hline
$0$ & $0$ &
$-0.082722332 - 0.097583569 i$ &
$-0.083118914 - 0.096503931 i$ &
$0.090377785 - 0.095076256 i$ &
$-0.083508801 - 0.093118746 i$ \\
$0$ & $1$ &
$0.089519326 - 0.097627872 i$ &
$0.089992550 - 0.096515050 i$ &
$-0.083385685 - 0.095080450 i$ &
$0.090677651 - 0.093148051 i$ \\
$1$ & $0$ &
$-0.22609173 - 0.08020018 i$ &
$-0.22698733 - 0.07982998 i$ &
$-0.22797191 - 0.07942590 i$ &
$-0.22904652 - 0.07898852 i$ \\
$1$ & $1$ &
$0.23295278 - 0.08060793 i$ &
$0.23383676 - 0.08024581 i$ &
$0.23480833 - 0.07985045 i$ &
$0.23586846 - 0.07942241 i$ \\
$10$ & $0$ &
$-1.6171105 - 0.0771263 i$ &
$-1.6172487 - 0.0771192 i$ &
$-1.6174002 - 0.0771115 i$ &
$-1.6175648 - 0.0771031 i$ \\
$10$ & $1$ &
$1.6237956 - 0.0771940 i$ &
$1.6239337 - 0.0771870 i$ &
$1.6240848 - 0.0771793 i$ &
$1.6242491 - 0.0771709 i$ \\
$30$ & $0$ &
$-4.7045873 - 0.0770910 i$ &
$-4.7046349 - 0.0770902 i$ &
$-4.7046871 - 0.0770892 i$ &
$-4.7047438 - 0.0770883 i$ \\
$30$ & $1$ &
$4.7112693 - 0.0771143 i$ &
$4.7113169 - 0.0771135 i$ &
$4.7113690 - 0.0771126 i$ &
$4.7114257 - 0.0771116 i$ \\
\hline
$\ell$  & $n$ & $mM = 0.14 $ & $mM = 0.15 $ & $mM = 0.16 $ & $mM = 0.17 $ \\
\hline
$0$ & $0$ &
$-0.083595886 - 0.090294353 i$ &
$-0.084138283 - 0.086226951 i$ &
$-0.086136987 - 0.081126205 i$ &
$-0.089956254 - 0.076194726 i$ \\
$0$ & $1$ &
$0.090998993 - 0.090478546 i$ &
$0.091723935 - 0.086797644 i$ &
$0.093580118 - 0.082225173 i$ &
$0.096965771 - 0.077590195 i$ \\
$1$ & $0$ &
$-0.23021228 - 0.07851861 i$ &
$-0.23147040 - 0.07801712 i$ &
$-0.23282210 - 0.07748521 i$ &
$-0.23426865 - 0.07692427 i$ \\
$1$ & $1$ &
$0.23701819 - 0.07896237 i$ &
$0.23825865 - 0.07847120 i$ &
$0.23959100 - 0.07794993 i$ &
$0.24101644 - 0.07739983 i$ \\
$10$ & $0$ &
$-1.6177425 - 0.0770941 i$ &
$-1.6179335 - 0.0770843 i$ &
$-1.6181376 - 0.0770739 i$ &
$-1.6183548 - 0.0770629 i$ \\
$10$ & $1$ &
$1.6244266 - 0.0771619 i$ &
$1.6246172 - 0.0771521 i$ &
$1.6248209 - 0.0771418 i$ &
$1.6250378 - 0.0771307 i$ \\
$30$ & $0$ &
$-4.7048050 - 0.0770872 i$ &
$-4.7048708 - 0.0770860 i$ &
$-4.7049411 - 0.0770848 i$ &
$-4.7050159 - 0.0770835 i$ \\
$30$ & $1$ &
$4.7114869 - 0.0771105 i$ &
$4.7115526 - 0.0771094 i$ &
$4.7116228 - 0.0771082 i$ &
$4.7116976 - 0.0771068 i$ \\
\hline
$\ell$ & $n$ & $mM = 0.18 $ & $mM = 0.19 $ & $mM = 0.20 $ & $mM = 0.21 $ \\
\hline
$0$ & $0$ &
$-0.094855902 - 0.072185081 i$ &
$-0.100181166 - 0.069050836 i$ &
$-0.105641081 - 0.066570034 i$ &
$-0.111125346 - 0.064563734 i$ \\
$0$ & $1$ &
$0.101453970 - 0.073607955 i$ &
$0.106482728 - 0.070389081 i$ &
$0.111735683 - 0.067795932 i$ &
$0.117070846 - 0.065678230 i$ \\
$1$ & $0$ &
$-0.23581128 - 0.07633593 i$ &
$-0.23745116 - 0.07572211 i$ &
$-0.23918934 - 0.07508499 i$ &
$-0.24102670 - 0.07442701 i$ \\
$1$ & $1$ &
$0.24253617 - 0.07682238 i$ &
$0.24415131 - 0.07621933 i$ &
$0.24586293 - 0.07559267 i$ &
$0.24767193 - 0.07494465 i$ \\
$10$ & $0$ &
$-1.6185852 - 0.0770511 i$ &
$-1.6188288 - 0.0770387 i$ &
$-1.6190856 - 0.0770257 i$ &
$-1.6193555 - 0.0770119 i$ \\
$10$ & $1$ &
$1.6252678 - 0.0771190 i$ &
$1.6255110 - 0.0771066 i$ &
$1.6257673 - 0.0770936 i$ &
$1.6260367 - 0.0770799 i$ \\
$30$ & $0$ &
$-4.7050953 - 0.0770821 i$ &
$-4.7051792 - 0.0770806 i$ &
$-4.7052676 - 0.0770791 i$ &
$-4.7053606 - 0.0770775 i$ \\
$30$ & $1$ &
$4.7117769 - 0.0771055 i$ &
$4.7118608 - 0.0771040 i$ &
$4.7119492 - 0.0771024 i$ &
$4.7120421 - 0.0771008 i$ \\
\hline
$\ell$ & $n$ & $mM = 0.22 $ & $mM = 0.23 $ & $mM = 0.24 $ & $mM = 0.25 $ \\
\hline
$0$ & $0$ &
$-0.116594182 - 0.062909138 i$ &
$-0.122035024 - 0.061522764 i$ &
$-0.127445993 - 0.060346491 i$ &
$-0.13282932 - 0.05933862 i$ \\
$0$ & $1$ &
$0.122427596 - 0.063921300 i$ &
$0.12778075 - 0.06244312 i$ &
$0.13312072 - 0.06118496 i$ &
$0.13844487 - 0.06010402 i$ \\
$1$ & $0$ &
$-0.24296392 - 0.07375084 i$ &
$-0.24500140 - 0.07305937 i$ &
$-0.24713924 - 0.07235566 i$ &
$-0.24937717 - 0.07164288 i$ \\
$1$ & $1$ &
$0.24957907 - 0.07427774 i$ &
$0.25158482 - 0.07359465 i$ &
$0.25368942 - 0.07289825 i$ &
$0.25589279 - 0.07219155 i$ \\
$10$ & $0$ &
$-1.6196386 - 0.0769975 i$ &
$-1.6199348 - 0.0769825 i$ &
$-1.6202442 - 0.0769667 i$ &
$-1.6205667 - 0.0769503 i$ \\
$10$ & $1$ &
$1.6263193 - 0.0770655 i$ &
$1.6266150 - 0.0770505 i$ &
$1.6269238 - 0.0770348 i$ &
$1.6272458 - 0.0770185 i$ \\
$30$ & $0$ &
$-4.7054581 - 0.0770757 i$ &
$-4.7055601 - 0.0770740 i$ &
$-4.7056667 - 0.0770721 i$ &
$-4.7057778 - 0.0770702 i$ \\
$30$ & $1$ &
$4.7121395 - 0.0770991 i$ &
$4.7122415 - 0.0770973 i$ &
$4.7123480 - 0.0770955 i$ &
$4.7124591 - 0.0770935 i$ \\
\hline
\end {tabular}}
\end {table}

\begin {table}[ht]
\caption {Dominant photon sphere QNFs ($\omega M$) for massive scalar fields with $qM=0.5$, $\ell = 0, 1, 10,$ and $30$ in the background of a RNdS black hole with
$\Lambda M^2= 0.04$, and $Q/M=0.1$.}
\label {T1}\centering
\scalebox{0.7}{
\begin {tabular} { | c | c | c | c | c | c |}
\hline
$\ell$ & $n$ & $mM = 0.02 $ & $mM = 0.03 $ & $mM = 0.04 $ & $mM = 0.05 $ \\\hline
$0 $ & $0$ &
$-0.066164805 - 0.100563562 i$ &
$-0.066422828 - 0.100435839 i$ &
$-0.066770125 - 0.100254130 i$ &
$-0.067192417 - 0.100014684 i$ \\
$0$ & $1$ &
$0.09962613 - 0.10191051 i$ &
$0.09987261 - 0.10171750 i$ &
$0.10020959 - 0.10144518 i$ &
$0.10062890 - 0.10109062 i$ \\

$1$ & $0$ &
$-0.20832444 - 0.08114240 i$ &
$-0.20853918 - 0.08104926 i$ &
$-0.20884015 - 0.08091889 i$ &
$-0.20922767 - 0.08075136 i$ \\

$1$ & $1$ &
$0.24287752 - 0.08299344 i$ &
$0.24307986 - 0.08291013 i$ &
$0.24336334 - 0.08279354 i$ &
$0.24372820 - 0.08264371 i$ \\

$10$ & $0$ &
$-1.6031234 - 0.0770228 i$ &
$-1.6031565 - 0.0770211 i$ &
$-1.6032027 - 0.0770187 i$ &
$-1.6032622 - 0.0770157 i$ \\
$10$ & $1$ &
$1.6365548 - 0.0773609 i$ &
$1.6365876 - 0.0773592 i$ &
$1.6366334 - 0.0773569 i$ &
$1.6366924 - 0.0773539 i$ \\

$30$ & $0$ &
$-4.6910115 - 0.0770480 i$ &
$-4.6910228 - 0.0770478 i$ &
$-4.6910387 - 0.0770476 i$ &
$-4.6910592 - 0.0770472 i$ \\

$30$ & $1$ &
$4.7244220 - 0.0771648 i$ &
$4.7244333 - 0.0771646 i$ &
$4.7244492 - 0.0771643 i$ &
$4.7244696 - 0.0771640 i$ \\\hline

$\ell$ & $n$ & $mM = 0.06 $ & $mM = 0.07 $ & $mM = 0.08 $ & $mM = 0.09 $ \\\hline
$0$ & $0$ &
$-0.067670716 - 0.099711308 i$ &
$-0.068180621 - 0.099333999 i$ &
$-0.068691169 - 0.098866708 i$ &
$-0.069163047 - 0.098283502 i$ \\

$0$ & $1$ &
$0.10111982 - 0.10064879 i$ &
$0.10166909 - 0.10011143 i$ &
$0.10226090 - 0.09946540 i$ &
$0.10287710 - 0.09869015 i$ \\

$1$ & $0$ &
$-0.20970218 - 0.08054674 i$ &
$-0.21026422 - 0.08030515 i$ &
$-0.21091445 - 0.08002675 i$ &
$-0.21165364 - 0.07971179 i$ \\

$1$ & $1$ &
$0.24417471 - 0.08246073 i$ &
$0.24470324 - 0.08224466 i$ &
$0.24531421 - 0.08199566 i$ &
$0.24600813 - 0.08171386 i$ \\

$10$ & $0$ &
$-1.6033349 - 0.0770119 i$ &
$-1.6034208 - 0.0770076 i$ &
$-1.6035199 - 0.0770025 i$ &
$-1.6036322 - 0.0769968 i$ \\

$10$ & $1$ &
$1.6367644 - 0.0773502 i$ &
$1.6368496 - 0.0773459 i$ &
$1.6369478 - 0.0773409 i$ &
$1.6370592 - 0.0773352 i$ \\

$30$ & $0$ &
$-4.6910841 - 0.0770468 i$ &
$-4.6911136 - 0.0770463 i$ &
$-4.6911477 - 0.0770457 i$ &
$-4.6911863 - 0.0770450 i$ \\

$30$ & $1$ &
$4.7244945 - 0.0771635 i$ &
$4.7245239 - 0.0771630 i$ &
$4.7245578 - 0.0771624 i$ &
$4.7245963 - 0.0771617 i$ \\\hline

$\ell$ & $n$ & $mM = 0.10 $ & $mM = 0.11 $ & $mM = 0.12 $ & $mM = 0.13 $ \\\hline
$0$ & $0$ &
$-0.069546021 - 0.097541533 i$ &
$-0.069775855 - 0.096567225 i$ &
$0.10469154 - 0.09517669 i$ &
$-0.069463673 - 0.093260243 i$ \\
$0$ & $1$ &
$0.10349820 - 0.09775392 i$ &
$0.10410613 - 0.09660777 i$ &
$-0.069773588 - 0.095226920 i$ &
$0.105272333 - 0.093348513 i$ \\

$1$ & $0$ &
$-0.21248266 - 0.07936058 i$ &
$-0.21340251 - 0.07897356 i$ &
$-0.21441432 - 0.07855132 i$ &
$-0.21551928 - 0.07809459 i$ \\
$1$ & $1$ &
$0.24678559 - 0.08139951 i$ &
$0.24764726 - 0.08105286 i$ &
$0.24859388 - 0.08067429 i$ &
$0.24962629 - 0.08026426 i$ \\

$10$ & $0$ &
$-1.6037578 - 0.0769904 i$ &
$-1.6038965 - 0.0769833 i$ &
$-1.6040485 - 0.0769755 i$ &
$-1.6042137 - 0.0769671 i$ \\

$10$ & $1$ &
$1.6371836 - 0.0773289 i$ &
$1.6373211 - 0.0773219 i$ &
$1.6374718 - 0.0773142 i$ &
$1.6376355 - 0.0773059 i$ \\

$30$ & $0$ &
$-4.6912294 - 0.0770442 i$ &
$-4.6912771 - 0.0770434 i$ &
$-4.6913293 - 0.0770425 i$ &
$-4.6913861 - 0.0770415 i$ \\

$30$ & $1$ &
$4.7246393 - 0.0771610 i$ &
$4.7246869 - 0.0771602 i$ &
$4.7247389 - 0.0771592 i$ &
$4.7247955 - 0.0771583 i$ \\\hline

$\ell$  & $n$ & $mM = 0.14 $ & $mM = 0.15 $ & $mM = 0.16 $ & $mM = 0.17 $ \\\hline
$0$ & $0$ &
$-0.068901335 - 0.090137834 i$ &
$-0.068882216 - 0.085042471 i$ &
$-0.071425233 - 0.078568931 i$ &
$-0.076395042 - 0.073176011 i$ \\

$0$ & $1$ &
$0.105936235 - 0.090970565 i$ &
$0.106915196 - 0.087895936 i$ &
$0.108612661 - 0.084154597 i$ &
$0.111366221 - 0.080132584 i$ \\

$1$ & $0$ &
$-0.21671870 - 0.07760430 i$ &
$-0.21801393 - 0.07708164 i$ &
$-0.21940634 - 0.07652801 i$ &
$-0.22089732 - 0.07594513 i$ \\

$1$ & $1$ &
$0.25074540 - 0.07982331 i$ &
$0.25195219 - 0.07935218 i$ &
$0.25324767 - 0.07885170 i$ &
$0.25463294 - 0.07832291 i$ \\

$10$ & $0$ &
$-1.6043921 - 0.0769580 i$ &
$-1.6045837 - 0.0769482 i$ &
$-1.6047885 - 0.0769378 i$ &
$-1.6050066 - 0.0769266 i$ \\

$10$ & $1$ &
$1.6378123 - 0.0772969 i$ &
$1.6380023 - 0.0772872 i$ &
$1.6382053 - 0.0772769 i$ &
$1.6384214 - 0.0772659 i$ \\

$30$ & $0$ &
$-4.6914474 - 0.0770404 i$ &
$-4.6915132 - 0.0770393 i$ &
$-4.6915836 - 0.0770380 i$ &
$-4.6916586 - 0.0770367 i$ \\

$30$ & $1$ &
$4.7248566 - 0.0771572 i$ &
$4.7249223 - 0.0771560 i$ &
$4.7249924 - 0.0771548 i$ &
$4.7250671 - 0.0771535 i$ \\\hline

$\ell$ & $n$ & $mM = 0.18 $ & $mM = 0.19 $ & $mM = 0.20 $ & $mM = 0.21 $ \\\hline
$0$ & $0$ &
$0.082148565 - 0.069289485 i$ &
$-0.088020870 - 0.066418277 i$ &
$-0.093844111 - 0.064204365 i$ &
$-0.099584873 - 0.062438467 i$ \\
$0$ & $1$ &
$0.115116766 - 0.076346488 i$ &
$0.119538303 - 0.073063895 i$ &
$0.12433711 - 0.07030406 i$ &
$0.12933399 - 0.06799254 i$ \\

$1$ & $0$ &
$-0.22248815 - 0.07533498 i$ &
$-0.22418004 - 0.07469986 i$ &
$-0.22597399 - 0.07404236 i$ &
$-0.22787077 - 0.07336532 i$ \\

$1$ & $1$ &
$0.25610906 - 0.07776706 i$ &
$0.25767711 - 0.07718557 i$ &
$0.25933812 - 0.07658011 i$ &
$0.26109302 - 0.07595258 i$ \\

$10$ & $0$ &
$-1.6052378 - 0.0769148 i$ &
$-1.6054823 - 0.0769024 i$ &
$-1.6057399 - 0.0768893 i$ &
$-1.6060108 - 0.0768755 i$ \\

$10$ & $1$ &
$1.6386506 - 0.0772542 i$ &
$1.6388929 - 0.0772419 i$ &
$1.6391483 - 0.0772289 i$ &
$1.6394168 - 0.0772153 i$ \\

$30$ & $0$ &
$-4.6917380 - 0.0770353 i$ &
$-4.6918220 - 0.0770338 i$ &
$-4.6919106 - 0.0770323 i$ &
$-4.6920037 - 0.0770307 i$ \\

$30$ & $1$ &
$4.7251464 - 0.0771521 i$ &
$4.7252301 - 0.0771507 i$ &
$4.7253184 - 0.0771491 i$ &
$4.7254112 - 0.0771475 i$ \\\hline

$\ell$ & $n$ & $mM = 0.22 $ & $mM = 0.23 $ & $mM = 0.24 $ & $mM = 0.25 $ \\\hline
$0$ & $0$ &
$-0.105245135 - 0.060994533 i$ &
$-0.110835581 - 0.059792157 i$ &
$-0.116367936 - 0.058777249 i$ &
$-0.121852810 - 0.057911760 i$ \\

$0$ & $1$ &
$0.13443255 - 0.06604434 i$ &
$0.13958253 - 0.06438768 i$ &
$0.14475782 - 0.06296650 i$ &
$0.14994495 - 0.06173776 i$ \\

$1$ & $0$ &
$-0.22987086 - 0.07267185 i$ &
$-0.23197441 - 0.07196521 i$ &
$-0.23418117 - 0.07124882 i$ &
$-0.23649051 - 0.07052613 i$ \\

$1$ & $1$ &
$0.26294261 - 0.07530510 i$ &
$0.26488754 - 0.07463999 i$ &
$0.26692823 - 0.07395978 i$ &
$0.26906485 - 0.07326715 i$ \\

$10$ & $0$ &
$-1.6062949 - 0.0768610 i$ &
$-1.6065922 - 0.0768459 i$ &
$-1.6069026 - 0.0768301 i$ &
$-1.6072263 - 0.0768136 i$ \\

$10$ & $1$ &
$1.6396984 - 0.0772010 i$ &
$1.6399930 - 0.0771860 i$ &
$1.6403008 - 0.0771704 i$ &
$1.6406216 - 0.0771541 i$ \\

$30$ & $0$ &
$-4.6921013 - 0.0770290 i$ &
$-4.6922035 - 0.0770272 i$ &
$-4.6923102 - 0.0770253 i$ &
$-4.6924214 - 0.0770233 i$ \\

$30$ & $1$ &
$4.7255085 - 0.0771458 i$ &
$4.7256104 - 0.0771440 i$ &
$4.7257168 - 0.0771421 i$ &
$4.7258277 - 0.0771402 i$ \\\hline

\end {tabular} }
\end {table}

\begin {table}[ht]
\caption {Dominant photon sphere QNFs ($\omega M$) for massive scalar fields with $qM=0.10$, $\ell = 0, 1, 10,$ and $30$ in the background of a RNdS black hole with $\Lambda M^2= 0.01$, and $Q/M=0.1$.}
\label {TA1}\centering
\scalebox{0.7}{
\begin {tabular} { | c | c | c | c | c | c |}
\hline
$\ell$ & $n$ & $mM = 0.02 $ & $mM = 0.03 $ & $mM = 0.04 $ & $mM = 0.05 $ \\\hline
$0 $ & $0$ &
$-0.10085470 - 0.10355718 i$ &
$-0.10098638 - 0.10306673 i$ &
$-0.10117481 - 0.10238707 i$ &
$-0.10142049 - 0.10153527 i$ \\
$0$ & $1$ &
$0.10833739 - 0.10475824 i$ &
$0.10847926 - 0.10430235 i$ &
$0.10867975 - 0.10366822 i$ &
$0.10893881 - 0.10286700 i$ \\

$1$ & $0$ &
$-0.27422146 - 0.09423365 i$ &
$-0.27444491 - 0.09410907 i$ &
$-0.27475785 - 0.09393451 i$ &
$-0.27516043 - 0.09370981 i$ \\
$1$ & $1$ &
$0.28109381 - 0.09483989 i$ &
$0.28131484 - 0.09471835 i$ &
$0.28162440 - 0.09454806 i$ &
$0.28202262 - 0.09432887 i$ \\

$10$ & $0$ &
$-1.9280093 - 0.0918701 i$ &
$-1.9280484 - 0.0918673 i$ &
$-1.9281031 - 0.0918633 i$ &
$-1.9281735 - 0.0918582 i$ \\
$10$ & $1$ &
$1.9346952 - 0.0919662 i$ &
$1.9347343 - 0.0919634 i$ &
$1.9347889 - 0.0919595 i$ &
$1.9348592 - 0.0919544 i$ \\

$30$ & $0$ &
$-5.6063683 - 0.0918493 i$ &
$-5.6063818 - 0.0918490 i$ &
$-5.6064007 - 0.0918485 i$ &
$-5.6064250 - 0.0918479 i$ \\

$30$ & $1$ &
$5.6130503 - 0.0918825 i$ &
$5.6130638 - 0.0918821 i$ &
$5.6130827 - 0.0918817 i$ &
$5.6131070 - 0.0918811 i$ \\\hline

$\ell$ & $n$ & $mM = 0.06 $ & $mM = 0.07 $ & $mM = 0.08 $ & $mM = 0.09 $ \\\hline
$0$ & $0$ &
$-0.10171364 - 0.10054993 i$ &
$-0.10201655 - 0.09950771 i$ &
$-0.10223006 - 0.09855530 i$ &
$-0.032463969 - 0.084917752 i$ \\

$0$ & $1$ &
$0.10925038 - 0.10192494 i$ &
$0.10959117 - 0.10089588 i$ &
$0.10989843 - 0.09988318 i$ &
$0.032072572 - 0.086232178 i$ \\

$1$ & $0$ &
$-0.27565280 - 0.09343478 i$ &
$-0.27623516 - 0.09310917 i$ &
$-0.276907746 - 0.092732683 i$ &
$-0.277670822 - 0.092304964 i$ \\

$1$ & $1$ &
$0.28250966 - 0.09406061 i$ &
$0.28308570 - 0.09374306 i$ &
$0.283750977 - 0.093375923 i$ &
$0.284505754 - 0.092958894 i$ \\

$10$ & $0$ &
$-1.9282596 - 0.0918520 i$ &
$-1.9283613 - 0.0918446 i$ &
$-1.92847867 - 0.09183615 i$ &
$-1.92861168 - 0.09182653 i$ \\

$10$ & $1$ &
$1.9349451 - 0.0919482 i$ &
$1.9350466 - 0.0919409 i$ &
$1.93516366 - 0.09193239 i$ &
$1.93529639 - 0.09192280 i$ \\

$30$ & $0$ &
$-5.60645469 - 0.09184718 i$ &
$-5.6064898 - 0.0918463 i$ &
$-5.60653032 - 0.09184530 i$ &
$-5.60657624 - 0.09184416 i$ \\

$30$ & $1$ &
$5.61313670 - 0.09188033 i$ &
$5.6131718 - 0.0918795 i$ &
$5.61321228 - 0.09187846 i$ &
$5.61325817 - 0.09187732 i$ \\\hline

$\ell$ & $n$ & $mM = 0.10 $ & $mM = 0.11 $ & $mM = 0.12 $ & $mM = 0.13 $ \\\hline
$0$ & $0$ &
$-0.060488872 - 0.080528724 i$ &
$-0.080300748 - 0.073194338 i$ &
$-0.093239147 - 0.065502234 i$ &
$-0.103118273 - 0.059956078 i$ \\
$0$ & $1$ &
$0.059696691 - 0.082663371 i$ &
$0.080269176 - 0.077376620 i$ &
$0.095392963 - 0.069677861 i$ &
$0.106119868 - 0.063326320 i$ \\
$1$ & $0$ &
$-0.278524691 - 0.091825598 i$ &
$-0.279469683 - 0.091294100 i$ &
$-0.280506154 - 0.090709905 i$ &
$-0.281634483 - 0.090072353 i$ \\
$1$ & $1$ &
$0.285350326 - 0.092491590 i$ &
$0.286285021 - 0.091973570 i$ &
$0.287310195 - 0.091404325 i$ &
$0.288426230 - 0.090783261 i$ \\

$10$ & $0$ &
$-1.92876034 - 0.09181577 i$ &
$-1.92892465 - 0.09180388 i$ &
$-1.92910461 - 0.09179085 i$ &
$-1.92930023 - 0.09177670 i$ \\

$10$ & $1$ &
$1.93544474 - 0.09191208 i$ &
$1.93560871 - 0.09190023 i$ &
$1.93578829 - 0.09188726 i$ &
$1.93598350 - 0.09187315 i$ \\

$30$ & $0$ &
$-5.60662757 - 0.09184288 i$ &
$-5.60668429 - 0.09184147 i$ &
$-5.60674642 - 0.09183993 i$ &
$-5.60681395 - 0.09183825 i$ \\

$30$ & $1$ &
$5.61330945 - 0.09187604 i$ &
$5.61336614 - 0.09187463 i$ &
$5.61342822 - 0.09187309 i$ &
$5.61349571 - 0.09187141 i$ \\\hline

$\ell$  & $n$ & $mM = 0.14 $ & $mM = 0.15 $ & $mM = 0.16 $ & $mM = 0.17 $ \\\hline
$0$ & $0$ &
$-0.1118370094 - 0.0558266753 i$ &
$-0.1199663676 - 0.0525864785 i$ &
$-0.1277421617 - 0.0499524062 i$ &
$-0.135286220 - 0.047759948 i$ \\

$0$ & $1$ &
$0.1151707083 - 0.0586156619 i$ &
$0.1234645461 - 0.0549609071 i$ &
$0.1313337061 - 0.0520134087 i$ &
$0.138934545 - 0.049572411 i$ \\

$1$ & $0$ &
$-0.282855063 - 0.089380667 i$ &
$-0.284168294 - 0.088633922 i$ &
$-0.285574575 - 0.087831001 i$ &
$-0.287074304 - 0.086970531 i$ \\

$1$ & $1$ &
$0.289633531 - 0.090109688 i$ &
$0.290932514 - 0.089382792 i$ &
$0.292323608 - 0.088601599 i$ &
$0.293807243 - 0.087764932 i$ \\

$10$ & $0$ &
$-1.92951149 - 0.09176141 i$ &
$-1.92973842 - 0.09174499 i$ &
$-1.92998100 - 0.09172744 I$ &
$-1.93023924 - 0.09170875 i$ \\

$10$ & $1$ &
$1.93619432 - 0.09185792 i$ &
$1.93642077 - 0.09184156 i$ &
$1.93666285 - 0.09182407 i$ &
$1.93692054 - 0.09180545 i$ \\

$30$ & $0$ &
$-5.60688689 - 0.09183643 i$ &
$-5.60696523 - 0.09183449 i$ &
$-5.60704897 - 0.09183240 i$ &
$-5.60713811 - 0.09183019 i$ \\

$30$ & $1$ &
$5.61356859 - 0.09186960 i$ &
$5.61364687 - 0.09186766 i$ &
$5.61373055 - 0.09186558 i$ &
$5.61381962 - 0.09186336 i$ \\\hline

$\ell$ & $n$ & $mM = 0.18 $ & $mM = 0.19 $ & $mM = 0.20 $ & $mM = 0.21 $ \\\hline
$0$ & $0$ &
$-0.142670555 - 0.045905307 i$ &
$-0.149941586 - 0.044318574 i$ &
$-0.157131057 - 0.042950193 i$ &
$-0.164261635 - 0.041763573 i$ \\
$0$ & $1$ &
$0.146354211 - 0.047513625 i$ &
$0.153646785 - 0.045754783 i$ &
$0.160848324 - 0.044238415 i$ &
$0.167984118 - 0.042922605 i$ \\

$1$ & $0$ &
$-0.288667881 - 0.086050788 i$ &
$-0.290355737 - 0.085069576 i$ &
$-0.292138398 - 0.084024054 i$ &
$-0.294016626 - 0.082910510 i$ \\

$1$ & $1$ &
$0.295383858 - 0.086871335 i$ &
$0.297053914 - 0.085918979 i$ &
$0.298817932 - 0.084905528 i$ &
$0.300676590 - 0.083827961 i$ \\

$10$ & $0$ &
$-1.93051314 - 0.09168893 i$ &
$-1.93080269 - 0.09166798 i$ &
$-1.93110792 - 0.09164590 i$ &
$-1.93142880 - 0.09162268 i$ \\

$10$ & $1$ &
$1.93719387 - 0.09178570 i$ &
$1.93748282 - 0.09176482 i$ &
$1.93778740 - 0.09174282 i$ &
$1.93810762 - 0.09171969 i$ \\

$30$ & $0$ &
$-5.60723265 - 0.09182784 i$ &
$-5.60733260 - 0.09182535 i$ &
$-5.60743795 - 0.09182273 i$ &
$-5.60754871 - 0.09181998 i$ \\

$30$ & $1$ &
$5.61391410 - 0.09186102 i$ &
$5.61401398 - 0.09185854 i$ &
$5.61411925 - 0.09185592 i$ &
$5.61422993 - 0.09185317 i$ \\\hline

$\ell$ & $n$ & $mM = 0.22 $ & $mM = 0.23 $ & $mM = 0.24 $ & $mM = 0.25 $ \\\hline
$0$ & $0$ &
$-0.171350024 - 0.040730779 i$ &
$-0.178408846 - 0.039829894 i$ &
$-0.185447815 - 0.039043316 i$ &
$-0.192474523 - 0.038356629 i$ \\

$0$ & $1$ &
$0.175072579 - 0.041775685 i$ &
$0.182127493 - 0.040773032 i$ &
$0.189159406 - 0.039895027 i$ &
$0.196176524 - 0.039125713 i$ \\

$1$ & $0$ &
$-0.295991675 - 0.081724096 i$ &
$-0.298065736 - 0.080458530 i$ &
$-0.300242672 - 0.079105846 i$ &
$-0.302529136 - 0.077656357 i$ \\

$1$ & $1$ &
$0.302630888 - 0.082682358 i$ &
$0.304682453 - 0.081463634 i$ &
$0.306834054 - 0.080165294 i$ &
$0.309090399 - 0.078779265 i$ \\

$10$ & $0$ &
$-1.93176535 - 0.09159833 i$ &
$-1.93211757 - 0.09157285 i$ &
$-1.93248546 - 0.09154624 i$ &
$-1.93286902 - 0.09151849 i$ \\

$10$ & $1$ &
$1.93844346 - 0.09169543 i$ &
$1.93879495 - 0.09167004 i$ &
$1.93916206 - 0.09164352 i$ &
$1.93954482 - 0.09161587 i$ \\

$30$ & $0$ &
$-5.60766487 - 0.09181709 i$ &
$-5.60778643 - 0.09181407 i$ &
$-5.60791339 - 0.09181091 i$ &
$-5.60804576 - 0.09180762 i$ \\

$30$ & $1$ &
$5.61434600 - 0.09185029 i$ &
$5.61446747 - 0.09184727 i$ &
$5.61459435 - 0.09184411 i$ &
$5.61472662 - 0.09184083 i$ \\\hline

\end {tabular} }
\end {table}

\begin {table}[ht]
\caption {Dominant photon sphere QNFs ($\omega M$) for massive scalar fields with $qM=0.5$, $\ell = 0, 1, 10,$ and $30$ in the background of a RNdS black hole with 
$\Lambda M^2= 0.01$, and $Q/M=0.1$.}
\label {TA2}\centering
\scalebox{0.7}{
\begin {tabular} { | c | c | c | c | c | c |}
\hline
$\ell$ & $n$ & $mM = 0.02 $ & $mM = 0.03 $ & $mM = 0.04 $ & $mM = 0.05 $ \\\hline
$0 $ & $0$ &
$-0.086155408 - 0.100795693 i$ &
$-0.086279942 - 0.100207966 i$ &
$-0.086466787 - 0.099404542 i$ &
$-0.086715719 - 0.098424355 i$ \\
$0$ & $1$ &
$0.12354177 - 0.10687582 i$ &
$0.12370546 - 0.10647128 i$ &
$0.12393434 - 0.10590609 i$ &
$0.12422757 - 0.10518466 i$ \\

$1$ & $0$ &
$-0.26061346 - 0.09298021 i$ &
$-0.26084186 - 0.09284923 i$ &
$-0.26116175 - 0.09266568 i$ &
$-0.261573262 - 0.092429383 i$ \\

$1$ & $1$ &
$0.29497145 - 0.09601229 i$ &
$0.29518777 - 0.09589655 i$ &
$0.29549073 - 0.09573439 i$ &
$0.295880452 - 0.095525697 i$ \\

$10$ & $0$ &
$-1.91465837 - 0.09167698 i$ &
$-1.91469765 - 0.09167413 i$ &
$-1.91475264 - 0.09167014 i$ &
$-1.91482335 - 0.09166501 i$ \\

$10$ & $1$ &
$1.94808816 - 0.09215757 i$ &
$1.94812703 - 0.09215477 i$ &
$1.94818145 - 0.09215085 i$ &
$1.94825143 - 0.09214581 i$ \\

$30$ & $0$ &
$-5.5930114 - 0.0917829 i$ &
$-5.59302488 - 0.09178258 i$ &
$-5.59304382 - 0.09178210 i$ &
$-5.59306817 - 0.09178150 i$ \\

$30$ & $1$ &
$5.6264217 - 0.0919487 i$ &
$5.62643519 - 0.09194834 i$ &
$5.62645406 - 0.09194788 i$ &
$5.62647832 - 0.09194727 i$ \\\hline

$\ell$ & $n$ & $mM = 0.06 $ & $mM = 0.07 $ & $mM = 0.08 $ & $mM = 0.09 $ \\\hline
$0$ & $0$ &
$-0.0869967490 - 0.0973452199 i$ &
$-0.0872091207 - 0.0963085290 i$ &
$-0.0871054515 - 0.0955917784 i$ &
$-0.0861682400 - 0.0959867025 i$ \\

$0$ & $1$ &
$0.124582048 - 0.104318473 i$ &
$0.124988017 - 0.103332429 i$ &
$0.125419579 - 0.102275387 i$ &
$0.125816459 - 0.101239714 i$ \\

$1$ & $0$ &
$-0.262076573 - 0.092140097 i$ &
$-0.262671886 - 0.091797530 i$ &
$-0.263359443 - 0.091401326 i$ &
$-0.264139518 - 0.090951058 i$ \\

$1$ & $1$ &
$0.296357077 - 0.095270320 i$ &
$0.296920791 - 0.094968069 i$ &
$0.297571810 - 0.094618715 i$ &
$0.298310382 - 0.094221986 i$ \\

$10$ & $0$ &
$-1.91490978 - 0.09165873 i$ &
$-1.91501191 - 0.09165132 i$ &
$-1.91512976 - 0.09164276 i$ &
$-1.91526333 - 0.09163307 i$ \\

$10$ & $1$ &
$1.94833695 - 0.09213965 i$ &
$1.94843802 - 0.09213237 i$ &
$1.94855465 - 0.09212396 i$ &
$1.94868682 - 0.09211444 i$ \\

$30$ & $0$ &
$-5.59309792 - 0.09178076 i$ &
$-5.59313309 - 0.09177988 i$ &
$-5.59317367 - 0.09177887 i$ &
$-5.59321965 - 0.09177773 i$ \\

$30$ & $1$ &
$5.62650797 - 0.09194654 i$ &
$5.62654301 - 0.09194567 i$ &
$5.62658344 - 0.09194466 i$ &
$5.62662926 - 0.09194353 i$ \\\hline

$\ell$ & $n$ & $mM = 0.10 $ & $mM = 0.11 $ & $mM = 0.12 $ & $mM = 0.13 $ \\\hline
$0$ & $0$ &
$-0.0621235053 - 0.0739272149 i$ &
$-0.0772534815 - 0.0646541162 i$ &
$-0.0877513198 - 0.0586112981 i$ &
$-0.0967844189 - 0.0543547019 i$ \\

$0$ & $1$ &
$0.0588028257 - 0.0859871183 i$ &
$0.0789004906 - 0.0831475318 i$ &
$0.0967720782 - 0.0786008919 i$ &
$0.1110665148 - 0.0716187576 i$ \\

$1$ & $0$ &
$-0.265012415 - 0.090446224 i$ &
$-0.265978460 - 0.089886232 i$ &
$-0.267037997 - 0.089270381 i$ &
$-0.268191377 - 0.088597844 i$ \\
$1$ & $1$ &

$0.299136792 - 0.093777569 i$ &
$0.300051356 - 0.093285099 i$ &
$0.301054422 - 0.092744158 i$ &
$0.302146370 - 0.092154263 i$ \\

$10$ & $0$ &
$-1.91541261 - 0.09162223 i$ &
$-1.91557762 - 0.09161025 i$ &
$-1.91575833 - 0.09159714 i$ &
$-1.9159548 - 0.0915829 i $ \\

$10$ & $1$ &
$1.94883455 - 0.09210380 i$ &
$1.94899783 - 0.09209203 i$ &
$1.94917667 - 0.09207915 i$ &
$1.9493711 - 0.0920651 i$ \\

$30$ & $0$ &
$-5.59327105 - 0.09177645 i$ &
$-5.59332786 - 0.09177504 i$ &
$-5.59339008 - 0.09177349 i$ &
$-5.59345771 - 0.09177180 i$ \\

$30$ & $1$ &
$5.62668047 - 0.09194226 i$ &
$5.62673708 - 0.09194085 i$ &
$5.62679907 - 0.09193931 i$ &
$5.62686646 - 0.09193764 i$ \\\hline

$\ell$  & $n$ & $mM = 0.14 $ & $mM = 0.15 $ & $mM = 0.16 $ & $mM = 0.17 $ \\\hline
$0$ & $0$ &
$-0.1051104647 - 0.0511126430 i$ &
$-0.1130215523 - 0.0485261044 i$ &
$-0.1206643872 - 0.0464017686 i$ &
$-0.1281241664 - 0.0446233574 i$ \\

$0$ & $1$ &
$0.1215495514 - 0.0653882790 i$ &
$0.130427708 - 0.060617227 i$ &
$0.138588160 - 0.056858445 i$ &
$0.146352731 - 0.053796706 i$ \\

$1$ & $0$ &
$-0.269438948 - 0.087867622 i$ &
$-0.270781045 - 0.087078501 i$ &
$-0.272217981 - 0.086228969 i$ &
$-0.273750053 - 0.085317106 i$ \\

$1$ & $1$ &
$0.303327606 - 0.091514860 i$ &
$0.304598565 - 0.090825308 i$ &
$0.305959700 - 0.090084855 i$ &
$0.307411483 - 0.089292615 i$ \\

$10$ & $0$ &
$-1.91616693 - 0.09156748 i$ &
$-1.91639481 - 0.09155094 i$ &
$-1.91663841 - 0.09153326 i$ &
$-1.91689774 - 0.09151444 i$ \\

$10$ & $1$ &
$1.94958101 - 0.09205002 i$ &
$1.94980651 - 0.09203378 i$ &
$1.95004757 - 0.09201641 i$ &
$1.95030420 - 0.09199793 i$ \\

$30$ & $0$ &
$-5.59353075 - 0.09176999 i$ &
$-5.59360920 - 0.09176803 i$ &
$-5.59369306 - 0.09176595 i$ &
$-5.59378233 - 0.09176373 i$ \\

$30$ & $1$ &
$5.62693923 - 0.09193583 i$ &
$5.62701740 - 0.09193389 i$ &
$5.62710096 - 0.09193182 i$ &
$5.62718991 - 0.09192961 i$ \\\hline

$\ell$ & $n$ & $mM = 0.18 $ & $mM = 0.19 $ & $mM = 0.20 $ & $mM = 0.21 $ \\\hline
$0$ & $0$ &
$-0.135455043 - 0.043115087 i$ &
$-0.142693616 - 0.041824496 i$ &
$-0.149865733 - 0.040713533 i$ &
$-0.156990267 - 0.039753540 i$ \\

$0$ & $1$ &
$0.153869790 - 0.051241837 i$ &
$0.161220703 - 0.049073344 i$ &
$0.168455459 - 0.047210278 i$ &
$0.175607214 - 0.045595492 i$ \\

$1$ & $0$ &
$-0.275377568 - 0.084340436 i$ &
$-0.277100922 - 0.083295711 i$ &
$-0.278920765 - 0.082178637 i$ &
$-0.280838319 - 0.080983531 i$ \\

$1$ & $1$ &
$0.308954405 - 0.088447520 i$ &
$0.310588976 - 0.087548268 i$ &
$0.312315743 - 0.086593241 i$ &
$0.314135325 - 0.085580399 i$ \\

$10$ & $0$ &
$-1.91717279 - 0.09149447 i$ &
$-1.91746356 - 0.09147337 i$ &
$-1.91777007 - 0.09145113 i$ &
$-1.91809231 - 0.09142774 i$ \\

$10$ & $1$ &
$1.95057638 - 0.09197832 i$ &
$1.95086413 - 0.09195759 i$ &
$1.95116744 - 0.09193575 i$ &
$1.95148632 - 0.09191278 i$ \\

$30$ & $0$ &
$-5.59387701 - 0.09176137 i$ &
$-5.59397711 - 0.09175888 i$ &
$-5.59408261 - 0.09175625 i$ &
$-5.59419353 - 0.09175349 i$ \\

$30$ & $1$ &
$5.62728425 - 0.09192727 i$ &
$5.62738398 - 0.09192479 i$ &
$5.62748910 - 0.09192218 i$ &
$5.62759962 - 0.09191944 i$ \\\hline

$\ell$ & $n$ & $mM = 0.22 $ & $mM = 0.23 $ & $mM = 0.24 $ & $mM = 0.25 $ \\\hline
$0$ & $0$ &
$-0.164081346 - 0.038922267 i$ &
$-0.171149763 - 0.038201991 i$ &
$-0.178203883 - 0.037578290 i$ &
$-0.185250268 - 0.037039214 i$ \\

$0$ & $1$ &
$0.182699180 - 0.044186971 i$ &
$0.189748258 - 0.042952782 i$ &
$0.196767125 - 0.041867974 i$ &
$0.203765500 - 0.040912595 i$ \\

$1$ & $0$ &
$-0.282855945 - 0.079702958 i$ &
$-0.284978077 - 0.078327422 i$ &
$-0.287212683 - 0.076845367 i$ &
$-0.289573274 - 0.075243911 i$ \\

$1$ & $1$ &
$0.316048491 - 0.084507139 i$ &
$0.318056297 - 0.083370124 i$ &
$0.320160327 - 0.082165077 i$ &
$0.322363095 - 0.080886581 i$ \\

$10$ & $0$ &
$-1.91843028 - 0.09140322 i$ &
$-1.91878398 - 0.09137755 i$ &
$-1.91915342 - 0.09135075 i$ &
$-1.91953859 - 0.09132280 i$ \\

$10$ & $1$ &
$1.95182076 - 0.09188869 i$ &
$1.95217077 - 0.09186349 i$ &
$1.95253636 - 0.09183716 i$ &
$1.95291752 - 0.09180971 i$ \\

$30$ & $0$ &
$-5.59430985 - 0.09175060 i$ &
$-5.59443159 - 0.09174757 i$ &
$-5.59455874 - 0.09174440 i$ &
$-5.59469130 - 0.09174111 i$ \\

$30$ & $1$ &
$5.62771552 - 0.09191656 i$ &
$5.62783682 - 0.09191355 i$ &
$5.62796351 - 0.09191041 i$ &
$5.62809559 - 0.09190713 i$ \\\hline

\end {tabular}}
\end {table}

\clearpage

\section{Near extremal modes}
\label{NE}

\begin{table}[h]
\caption {The QNFs $\omega M$ for massive ($mM =0.05$) scalar fields in the background of RNdS black holes with $\Lambda M^2= 0.01 $, and different values of $\ell$, and $q M$. Here, for $Q/M =0.999$, $\Delta r /M \approx 0.148$, and for $Q/M =1$, $\Delta r/M \approx 0.118$, with $\Delta r/M=(r_H-r_C)/M $. For $Q/M =1.002$ two complex roots appear for $f(r)$. "..." means that it is not possible to guarantee the convergence of the QNFs with 270 Chebyshev polynomials.}
\label {NE1}\centering
\scalebox{0.7} {
\begin {tabular} { | c | c | c | c | }
\hline
${Q/M=0.999}$ & $qM = 0$ & $qM = 0.1$  & $qM = 0.2$ \\\hline
$\omega M(\ell=0)$ &
$-0.015076441 i$ &
$0.005152360 - 0.014608818 i$ &
$0.010813814 - 0.013491279 i$    \\\hline
$\omega M(\ell=0)$ &
$-0.073785686 i$ &
$0.102487111 - 0.073674244 i$ &
$0.21016237 - 0.07194639 i$   \\\hline
$\omega M(\ell=0)$ &
$0.12888550 - 0.09470544 i$ &
$0.18034802 - 0.09117840 i$ &
$-0.045716001 - 0.081203854 i$   \\\hline
$\omega M(\ell=0)$ &
$-0.12888550 - 0.09470544 i$ &
$-0.082281358 - 0.091779905 i$ &
$0.23392927 - 0.08357275 i$   \\\hline\hline
$\omega M(\ell=1)$ &
$-0.073126720 i$ &
$0.003130990 - 0.073052324 i$ &
$0.006255023 - 0.072828727 i$    \\\hline
$\omega M(\ell=1)$ &
$-0.36627990 - 0.08741163 i$ &
$0.41774396 - 0.08685347 i$ &
$0.47213504 - 0.08531751 i$   \\\hline
$\omega M(\ell=1)$ &
$0.36627990 - 0.08741163 i$ &
$-0.31773646 - 0.08698589 i$ &
$-0.27212075 - 0.08558335 i$   \\\hline
$\omega M(\ell=1)$ &
$-0.14661132 i$ &
$0.10081482 - 0.14638474 i$ &
$0.20165318 - 0.14570340 i$    \\\hline\hline
$\omega M(\ell=10)$ &
$-2.5515929 - 0.0861901 i$ &
$-2.5019670 - 0.0861703 i$ &
$-2.4527994 - 0.0861289 i$    \\\hline
$\omega M(\ell=10)$ &
$2.5515929 - 0.0861901 i$ &
$2.6016771 - 0.0861883 i$ &
$2.6522194 - 0.0861648 i$   \\\hline
$\omega M(\ell=10)$ &
$-2.5475307 - 0.2587064 i$ &
$-2.4978986 - 0.2586476 i$ &
$-2.4487196 - 0.2585241 i$  \\\hline
$\omega M(\ell=10)$ &
$2.5475307 - 0.2587064 i$ &
$2.5976159 - 0.2587006 i$ &
$2.6481541 - 0.2586303 i$   \\\hline\hline
${Q/M=1.000}$ & $qM = 0$ & $qM = 0.1$  & $qM = 0.2$ 
\\\hline
$\omega M(\ell=0)$ &
$-0.015079204 i$ &
$0.005157715 - 0.014606878 i$ &
$0.010826180 - 0.013487686 i$ 
\\\hline
$\omega M(\ell=0)$ &
$-0.057887183 i$ &
$0.101410339 - 0.057473594 i $ &
$0.20465622 - 0.05499399 i$ 
\\\hline
$\omega M(\ell=0)$ &
$0.12888859 - 0.09470986 i$ &
$0.18041968 - 0.09112707 i$ &
$-0.045656086 - 0.081183740 i$ 
\\\hline
$\omega M(\ell=0)$ &
$-0.12888859 - 0.09470986 i$ &
$-0.082238014 - 0.091777948 i$ &
$0.23696903 - 0.08205709 i$ 
\\\hline \hline
$\omega M(\ell=1)$ &
$-0.073126749 i$ &
$0.003134130 - 0.073052202 i$ &
$0.006261284 - 0.072828152 i$ 
\\\hline
$\omega M(\ell=1)$ &
$0.36658358-0.08727065 i$ &
$0.41817256 - 0.08665059 i$ &
$0.47270653 - 0.08502996 i$ 
\\\hline
$\omega M(\ell=1)$ &
$-0.36658358-0.08727065 i$ &
$-0.31793195 - 0.08688899 i$ &
$-0.27222360 - 0.08551647 i$ 
\\\hline
$\omega M(\ell=1)$ &
$-0.115517389 i$ &
$0.10032998 - 0.11533209 i$ &
$0.20067585 - 0.11477468 i$ 
\\\hline\hline
$\omega M(\ell=2)$ &
$0.60908866 - 0.08646252 i$ &
$0.66008249 - 0.08624493 i$ &
$0.71295318 - 0.08564873 i$ 
\\\hline
$\omega M(\ell=2)$ &
$-0.60908866 - 0.08646252 i$ &
$-0.55997098 - 0.08630072 i$ &
$-0.51272938 - 0.08576110 i$ 
\\\hline
$\omega M(\ell=2)$ &
$-0.130962416 i$ &
$0.002506561 - 0.130920735 i$ &
$0.005009117 - 0.130795675 i$ 
\\\hline
$\omega M(\ell=2)$ &
$-0.17323372 i$ &
$0.10019906 - 0.17312038 i$ &
$0.20040283 - 0.17277992 i$ 
\\\hline\hline
$\omega M(\ell=10)$ &
$2.5542682-0.0860254 i$ &
$2.6045029 - 0.0860139 i$ &
$2.6552000 - 0.0859804 i$ 
\\\hline
$\omega M(\ell=10)$ &
$-2.5542682-0.0860254 i$ &
$-2.5044958 - 0.0860147 i$ &
$...$ 
\\\hline
$\omega M(\ell=10)$ &
$-0.59303982 i$ &
$0.00127148 - 0.59303029 i$ &
$0.00254240 - 0.59300169 i$ 
\\\hline
$\omega M(\ell=10)$ &
$...$ &
$...$ &
$...$ 
\\\hline
\end {tabular}
}
\end{table}

\begin{table}[H]
\caption {The QNFs $\omega M$ for massive ($m M=0.05$) scalar fields in the background of RNdS black holes with $\Lambda M^2= 0.04 $, and different values of $\ell$, and $q M$. Here, for $Q/M=1$, $\Delta r/M \approx 0.248$, and for $Q/M=1.004$, $\Delta r /M \approx 0.163$, with $\Delta r/M =(r_H-r_C)/M$. For $Q/M=1.008$ two complex roots appear for $f(r)$.}
\label {NE2}\centering
\scalebox{0.7} {
\begin {tabular} { | c | c | c | c | }
\hline
${Q/M=1.000}$ & $qM = 0$ & $qM = 0.1$  & $qM = 0.2$ \\\hline
$\omega M (\ell=0)$ &
$-0.0062885575 i$ &
$0.014472842 - 0.005654632 i$ &
$0.030032717 - 0.004819872 i$ 
\\\hline
$\omega  M (\ell=0)$ &
$0.11074559 - 0.09520460 i$ &
$0.16061118 - 0.08531414 i$ &
$0.22416137 - 0.06853398 i$ 
\\\hline
$\omega M (\ell=0)$ &
$-0.11074559 - 0.09520460 i$ &
$-0.068172088 - 0.099368915 i$ &
$-0.033349848 - 0.100559182 i$ 
\\\hline
$\omega M (\ell=0)$ &
$-0.116457377 i$ &
$0.10364297 - 0.11968586 i$ &
$0.19901225 - 0.12864834 i$ 
\\\hline
$\omega M (\ell=0)$ &
$-0.24214850 i$ &
$0.10524393 - 0.23691799 i$ &
$-0.01665146 - 0.22014849 i$ 
\\\hline
$\omega M (\ell=0)$ &
$-0.26154517 i$ &
$-0.02627781 - 0.23883927 i$ &
$0.19856772 - 0.22731204 i$ 
\\\hline\hline

$\omega M (\ell=1)$ &
$-0.32961550 - 0.08084325 i$ &
$0.38168229 - 0.07996375 i$ &
$0.43648729 - 0.07836442 i$ 
\\\hline
$\omega M (\ell=1)$ &
$0.32961550 - 0.08084325 i$ &
$-0.28026629 - 0.08100233 i$ &
$-0.23361439 - 0.08046986 i$ 
\\\hline

$\omega M (\ell=1)$ &
$-0.122285898 i$ &
$0.007552960 - 0.122133180 i$ &
$0.015091187 - 0.121675160 i$   
\\\hline

$\omega M (\ell=1)$ &
$-0.23103095 i$ &
$0.10126133 - 0.23069641 i$ &
$0.20256651 - 0.22969902 i$ 
\\\hline

$\omega M (\ell=1)$ &
$0.31077713 - 0.24629967 i$ &
$0.36136078 - 0.24440650 i$ &
$0.41376972 - 0.24044550 i$ 
\\\hline

$\omega M (\ell=1)$ &
$-0.31077713 - 0.24629967 i$ &
$-0.26204705 - 0.24616812 i$ &
$-0.21527645 - 0.24399911 i$
\\\hline\hline

$\omega M (\ell=10)$ &
$2.3277451 - 0.0784452 i$ &
$2.3779731 - 0.0784354 i$ &
$2.4286226 - 0.0784088 i$   
\\\hline
$\omega M (\ell=10)$ &
$-2.3277451 - 0.0784452 i$ &
$-2.2779387 - 0.0784383 i$ &
$-2.2285539 - 0.0784147 i$ 
\\\hline

$\omega M (\ell=10)$ &
$2.3246377 - 0.2354234 i$ &
$2.3748638 - 0.2353939 i$ &
$2.4255085 - 0.2353143 i$ 
\\\hline

$\omega M (\ell=10)$ &
$-2.3246377 - 0.2354234 i$ &
$-2.2748302 - 0.2354027 i$ &
$-2.2254414 - 0.2353319 i$ 
\\\hline

$\omega M (\ell=10)$ &
$2.3184245 - 0.3926658 i$ &
$2.3686468 - 0.3926168 i$ &
$2.4192817 - 0.3924846 i$  
\\\hline

$\omega M (\ell=10)$ &
$-2.3184245 - 0.3926658 i$ &
$-2.2686149 - 0.3926314 i$ &
$-2.2192181 - 0.3925137 i$ 
\\\hline\hline

${Q/M=1.004}$ & $qM = 0$ & $qM = 0.1$  & $qM = 0.2$ 
\\\hline
$\omega M (\ell=0)$ &
$-0.0063034212 i$ &
$0.014529220 - 0.005629515 i$ &
$0.030165920 - 0.004800559 i$ 
\\\hline
$\omega M (\ell=0)$ &
$-0.072957479 i$ &
$0.102298742 - 0.072831564 i$ &
$0.21321901 - 0.06169199 i$ 
\\\hline
$\omega M (\ell=0)$ &
$0.11089045 - 0.09537728 i$ &
$0.16094196 - 0.08715735 i$ &
$0.21160701 - 0.08695099 i$ 
\\\hline
$\omega M (\ell=0)$ &
$-0.11089045 - 0.09537728 i$ &
$-0.068073975 - 0.099456273 i$ &
$-0.033136352 - 0.100630942 i$ 
\\\hline
$\omega M (\ell=0)$ &
$-0.15069275 i$ &
$0.10042711 - 0.15214780 i$ &
$0.19665055 - 0.15458727 i$ 
\\\hline
$\omega M (\ell=0)$ &
$-0.22863879 i$ &
$0.10191145 - 0.22662062 i$ &
$-0.01654354 - 0.22015356 i$ 
\\\hline\hline

$\omega M (\ell=1)$ &
$0.33107901 - 0.08029806 i$ &
$0.38368887 - 0.07917121 i$ &
$0.43912084 - 0.07723394 i$ 
\\\hline
$\omega M (\ell=1)$ &
$-0.33107901 - 0.08029806 i$ &
$-0.28126124 - 0.08063443 i$ &
$-0.23420783 - 0.08022482 i$ 
\\\hline

$\omega M (\ell=1)$ &
$-0.122287640 i$ &
$0.007583360 - 0.122133553 i$ &
$0.015151842 - 0.121671692 i$
\\\hline

$\omega M (\ell=1)$ &
$-0.14814440 i$ &
$0.09937589 - 0.14789775 i$ &
$0.19878038 - 0.14715619 i$  
\\\hline

$\omega M (\ell=1)$ &
$-0.22266534 i$ &
$0.10042705 - 0.22246201 i$ &
$0.20088650 - 0.22185695 i$   
\\\hline

$\omega M (\ell=1)$ &
$0.31128827 - 0.24528678 i$ &
$0.36196809 - 0.24307214 i$ &
$0.41438786 - 0.23873066 i$ 
\\\hline\hline

$\omega M (\ell=10)$ &
$2.3397380 - 0.0778733 i$ &
$2.3905844 - 0.0778286 i$ &
$2.4418691 - 0.0777654 i$ 
\\\hline
$\omega M (\ell=10)$ &
$-2.3397380 - 0.0778733 i$ &
$-2.2893295 - 0.0778996 i$ &
$-2.2393584 - 0.0779075 i$ 
\\\hline

$\omega M (\ell=10)$ &
$2.3364674 - 0.2337186 i$ &
$2.3872992 - 0.2335857 i$ &
$2.4385655 - 0.2333975 i$ 
\\\hline

$\omega M (\ell=10)$ &
$-2.3364674 - 0.2337186 i$ &
$-2.2860698 - 0.2337964 i$ &
$-2.2361060 - 0.2338192 i$ 
\\\hline

$\omega M (\ell=10)$ &
$2.3299308 - 0.3898613 i$ &
$2.3807337 - 0.3896440 i$ &
$2.4319638 - 0.3893350 i$
\\\hline

$\omega M (\ell=10)$ &
$-2.3299308 - 0.3898613 i$ &
$-2.2795547 - 0.3899872 i$ &
$-2.2296050 - 0.3900221 i$
\\\hline

\end {tabular}
}
\end{table}

\begin{table}[H]
\caption {The QNFs $\omega M$ for massive scalar fields in the background of RNdS black holes with $\Lambda M^2= 0.01 $, and different values of $\ell$, and $q M$. Here, for $Q/M =0.999$, $\Delta r /M \approx 0.148$.}
\label {TNE1}\centering
\scalebox{0.7} {

\begin {tabular} { | c | c | c | c |}
\hline
${mM=0.05}$ & $qM = 0$ & $qM = 0.1$  & $qM = 0.2$  \\\hline
$\omega M(\ell=0)$ &
$-0.015076441 i$ &
$0.005152360 - 0.014608818 i$ &
$0.010813814 - 0.013491279 i$    \\\hline
$\omega M(\ell=0)$ &
$-0.073785686 i$ &
$0.102487111 - 0.073674244 i$ &
$0.21016237 - 0.07194639 i$   \\\hline
$\omega M(\ell=0)$ &
$0.12888550 - 0.09470544 i$ &
$0.18034802 - 0.09117840 i$ &
$-0.045716001 - 0.081203854 i$   \\\hline
$\omega M(\ell=0)$ &
$-0.12888550 - 0.09470544 i$ &
$-0.082281358 - 0.091779905 i$ &
$0.23392927 - 0.08357275 i$   \\\hline\hline
$\omega M(\ell=1)$ &
$-0.073126720 i$ &
$0.003130990 - 0.073052324 i$ &
$0.006255023 - 0.072828727 i$    \\\hline
$\omega M(\ell=1)$ &
$-0.36627990 - 0.08741163 i$ &
$0.41774396 - 0.08685347 i$ &
$0.47213504 - 0.08531751 i$   \\\hline
$\omega M(\ell=1)$ &
$0.36627990 - 0.08741163 i$ &
$-0.31773646 - 0.08698589 i$ &
$-0.27212075 - 0.08558335 i$   \\\hline
$\omega M(\ell=1)$ &
$-0.14661132 i$ &
$0.10081482 - 0.14638474 i$ &
$0.20165318 - 0.14570340 i$    \\\hline\hline
$\omega M(\ell=10)$ &
$-2.5515929 - 0.0861901 i$ &
$-2.5019670 - 0.0861703 i$ &
$-2.4527994 - 0.0861289 i$    \\\hline
$\omega M(\ell=10)$ &
$2.5515929 - 0.0861901 i$ &
$2.6016771 - 0.0861883 i$ &
$2.6522194 - 0.0861648 i$   \\\hline
$\omega M(\ell=10)$ &
$-2.5475307 - 0.2587064 i$ &
$-2.4978986 - 0.2586476 i$ &
$-2.4487196 - 0.2585241 i$  \\\hline
$\omega M(\ell=10)$ &
$2.5475307 - 0.2587064 i$ &
$2.5976159 - 0.2587006 i$ &
$2.6481541 - 0.2586303 i$   \\\hline\hline
${mM=0.5}$ & $qM = 0$ & $qM = 0.1$  & $qM = 0.2$  \\\hline
$\omega M(\ell=0)$ &
$0.38043865 - 0.03372638 i$ &
$0.39907548 - 0.03287854 i$ &
$0.42035909 - 0.03241380 i$    \\\hline
$\omega M(\ell=0)$ &
$-0.38043865 - 0.03372638 i$ &
$-0.36369118 - 0.03458732 i$ &
$-0.34829539 - 0.03530051 i$    \\\hline
$\omega M(\ell=0)$ &
$-0.088290287 i$ &
$0.100788526 - 0.087793501 i$ &
$0.20168777 - 0.08628376 i$    \\\hline
$\omega M(\ell=0)$ &
$0.38658447 - 0.10410551 i$ &
$0.40601457 - 0.10325709 i$ &
$0.42708851 - 0.10252671 i$    \\\hline\hline
$\omega M(\ell=1)$ &
$-0.45192323 - 0.04532902 i$ &
$-0.42360814 - 0.03801860 i$ &
$-0.39905840 - 0.03390202 i$    \\\hline
$\omega M(\ell=1)$ &
$0.45192323 - 0.04532902 i$ &
$0.48759377 - 0.05674254 i$ &
$0.53438034 - 0.06577908 i$    \\\hline
$\omega M(\ell=1)$ &
$0.44763864 - 0.10256353 i$ &
$0.46550600 - 0.10180503 i$ &
$-0.40556997 - 0.10045899 i$    \\\hline
$\omega M(\ell=1)$ &
$-0.44763864 - 0.10256353 i$ &
$-0.42657950 - 0.10185139 i$ &
$0.47629753 - 0.10561101 i$    \\\hline\hline
$\omega M(\ell=10)$ &
$-2.5630934 - 0.0856336 i$ &
$-2.5136806 - 0.0855902 i$ &
$-2.4647301 - 0.0855244 i$    \\\hline
$\omega M(\ell=10)$ &
$2.5630934 - 0.0856336 i$ &
$2.6129682 - 0.0856547 i $ &
$2.6633049 - 0.0856536 i$    \\\hline
$\omega M(\ell=10)$ &
$-2.5588917 - 0.2570465 i$ &
$-2.5094626 - 0.2569179 i$ &
$-2.4604902 - 0.2567227 i$    \\\hline
$\omega M(\ell=10)$ &
$2.5588917 - 0.2570465 i$ &
$2.6087773 - 0.2571087 i$ &
$2.6591193 - 0.2571044 i$    \\\hline
\end {tabular}
}
\end{table}\leavevmode\newline

\begin{table}[H]
\caption {The QNFs $\omega M$ for massive scalar fields in the background of RNdS black holes with $\Lambda M^2= 0.04 $, and different values of $\ell$, and $q M$. Here, $Q/M=1.004$, $\Delta r\approx 0.163$, with $\Delta r=r_H-r_C$.}
\label {TNE2}\centering
\scalebox{0.7} {

\begin {tabular} { | c | c | c | c | }
\hline
${mM=0.05}$ & $qM = 0$ & $qM = 0.1$  & $qM = 0.2$ \\\hline
$\omega M (\ell=0)$ &
$-0.0063034212 i$ &
$0.014529220 - 0.005629515 i$ &
$0.030165920 - 0.004800559 i$  \\\hline
$\omega M (\ell=0)$ &
$-0.072957479 i$ &
$0.102298742 - 0.072831564 i$ &
$0.21321901 - 0.06169199 i$   \\\hline
$\omega M (\ell=0)$ &
$0.11089045 - 0.09537728 i$ &
$0.16094196 - 0.08715735 i$ &
$0.21160701 - 0.08695099 i$   \\\hline
$\omega M (\ell=0)$ &
$-0.11089045 - 0.09537728 i$ &
$-0.068073975 - 0.099456273 i$ &
$-0.033136352 - 0.100630942 i$  \\\hline
$\omega M (\ell=0)$ &
$-0.15069275 i$ &
$0.10042711 - 0.15214780 i$ &
$0.19665055 - 0.15458727 i$   \\\hline
$\omega M (\ell=0)$ &
$-0.22863879 i$ &
$0.10191145 - 0.22662062 i$ &
$-0.01654354 - 0.22015356 i$  \\\hline\hline

$\omega M (\ell=1)$ &
$0.33107901 - 0.08029806 i$ &
$0.38368887 - 0.07917121 i$ &
$0.43912084 - 0.07723394 i$  \\\hline
$\omega M (\ell=1)$ &
$-0.33107901 - 0.08029806 i$ &
$-0.28126124 - 0.08063443 i$ &
$-0.23420783 - 0.08022482 i$   \\\hline

$\omega M (\ell=1)$ &
$-0.122287640 i$ &
$0.007583360 - 0.122133553 i$ &
$0.015151842 - 0.121671692 i$  \\\hline

$\omega M (\ell=1)$ &
$-0.14814440 i$ &
$0.09937589 - 0.14789775 i$ &
$0.19878038 - 0.14715619 i$   \\\hline

$\omega M (\ell=1)$ &
$-0.22266534 i$ &
$0.10042705 - 0.22246201 i$ &
$0.20088650 - 0.22185695 i$   \\\hline

$\omega M (\ell=1)$ &
$0.31128827 - 0.24528678 i$ &
$0.36196809 - 0.24307214 i$ &
$0.41438786 - 0.23873066 i$   \\\hline\hline

$\omega M (\ell=10)$ &
$2.3397380 - 0.0778733 i$ &
$2.3905844 - 0.0778286 i$ &
$2.4418691 - 0.0777654 i$   \\\hline
$\omega M (\ell=10)$ &
$-2.3397380 - 0.0778733 i$ &
$-2.2893295 - 0.0778996 i$ &
$-2.2393584 - 0.0779075 i$   \\\hline

$\omega M (\ell=10)$ &
$2.3364674 - 0.2337186 i$ &
$2.3872992 - 0.2335857 i$ &
$2.4385655 - 0.2333975 i$    \\\hline

$\omega M (\ell=10)$ &
$-2.3364674 - 0.2337186 i$ &
$-2.2860698 - 0.2337964 i$ &
$-2.2361060 - 0.2338192 i$   \\\hline

$\omega M (\ell=10)$ &
$2.3299308 - 0.3898613 i$ &
$2.3807337 - 0.3896440 i$ &
$2.4319638 - 0.3893350 i$  \\\hline

$\omega M (\ell=10)$ &
$-2.3299308 - 0.3898613 i$ &
$-2.2795547 - 0.3899872 i$ &
$-2.2296050 - 0.3900221 i$   \\\hline\hline

${mM=0.5}$ & $qM = 0$ & $qM = 0.1$  & $qM = 0.2$  \\\hline
$\omega M  (\ell=0)$ &
$0.29680494 - 0.05239528 i$ &
$0.32758840 - 0.05157525 i$ &
$0.36178412 - 0.05067970 i$  \\\hline
$\omega M (\ell=0)$ &
$-0.29680494 - 0.05239528 i$ &
$-0.26855115 - 0.05314341 i$ &
$-0.24223135 - 0.05379313 i$  \\\hline
$\omega M (\ell=0)$ &
$-0.088661542 i$ &
$0.099384933 - 0.088120920 i$ &
$0.19891007 - 0.08648319 i$  \\\hline
$\omega M (\ell=0)$ &
$0.30148344 - 0.16031946 i$ &
$0.33056528 - 0.15814880 i$ &
$0.36095138 - 0.15524170 i$  \\\hline
$\omega M (\ell=0)$ &
$-0.30148344 - 0.16031946 i$ &
$-0.27378515 - 0.16199674 i$ &
$0.20270242 - 0.16181516 i$  \\\hline
$\omega M (\ell=0)$ &
$-0.16337376 i$ &
$0.10125203 - 0.16297572 i$ &
$-0.24741207 - 0.16332362 i$  \\\hline\hline

$\omega M (\ell=1)$ &
$-0.40765524 - 0.06003450 i$ &
$-0.36821353 - 0.05759284 i$ &
$-0.33208705 - 0.05552967 i$  \\\hline
$\omega M (\ell=1)$ &
$0.40765524 - 0.06003450 i$ &
$0.45089846 - 0.06252130 i$ &
$0.49821172 - 0.06452569 I$  \\\hline

$\omega M (\ell=1)$ &
$-0.15420236 i$ &
$0.09926816 - 0.15396332 i$ &
$0.19856024 - 0.15324502 i$  \\\hline

$\omega M (\ell=1)$ &
$-0.39234255 - 0.16752691 i$ &
$-0.35991155 - 0.16367446 i$ &
$-0.32881920 - 0.16113081 i$  \\\hline

$\omega M (\ell=1)$ &
$0.39234255 - 0.16752691 i$ &
$0.42671404 - 0.17361446 i$ &
$0.46507897 - 0.18325566 i$  \\\hline

$\omega M (\ell=1)$ &
$-0.22833331 i$ &
$0.10027124 - 0.22812889 i$ &
$0.20057408 - 0.22751680 i$  \\\hline\hline

$\omega M (\ell=10)$ &
$2.3500769 - 0.0774728 i$ &
$2.4007457 - 0.0774434 i$ &
$2.4518552 - 0.0773953 i$  \\\hline
$\omega M (\ell=10)$ &
$-2.3500769 - 0.0774728 i$ &
$-2.2998486 - 0.0774834 i$ &
$-2.2500604 - 0.0774754 i$  \\\hline

$\omega M (\ell=10)$ &
$2.3467409 - 0.2325179 i$ &
$2.3973996 - 0.2324309 i$ &
$2.4484952 - 0.2322878 i$  \\\hline
$\omega M (\ell=10)$ &
$-2.3467409 - 0.2325179 i$ &
$-2.2965190 - 0.2325488 i$ &
$-2.2467334 - 0.2325240 i$  \\\hline

$\omega M (\ell=10)$ &
$2.3400720 - 0.3878630 i$ &
$2.3907109 - 0.3877215 i$ &
$2.4417792 - 0.3874869 i$  \\\hline

$\omega M (\ell=10)$ &
$-2.3400720 - 0.3878630 i$ &
$-2.2898622 - 0.3879117 i$ &
$-2.2400815 - 0.3878678 i$  \\\hline

\end {tabular}
}
\end{table}\leavevmode\newline


\clearpage

\begin{thebibliography}{999}



\bibitem{Abbott:2016blz}
 LIGO Scientific and Virgo Collaborations collaboration, B.~P. Abbott
  et~al., Observation of Gravitational Waves from a Binary Black Hole
  Merger,  Phys. Rev. Lett. 116 (2016) 061102.

\bibitem{Abbott:2016nmj}
 VGW151226: Observation of Gravitational Waves from a 22-Solar-Mass
  Binary Black Hole Coalescence,
  Phys. Rev. Lett. 116 (2016) 241103.

\bibitem{Abbott:2017vtc}
{\scshape VIRGO, LIGO Scientific} collaboration, B.~P. Abbott et~al.,
  GW170104: Observation of a 50-Solar-Mass Binary Black Hole Coalescence
  at Redshift 0.2,
  Phys. Rev. Lett.  118 (2017) 221101.

\bibitem{Abbott:2017oio}
 Virgo, LIGO Scientific collaboration, B.~P. Abbott et~al.,
  GW170814: A Three-Detector Observation of Gravitational Waves from a
  Binary Black Hole Coalescence,
  Phys. Rev. Lett. 119 (2017) 141101.

\bibitem{TheLIGOScientific:2017qsa}
 Virgo, LIGO Scientific collaboration, B.~P. Abbott et~al.,
  GW170817: Observation of Gravitational Waves from a Binary Neutron
  Star Inspiral,
  Phys. Rev. Lett. 119 (2017) 161101.

%\cite{Regge:1957td}
\bibitem{Regge:1957td}
T.~Regge and J.~A.~Wheeler,
``Stability of a Schwarzschild singularity,''
Phys. Rev. \textbf{108} (1957), 1063-1069.
%doi:10.1103/PhysRev.108.1063
%1742 citations counted in INSPIRE as of 28 Mar 2022


%\cite{Zerilli:1970wzz}
\bibitem{Zerilli:1970wzz}
F.~J.~Zerilli,
``Gravitational field of a particle falling in a schwarzschild geometry analyzed in tensor harmonics,''
Phys. Rev. D \textbf{2} (1970), 2141-2160
%doi:10.1103/PhysRevD.2.2141
%669 citations counted in INSPIRE as of 26 Jul 2021

%\cite{Nollert:1999ji}
\bibitem{Nollert:1999ji}
H.~P.~Nollert,
``TOPICAL REVIEW: Quasinormal modes: the characteristic `sound' of black holes and neutron stars,''
Class. Quant. Grav. \textbf{16} (1999), R159-R216
%doi:10.1088/0264-9381/16/12/201
%634 citations counted in INSPIRE as of 26 Jul 2021


%\cite{Berti:2009kk}
\bibitem{Berti:2009kk}
E.~Berti, V.~Cardoso and A.~O.~Starinets,
``Quasinormal modes of black holes and black branes,''
Class. Quant. Grav. \textbf{26} (2009), 163001
%doi:10.1088/0264-9381/26/16/163001
[arXiv:0905.2975 [gr-qc]].
%1118 citations counted in INSPIRE as of 26 Jul 2021

%\cite{Konoplya:2011qq}
\bibitem{Konoplya:2011qq}
R.~A.~Konoplya and A.~Zhidenko,
``Quasinormal modes of black holes: From astrophysics to string theory,''
Rev. Mod. Phys. \textbf{83} (2011), 793-836
%doi:10.1103/RevModPhys.83.793
[arXiv:1102.4014 [gr-qc]].
%647 citations counted in INSPIRE as of 26 Jul 2021


%\cite{Kokkotas:1999bd}
\bibitem{Kokkotas:1999bd}
K.~D.~Kokkotas and B.~G.~Schmidt,
``Quasinormal modes of stars and black holes,''
Living Rev. Rel. \textbf{2} (1999), 2
%doi:10.12942/lrr-1999-2
[arXiv:gr-qc/9909058 [gr-qc]].
%1120 citations counted in INSPIRE as of 26 Jul 2021

%\cite{LIGOScientific:2016aoc}
\bibitem{LIGOScientific:2016aoc}
B.~P.~Abbott \textit{et al.} [LIGO Scientific and Virgo],
``Observation of Gravitational Waves from a Binary Black Hole Merger,''
Phys. Rev. Lett. \textbf{116} (2016) no.6, 061102
%doi:10.1103/PhysRevLett.116.061102
[arXiv:1602.03837 [gr-qc]].
%6684 citations counted in INSPIRE as of 26 Jul 2021

%\cite{Konoplya:2004wg}
\bibitem{Konoplya:2004wg}
R.~A.~Konoplya and A.~V.~Zhidenko,
``Decay of massive scalar field in a Schwarzschild background,''
Phys. Lett. B \textbf{609} (2005), 377-384
%doi:10.1016/j.physletb.2005.01.078
[arXiv:gr-qc/0411059 [gr-qc]].
%93 citations counted in INSPIRE as of 26 Jul 2021

%\cite{Konoplya:2006br}
\bibitem{Konoplya:2006br}
R.~A.~Konoplya and A.~Zhidenko,
``Stability and quasinormal modes of the massive scalar field around Kerr black holes,''
Phys. Rev. D \textbf{73} (2006), 124040
%doi:10.1103/PhysRevD.73.124040
[arXiv:gr-qc/0605013 [gr-qc]].
%89 citations counted in INSPIRE as of 26 Jul 2021


%\cite{Dolan:2007mj}
\bibitem{Dolan:2007mj}
S.~R.~Dolan,
``Instability of the massive Klein-Gordon field on the Kerr spacetime,''
Phys. Rev. D \textbf{76} (2007), 084001
%doi:10.1103/PhysRevD.76.084001
[arXiv:0705.2880 [gr-qc]].
%279 citations counted in INSPIRE as of 26 Jul 2021

%\cite{Tattersall:2018nve}
\bibitem{Tattersall:2018nve}
O.~J.~Tattersall and P.~G.~Ferreira,
``Quasinormal modes of black holes in Horndeski gravity,''
Phys. Rev. D \textbf{97} (2018) no.10, 104047
%doi:10.1103/PhysRevD.97.104047
[arXiv:1804.08950 [gr-qc]].
%44 citations counted in INSPIRE as of 26 Jul 2021



%\cite{Lagos:2020oek}
\bibitem{Lagos:2020oek}
M.~Lagos, P.~G.~Ferreira and O.~J.~Tattersall,
``Anomalous decay rate of quasinormal modes,''
Phys. Rev. D \textbf{101} (2020) no.8, 084018
%doi:10.1103/PhysRevD.101.084018
[arXiv:2002.01897 [gr-qc]].
%9 citations counted in INSPIRE as of 26 Jul 2021



%\cite{Aragon:2020tvq}
\bibitem{Aragon:2020tvq}
A.~Arag\'on, P.~A.~Gonz\'alez, E.~Papantonopoulos and Y.~V\'asquez,
``Anomalous decay rate of quasinormal modes in Schwarzschild-dS and Schwarzschild-AdS black holes,''
JHEP \textbf{08} (2020), 120
%doi:10.1007/JHEP08(2020)120
[arXiv:2004.09386 [gr-qc]].
%7 citations counted in INSPIRE as of 26 Jul 2021

%\cite{Aragon:2020teq}
\bibitem{Aragon:2020teq}
A.~Arag\'on, R.~B\'ecar, P.~A.~Gonz\'alez and Y.~V\'asquez,
``Massive Dirac quasinormal modes in Schwarzschild\textendash{}de Sitter black holes: Anomalous decay rate and fine structure,''
Phys. Rev. D \textbf{103} (2021) no.6, 064006
%doi:10.1103/PhysRevD.103.064006
[arXiv:2009.09436 [gr-qc]].
%4 citations counted in INSPIRE as of 26 Jul 2021

%\cite{Destounis:2020pjk}
\bibitem{Destounis:2020pjk}
K.~Destounis, R.~D.~B.~Fontana and F.~C.~Mena,
``Accelerating black holes: quasinormal modes and late-time tails,''
Phys. Rev. D \textbf{102} (2020) no.4, 044005
%doi:10.1103/PhysRevD.102.044005
[arXiv:2005.03028 [gr-qc]].
%10 citations counted in INSPIRE as of 16 Aug 2021




%\cite{Aragon:2020xtm}
\bibitem{Aragon:2020xtm}
A.~Arag\'on, P.~A.~Gonz\'alez, E.~Papantonopoulos and Y.~V\'asquez,
``Quasinormal modes and their anomalous behavior for black holes in $f(R)$ gravity,''
Eur. Phys. J. C \textbf{81} (2021) no.5, 407
%doi:10.1140/epjc/s10052-021-09193-7
[arXiv:2005.11179 [gr-qc]].
%7 citations counted in INSPIRE as of 26 Jul 2021


%\cite{Fontana:2020syy}
\bibitem{Fontana:2020syy}
R.~D.~B.~Fontana, P.~A.~Gonz\'alez, E.~Papantonopoulos and Y.~V\'asquez,
``Anomalous decay rate of quasinormal modes in Reissner-Nordstr\"om black holes,''
Phys. Rev. D \textbf{103} (2021) no.6, 064005
%doi:10.1103/PhysRevD.103.064005
[arXiv:2011.10620 [gr-qc]].
%4 citations counted in INSPIRE as of 30 Jul 2021



%\cite{Aragon:2021ogo}
\bibitem{Aragon:2021ogo}
A.~Arag\'on, P.~A.~Gonz\'alez, J.~Saavedra and Y.~V\'asquez,
``Scalar quasinormal modes for $2+1$-dimensional Coulomb-like AdS black holes from nonlinear electrodynamics,''
Gen. Rel. Grav. \textbf{53} (2021) no.10, 91
%doi:10.1007/s10714-021-02864-6
[arXiv:2104.08603 [gr-qc]].
%3 citations counted in INSPIRE as of 28 Mar 2022

 \bibitem{Penrose:1964wq}
  R.~Penrose,
 ``Gravitational collapse and space-time singularities,''
  Phys.\ Rev.\ Lett.\  {\bf 14}, 57 (1965).
%  doi:10.1103/PhysRevLett.14.57

\bibitem{Penrose:1969pc}
  R.~Penrose,
  ``Gravitational collapse: The role of general relativity,''
  Riv.\ Nuovo Cim.\  {\bf 1}, 252 (1969)
  [Gen.\ Rel.\ Grav.\  {\bf 34}, 1141 (2002)].

%\cite{Cardoso:2017soq}
\bibitem{Cardoso:2017soq}
V.~Cardoso, J.~L.~Costa, K.~Destounis, P.~Hintz and A.~Jansen,
``Quasinormal modes and Strong Cosmic Censorship,''
Phys. Rev. Lett. \textbf{120} (2018) no.3, 031103
%doi:10.1103/PhysRevLett.120.031103
[arXiv:1711.10502 [gr-qc]].
%129 citations counted in INSPIRE as of 16 Aug 2021




%\cite{Cardoso:2018nvb}
\bibitem{Cardoso:2018nvb}
V.~Cardoso, J.~L.~Costa, K.~Destounis, P.~Hintz and A.~Jansen,
``Strong cosmic censorship in charged black-hole spacetimes: still subtle,''
Phys. Rev. D \textbf{98} (2018) no.10, 104007
%doi:10.1103/PhysRevD.98.104007
[arXiv:1808.03631 [gr-qc]].
%52 citations counted in INSPIRE as of 16 Aug 2021




%\cite{Zhu:2014sya}
\bibitem{Zhu:2014sya}
Z.~Zhu, S.~J.~Zhang, C.~E.~Pellicer, B.~Wang and E.~Abdalla,
``Stability of Reissner-Nordstr\"om black hole in de Sitter background under charged scalar perturbation,''
Phys. Rev. D \textbf{90} (2014) no.4, 044042
%doi:10.1103/PhysRevD.90.044042
[arXiv:1405.4931 [hep-th]].
%56 citations counted in INSPIRE as of 26 Jul 2021

%\cite{Destounis:2019hca}
\bibitem{Destounis:2019hca}
K.~Destounis,
``Superradiant instability of charged scalar fields in higher-dimensional Reissner-Nordstr\"om-de Sitter black holes,''
Phys. Rev. D \textbf{100} (2019) no.4, 044054
%doi:10.1103/PhysRevD.100.044054
[arXiv:1908.06117 [gr-qc]].
%24 citations counted in INSPIRE as of 28 Mar 2022

\bibitem{Boyd}
J. P. Boyd, Chebyshev and Fourier Spectral Methods. Dover Books on Mathematics. Dover Publications, Mineola, NY, second ed., 2001.


\bibitem{WM}
 Wolfram “Mathematica 10.” http://www.wolfram.com, 2015.

%\cite{Gonzalez:2021vwp}
\bibitem{Gonzalez:2021vwp}
P.~A.~Gonz\'alez, \'A.~Rinc\'on, J.~Saavedra and Y.~V\'asquez,
``Superradiant instability and charged scalar quasinormal modes for (2+1)-dimensional Coulomb-like AdS black holes from nonlinear electrodynamics,''
Phys. Rev. D \textbf{104} (2021) no.8, 084047
%doi:10.1103/PhysRevD.104.084047
[arXiv:2107.08611 [gr-qc]].
%3 citations counted in INSPIRE as of 28 Mar 2022

\bibitem{Bekenstein}
%\cite{Bekenstein:1973mi}
%\bibitem{Bekenstein:1973mi}
  J.~D.~Bekenstein,
 ``Extraction of energy and charge from a black hole,''
  Phys. Rev. D {\bf 7}, 949 (1973).
  %doi:10.1103/PhysRevD.7.949
  %%CITATION = doi:10.1103/PhysRevD.7.949;%%
  %175 citations counted in INSPIRE as of 22 Jan 2017

\bibitem{Mashhoon}
B. Mashhoon,
``Quasi-normal modes of a black hole,''
Third Marcel Grossmann Meeting on General Relativity 1983.
%        month = jan,
%        pages = {599-608},



%\cite{Schutz:1985zz}
\bibitem{Schutz:1985zz}
  B.~F.~Schutz and C.~M.~Will,
  ``Black Hole Normal Modes: A Semianalytic Approach,''
  Astrophys. J. Lett.  {\bf 291}, L33 (1985).
 % doi:10.1086/184453
  %%CITATION = doi:10.1086/184453;%%
  %358 citations counted in INSPIRE as of 22 Jun 2020





%\cite{Iyer:1986np}
\bibitem{Iyer:1986np}
  S.~Iyer and C.~M.~Will,
  ``Black Hole Normal Modes: A {WKB} Approach. 1. Foundations and Application of a Higher Order {WKB} Analysis of Potential Barrier Scattering,''
  Phys. Rev. D {\bf 35}, 3621 (1987).
  %doi:10.1103/PhysRevD.35.3621
  %%CITATION = doi:10.1103/PhysRevD.35.3621;%%
  %442 citations counted in INSPIRE as of 22 Jun 2020



%\cite{Konoplya:2003ii}
\bibitem{Konoplya:2003ii}
  R.~A.~Konoplya,
  ``Quasinormal behavior of the d-dimensional Schwarzschild black hole and higher order WKB approach,''
   Phys. Rev. D {\bf 68}, 024018 (2003)
 % doi:10.1103/PhysRevD.68.024018
  [gr-qc/0303052].
  %%CITATION = doi:10.1103/PhysRevD.68.024018;%%
  %392 citations counted in INSPIRE as of 22 Jun 2020

%\cite{Matyjasek:2017psv}
\bibitem{Matyjasek:2017psv}
  J.~Matyjasek and M.~Opala,
  ``Quasinormal modes of black holes. The improved semianalytic approach,''
   Phys. Rev. D {\bf 96}, no. 2, 024011 (2017)
 % doi:10.1103/PhysRevD.96.024011
  [arXiv:1704.00361 [gr-qc]].
  %%CITATION = doi:10.1103/PhysRevD.96.024011;%%
  %50 citations counted in INSPIRE as of 22 Jun 2020


%\cite{Konoplya:2019hlu}
\bibitem{Konoplya:2019hlu}
  R.~A.~Konoplya, A.~Zhidenko and A.~F.~Zinhailo,
  ``Higher order WKB formula for quasinormal modes and grey-body factors: recipes for quick and accurate calculations,''
   Class. Quant. Grav. {\bf 36}, 155002 (2019)
 % doi:10.1088/1361-6382/ab2e25
  [arXiv:1904.10333 [gr-qc]].
  %%CITATION = doi:10.1088/1361-6382/ab2e25;%%
  %39 citations counted in INSPIRE as of 22 Jun 2020

%\cite{Hatsuda:2019eoj}
\bibitem{Hatsuda:2019eoj}
Y.~Hatsuda,
``Quasinormal modes of black holes and Borel summation,''
Phys. Rev. D \textbf{101} (2020) no.2, 024008
%doi:10.1103/PhysRevD.101.024008
[arXiv:1906.07232 [gr-qc]].
%35 citations counted in INSPIRE as of 30 Aug 2021

%\cite{Destounis:2018qnb}
\bibitem{Destounis:2018qnb}
K.~Destounis,
``Charged Fermions and Strong Cosmic Censorship,''
Phys. Lett. B \textbf{795} (2019), 211-219
%doi:10.1016/j.physletb.2019.06.015
[arXiv:1811.10629 [gr-qc]].
%28 citations counted in INSPIRE as of 16 Aug 2021

\bibitem{Berti:2003ud}
E.~Berti and K.~D.~Kokkotas,
``Quasinormal modes of Reissner-Nordstr\"om-anti-de Sitter black holes: Scalar, electromagnetic and gravitational perturbations,''
Phys. Rev. D \textbf{67} (2003), 064020
%doi:10.1103/PhysRevD.67.064020
[arXiv:gr-qc/0301052 [gr-qc]].

%\cite{Richartz:2014jla}
\bibitem{Richartz:2014jla}
M.~Richartz and D.~Giugno,
``Quasinormal modes of charged fields around a Reissner-Nordstr\"om black hole,''
Phys. Rev. D \textbf{90} (2014) no.12, 124011
%doi:10.1103/PhysRevD.90.124011
[arXiv:1409.7440 [gr-qc]].
%31 citations counted in INSPIRE as of 16 Aug 2021

%\cite{Richartz:2015saa}
\bibitem{Richartz:2015saa}
M.~Richartz,
``Quasinormal modes of extremal black holes,''
Phys. Rev. D \textbf{93} (2016) no.6, 064062
%doi:10.1103/PhysRevD.93.064062
[arXiv:1509.04260 [gr-qc]].
%24 citations counted in INSPIRE as of 16 Aug 2021


%\cite{Panotopoulos:2019tyg}
\bibitem{Panotopoulos:2019tyg}
G.~Panotopoulos,
``Charged scalar fields around Einstein-power-Maxwell black holes,''
Gen. Rel. Grav. \textbf{51} (2019) no.6, 76.
%doi:10.1007/s10714-019-2560-z
%9 citations counted in INSPIRE as of 28 Mar 2022

%\cite{Liu:2019lon}
\bibitem{Liu:2019lon}
H.~Liu, Z.~Tang, K.~Destounis, B.~Wang, E.~Papantonopoulos and H.~Zhang,
``Strong Cosmic Censorship in higher-dimensional Reissner-Nordstr\"om-de Sitter spacetime,''
JHEP \textbf{03} (2019), 187
%doi:10.1007/JHEP03(2019)187
[arXiv:1902.01865 [gr-qc]].
%29 citations counted in INSPIRE as of 16 Aug 2021







%\cite{Destounis:2019omd}
\bibitem{Destounis:2019omd}
K.~Destounis, R.~D.~B.~Fontana, F.~C.~Mena and E.~Papantonopoulos,
``Strong Cosmic Censorship in Horndeski Theory,''
JHEP \textbf{10} (2019), 280
%doi:10.1007/JHEP10(2019)280
[arXiv:1908.09842 [gr-qc]].
%15 citations counted in INSPIRE as of 16 Aug 2021




%\cite{Destounis:2020yav}
\bibitem{Destounis:2020yav}
K.~Destounis, R.~D.~B.~Fontana and F.~C.~Mena,
``Stability of the Cauchy horizon in accelerating black-hole spacetimes,''
Phys. Rev. D \textbf{102} (2020) no.10, 104037
%doi:10.1103/PhysRevD.102.104037
[arXiv:2006.01152 [gr-qc]].
%8 citations counted in INSPIRE as of 16 Aug 2021

%\cite{Kolyvaris:2010yyf}
\bibitem{Kolyvaris:2010yyf}
T.~Kolyvaris, G.~Koutsoumbas, E.~Papantonopoulos and G.~Siopsis,
``A New Class of Exact Hairy Black Hole Solutions,''
Gen. Rel. Grav. \textbf{43}, 163-180 (2011)
%doi:10.1007/s10714-010-1079-0
[arXiv:0911.1711 [hep-th]].
%53 citations counted in INSPIRE as of 28 Feb 2022

%\cite{Kolyvaris:2011fk}
\bibitem{Kolyvaris:2011fk}
T.~Kolyvaris, G.~Koutsoumbas, E.~Papantonopoulos and G.~Siopsis,
``Scalar Hair from a Derivative Coupling of a Scalar Field to the Einstein Tensor,''
Class. Quant. Grav. \textbf{29}, 205011 (2012)
%doi:10.1088/0264-9381/29/20/205011
[arXiv:1111.0263 [gr-qc]].
%87 citations counted in INSPIRE as of 28 Feb 2022

%\cite{Rinaldi:2012vy}
	\bibitem{Rinaldi:2012vy}
	M.~Rinaldi,
	``Black holes with non-minimal derivative coupling,''
	Phys.\ Rev.\ D {\bf 86}, 084048 (2012)
	% doi:10.1103/PhysRevD.86.084048
	[arXiv:1208.0103 [gr-qc]].
	%%CITATION = doi:10.1103/PhysRevD.86.084048;%%
	%154 citations counted in INSPIRE as of 10 Jan 2020

%\cite{Gonzalez:2013aca}
\bibitem{Gonzalez:2013aca}
P.~A.~Gonz\'alez, E.~Papantonopoulos, J.~Saavedra and Y.~V\'asquez,
``Four-Dimensional Asymptotically AdS Black Holes with Scalar Hair,''
JHEP \textbf{12}, 021 (2013)
%doi:10.1007/JHEP12(2013)021
[arXiv:1309.2161 [gr-qc]].
%59 citations counted in INSPIRE as of 28 Feb 2022

%\cite{Charmousis:2014zaa}
\bibitem{Charmousis:2014zaa}
C.~Charmousis, T.~Kolyvaris, E.~Papantonopoulos and M.~Tsoukalas,
``Black Holes in Bi-scalar Extensions of Horndeski Theories,''
JHEP \textbf{07}, 085 (2014)
%doi:10.1007/JHEP07(2014)085
[arXiv:1404.1024 [gr-qc]].
%101 citations counted in INSPIRE as of 28 Feb 2022

%\cite{Gonzalez:2014tga}
\bibitem{Gonzalez:2014tga}
P.~A.~Gonz\'alez, E.~Papantonopoulos, J.~Saavedra and Y.~V\'asquez,
``Extremal Hairy Black Holes,''
JHEP \textbf{11}, 011 (2014)
%doi:10.1007/JHEP11(2014)011
[arXiv:1408.7009 [gr-qc]].
%15 citations counted in INSPIRE as of 28 Feb 2022





\end{thebibliography}
\end{document}